\documentclass[a4paper,fleqn,usenatbib]{mnras}

\usepackage[T1]{fontenc}
\usepackage{ae,aecompl}

\usepackage{graphicx}	% Including figure files
\usepackage{amsmath}	% Advanced maths commands
\usepackage{amssymb}	% Extra maths symbols
\usepackage{url}
\usepackage{epstopdf}
\usepackage{float}
\usepackage{longtable}
\usepackage{caption}
\usepackage{subcaption}
\usepackage{mathtools}
\usepackage{tablefootnote}
\usepackage{booktabs}
\usepackage{makecell}
\usepackage{geometry}
\usepackage{pdflscape}
\title[Thermal component in the early X-ray afterglows of GRBs]{Thermal components in the early X-ray afterglows of GRBs: likely cocoon emission and constraints on the progenitors}

\author[V. Valan]{
Vlasta Valan,$^{1}$\thanks{E-mail: vlasta2@kth.se} 
Josefin Larsson$^{1}$
Bj\"{o}rn Ahlgren$^{1}$
\\
$^{1}$KTH, Department of Physics, and the Oskar Klein Centre, AlbaNova, SE 106-91 Stockholm, Sweden}

\date{Accepted XXX. Received YYY; in original form ZZZ}

\pubyear{2017}

\begin{document}
\label{firstpage}
\pagerange{\pageref{firstpage}--\pageref{lastpage}}
\maketitle

\begin{abstract}
The early X-ray afterglows of gamma-ray bursts (GRBs) are usually well described by absorbed power laws. However, in some cases, additional thermal components have been identified. The origin of this emission is debated, with proposed explanations including supernova shock breakout, emission from a cocoon surrounding the jet, as well as emission from the jet itself. A larger sample of detections is needed in order to place constraints on these different models. Here we present a time-resolved spectral analysis of 74 GRBs observed by \textit{Swift} XRT in a search for thermal components. We report six detections in our sample, and also confirm an additional three cases that were previously reported in the literature. The majority of these bursts have a narrow range of blackbody radii around $\sim 2 \times 10^{12}$~cm, despite having a large range of luminosities ($L_{\mathrm{peak}} \sim 10^{47} - 10^{51}\  \mathrm{erg\ s^{-1}}$). This points to an origin connected to the progenitor stars, and we suggest that emission from a cocoon breaking out from a thick wind may explain the observations. For two of the bursts in the sample, an explanation in terms of late prompt emission from the jet is instead more likely. We also find that these thermal components are preferentially detected when the X-ray luminosity is low, which suggests that they may be hidden by bright afterglows in the majority of GRBs. 
\end{abstract}


%%%%%%%%%%%%%%%%%%%%%%%%%%%%%%%%%%%%%%%%%%%%%%%%%%

%%%%%%%%%%%%%%%%% BODY OF PAPER %%%%%%%%%%%%%%%%%%

\section{Introduction}
\label{intro}

Gamma-ray bursts (GRBs) are very energetic explosions in distant galaxies.They are divided into two categories based on the duration of their gamma-ray emission: short with duration shorter than $2$ s and long with duration longer than $2$ s. 
Long GRBs are connected with the core-collapse of a massive star. Spectroscopic signatures of SNe have been observed together with  GRBs and a number of spectroscopic and photometric connections between the two phenomena have been made (for review see e.g., \citealt{cano}). All of the SNe connected with GRBs up to date are of Type Ic broad-lined with signatures of neither H nor He.

There are two main phases of GRB emission (see \citealt{kumar2015} for a detailed review). The prompt emission, which is usually observed in gamma-rays, is thought to originate from a relativistic jet. It typically lasts seconds to minutes and is characterised by variable light curves. The afterglow phase, on the other hand, lasts up to years and comes from the interaction between the jet and the circum stellar medium (CSM). It is observed from X-rays to radio wavelengths and is characterized by a more smooth behaviour. The \textit{Swift} satellite, launched in 2004, is the main facility for observing the early X-ray emission of GRBs. The X-ray Telescope (XRT; \citealt{Burrows2005}) observes in soft X-rays (energy range 0.3--10~keV) and the observations typically start $\sim$ 100~s after the trigger and last up to $>$ 1000~s. The canonical XRT light curve is composed of four parts: a steep decay, a shallow decay, a normal decay and a post-jet break component \citep{Zhang2006}. However, all four components are visible for less than half of the observed GRBs \citep{Evans}. 

Only the early parts of the light curve are relevant for the work presented in this paper. The steep decay phase has been suggested to have an origin in high-latitude emission from the jet, as this emission smoothly joins with the tail of the prompt emission (e.g., \citealt{Barthelmy2005}). However, it has been shown that pure high-latitude emission is inconsistent with the observations and that this phase is most likely a combination of the afterglow and the late-time prompt emission \citep{Obrien2006}. The steep decay phase usually transitions to a shallow phase after about 100--1000~s \citep{Zhang2006}. This phase has been suggested to be due to continued energy injection from the central engine \citep{Zhang2001, Zhang2002,Zhang2006}.  Another characteristic feature of the early X-ray emission is flares. This is seen in a large fraction of GRBs and has been explained as the late-time activity of the central engine \citep{Burrows2007}. Much of the diversity in the X-ray light curves of GRBs can likely be explained by varying contributions from prompt and afterglow (i.e. external shock) emission \citep{Willingale2007}. The spectra of the early X-ray light curves are usually well described by power laws with spectral indices ($\beta$) in the range 0.5 -- 1.5 (\citealt{Racusin2009} and their Fig. 2). Here $\beta$ is defined from $F \propto E^{-\beta}$, where $F$ is the flux and $E$ is the energy. The spectral index is related to the photon index ($\Gamma$) as $\beta = \Gamma -1$. The afterglow emission from GRBs is well described by synchrotron emission, which gives rise to a power-law spectrum when obsreved in a limited energy range (see \citealt{Sari1998} and references therein). The emission processes giving rise to the prompt emission are still being investigated (see e.g. \citealt{kumar2015}).
  
In 2006, observations of the nearby, low-luminosity GRB 060218 revealed a thermal component in the early X-ray light curve \citep{campana}. Following this detection, about 15 other GRBs have been reported to have thermal components in their X-ray light curves \citep{page2011, Thone2011, starling2011, sparre,starling2012,mette,Nappo2016}. An origin of this component due to shock breakout is appealing for GRBs with associated SNe, and was suggested for the case of  GRB 060218  \citep{campana}. However, this explanation faces difficulties in terms of the high observed luminosities for all the other reported cases ($\mathrm{L_{BB}} > 10^{47} \mathrm{erg \ s^{-1}}$). Other possible explanations are that the thermal component originates from the cocoon that surrounds the jet \citep{meszaros2001, Peer2006, ghisellini2007b,starling2012}  or the jet itself \citep{mette,Irwin2016,Nappo2016}. Understanding the origin of this emission is important since it can provide new insights into the progenitor of GRBs, their jets and the connection to SNe.

To evaluate different models for the origin of the blackbody component, a larger sample of detections is needed, as well as accurate time-resolved measurements. To this end we here perform a systematic study of 74 GRBs observed by  \textit{Swift} XRT between the beginning of 2011 and the end of 2015. We compare a simple absorbed power-law model, which in most cases describes this phase of GRBs well, with an absorbed power-law plus blackbody model. Using Monte-Carlo simulations to assess the significance of the blackbody, we identify 6 new cases. We also investigate the prompt emission where \textit{Swift} Burst alert telescope (BAT) and/or \textit{Fermi} Gamma-ray burst monitor (GBM; \citealt{gbm}) data are available, as well as the conditions needed for detecting a thermal component.

This paper is organized as follows. In Section 2, the sample selection is presented, followed by data reduction and analysis methods in Section 3 and Section 4, respectively. We present our results in Section 5, address GRBs with previous detections of thermal components in Section 6, while Section 7 and Section 8 are devoted to discussion and conclusions, respectively. In all of our calculations we assume a flat Universe with recent cosmological parameters from Planck ($H_{0} = 67.3 \ \mathrm{km \ s^{-1}\  Mpc^{-1}}$, $\Omega_{M} = 0.315$, $\Omega_{\Lambda} = 0.685$, \citealt{planck2014}). All given error-bars are $1 \sigma$.

\section{Sample selection}
\label{sample}

GRBs that meet the following criteria were included in the sample:

\begin{itemize}

\item Known spectroscopic redshift
\item \textit{Swift} XRT window timing (WT) mode data available
\item Observed between 2011-01-01 and 2015-12-31
\item Observed time-averaged WT mode flux ($F_{\mathrm{av, 0.3-10~keV}}$) higher than $2 \ \times \ 10^{-10} \ \mathrm{erg \ cm^{-2} \ s^{-1}}$

\end{itemize}

The resulting sample (see Table \ref{bursts}) includes 74 GRBs covering a redshift range $0.282 - 6.32$ and a flux range $2.0 \ \times \ 10^{-10} \ \mathrm{erg \ cm^{-2} \ s^{-1}}  \leq \mathrm{F_{av, 0.3-10~keV}} \leq \ 1.5 \ \times \ 10^{-8}  \ \mathrm{erg \ cm^{-2} \ s^{-1}} $. GRBs observed before 2011 were not considered since these bursts have already been included in previous systematic searches for blackbody components (see eg. \citealt{starling2012, mette}).  The flux limit of $F_{\mathrm{av, 0.3-10~keV}} > 2 \ \times \ 10^{-10} \ \mathrm{erg \ cm^{-2} \ s^{-1}}$ was set after initial testing, which showed that we were not able to obtain meaningful constraints from fainter bursts.

All GRBs that meet the above criteria are long, including four ultra-long GRBs:  GRB 111209A, GRB 121027A, GRB 130925A and GRB 141121A. The sample also includes six GRBs with associated SN  as reported in \cite{cano} (GRB~111209A, GRB~111228A, GRB~120422A, GRB~120729A, GRB~130427A, GRB~150818A) and one GRB with a hint of a SN bump in the late optical light curve: GRB~151027A \citep{Nappo2016}. Out of these bursts, GRB~130427A was excluded from further analysis since it was observed during an attitude anomaly of the spacecraft.   
 
\begin{table}
    \centering
    \caption{Analysed GRBs with spectroscopic redshifts. Values of redshifts are taken from this webpage.\tablefootnote{\url{http://www.mpe.mpg.de/~jcg/grbgen.html}} They are usually reported without error bars.} 
    \label{bursts}
    \begin{tabular}{c c|c c } % four columns, alignment for each
        \hline
        GRB  & z & GRB & z \\ 
        \hline
        110422A & 1.77 & 130606A & 5.913 \\
        110503A & 1.613 & 130701A & 1.155 \\
        110709B & 2.09 & 130907A & 1.238 \\
        110715A & 0.82 & 130925A	& 0.347 \\
        110726A	 & 2.7 & 131030A	& 1.295 \\
        110731A & 2.83 & 131103A	& 0.599 \\
        110801A & 1.858 & 131117A	& 4.042 \\ 
        110808A & 1.348 & 140114A	& 3 \\
        111008A	& 4.9898 & 140206A	& 2.73 \\
		111107A	& 2.893 & 140301A	& 1.416 \\
		111123A	& 3.1516 & 140304A	& 5.3 \\
		111209A	& 0.677 & 140318A	& 1.02 \\
		111215A	& 2.06 & 140419A	& 3.956 \\
		111225A	& 0.297 & 140430A	& 1.6 \\
		111228A	& 0.714 & 140506A	& 0.889 \\
		120119A	& 1.728 & 140512A	& 0.725 \\
		120308A	& 4.5 & 140518A	& 4.707 \\
		120326A	& 1.798 & 140614A	& 4.233 \\
		120327A	& 2.813 & 140703A	& 3.14 \\
		120404A	& 2.876 & 141026A	& 3.35 \\
		120422A	& 0.283 & 141109A	& 2.993 \\
		120521C	& 6 & 141121A	& 1.470 \\
		120711A	& 1.405 & 141220A	& 1.3195 \\
        120729A	& 0.8 & 141221A	& 1.452 \\
		120805A	& 3.1 & 141225A	& 0.915 \\
		120811C	& 2.671 & 150314A	& 1.758 \\
		120922A	& 3.1 & 150323A	& 0.593 \\
		121024A	& 2.298 & 150403A	& 2.06 \\
		121027A	& 1.773 & 150727A	& 0.313 \\
		121128A	& 2.2 & 150818A	& 0.282 \\
		121211A	& 1.023 & 150821A	& 0.755 \\
		121229A	& 2.707 & 150910A	& 1.359 \\
		130418A	& 1.218 & 150915A	& 1.968 \\
		130427B	& 2.78 & 151021A	& 2.33 \\
		130505A	& 2.27 & 151027A	& 0.81 \\
		130514A	& 3.6 & 151111A	& 3.5 \\
		130604A	& 1.06 & & \\		
        \hline
    \end{tabular}
\end{table}

\section{Data reduction} \label{datareduction}
\subsection{\bf XRT data}
\textit{Swift} XRT has two modes of operation: WT and photon counting (PC) mode. WT mode has better time resolution and is used when the count rate is high (generally in the beginning of the XRT observation), while the PC mode is used for lower count rates and has a lower time resolution. We focus our analysis on WT data, but also extract PC spectra for a subset of GRBs in order to compare methods for determining columns densities (see Section \ref{analysis}).

 The XRT observational data were downloaded from the UK \textit{Swift} Science Data Centre XRT GRB repository\footnote{\url{www.swift.ac.uk/xrt_live_cat}} in January 2016 as locally reprocessed XRT data. The data reduction was performed using {\sc HEASoft} version 6.18, following \cite{Evans} with updates according to the latest release notes\footnote{\url{http://www.swift.ac.uk/analysis/Gain_RMF_releases.html}}. In order to perform a time-resolved analysis we first binned the light curves with Bayesian blocks \citep{scargle}, using the routine {\sc battblocks} with its default settings. We then extracted spectra in each of the blocks using the standard {\sc FTOOLs} \textit{Swift} package {\sc xrtproducts}, providing the gti file containing the start and stop times of the Bayesian blocks as input. Source and background spectra were extracted from boxes of equal height. All XRT spectra were grouped to contain a minimum of 20 counts per bin in order for $\chi^2$ statistics to be used. This is needed when performing joint fits with BAT data, which have Gaussian errors.

\paragraph*{{\bf Pile-up}} WT data with count rates above 100 counts~s$^{-1}$ are likely to be affected by pile-up. In time intervals where the count rate exceeded this value we used the procedure described in Appendix A of \cite{pileup} to correct for it. In some cases the count rate still exceeded 100 counts~s$^{-1}$ after excluding the number of pixels from the core as described in \cite{pileup}. For those time intervals we excluded additional pixels until the count rate dropped below the limit. Table \ref{pile-up} in Appendix \ref{pileupapp} lists all GRBs that were affected by pile-up and the number of pixels excluded from each time interval. To check our results we also extracted time-resolved spectra of five GRBs that suffered from pile-up directly from the UKSSDC webpage, where the pile-up correction is performed automatically using a slightly different procedure that involves fitting a King function \citep{Moretti2005} to the PSF and marking the point at which the function no longer fits the data (for details see the pile-up walk through guide on the Swift webpage\footnote{\url{http://www.swift.ac.uk/analysis/xrt/pileup.php}}). The results from fitting these spectra were found to be compatible within the error-bars with the fits to our manually corrected spectra. 

\paragraph*{{\bf Redistribution issues}} For the majority of the bursts we used standard redistribution matrix files (RMFs) (the appropriate RMF file for each burst was selected based on Table P1 in the "Swift Calibration release note: $\mathrm{SWIFT-XRT-CALDB-09-v19}$") in our analysis. However, we note that there are uncertainties associated with the instrument response for some observations, as described in the "XRT Calibration Digest''\footnote{\url{http://www.swift.ac.uk/analysis/xrt/digest_cal.php}}. One known problem is that moderately to highly absorbed sources are prone to redistribution issues at low energies (seen as a bump in the spectra below $\sim 1$~keV). For GRBs where the absorbing column density derived from fits to a power-law model was above $0.5 \times 10^{22} \ \rm{cm^{-2}}$ we performed additional checks for these redistribution issues. A total of 15 GRBs had column densities exceeding this value. We extracted the spectra of these GRBs using Grade 0 events only, and compared them to the spectra extracted using Grade 0-2 events. The comparison of the spectra did not reveal any problems with a bump at low energies (an example is shown in Fig. \ref{nobump} in Appendix \ref{rmfissues}) and we therefore continued to use the spectra produced using Grade 0-2 events.

Another redistribution effect that is important in moderately and highly absorbed sources is the appearance of a 'turn up' at energies below $\sim 0.6$~keV. In total 12 GRBs (out of the 15 GRBs with absorbing column densities higher than $0.5 \times 10^{22} \ \rm{cm^{-2}}$) showed this turn-up issue. To account for this we selected position-dependent RMFs for the relevant epochs and compared the fit statistics obtained when fitting an absorbed power law using different RMFs. The RMF that gave the lowest fit statistic was then used in the analysis. A table including the fit statistics for different RMFs is provided in Appendix \ref{rmfissues} as Table \ref{RMFs}. The position dependent RMFs assume that the center of the source lies on (psf1\footnote{psf - point spread function}), 2.5 pixels away (psf2) or 5 pixels away (psf3) from the 10 pixel boundary, while the standard RMF assumes a uniform density distribution for the source. We note that the effects of using the position-dependent RMFs are relatively small, with all the best-fitting parameters being compatible within the one-sigma error bars.

\subsection{\bf BAT data} For some GRBs in our sample, \textit{Swift} BAT data were available during the beginning of the XRT observations. These data were analysed using the Burst analyser software described in \cite{BAT} and spectra were produced following the procedure described in "The SWIFT BAT Software Guide."\footnote{\url{http://swift.gsfc.nasa.gov/analysis/bat_swguide_v6_3.pdf}}

\section{Analysis} \label{analysis}

The spectral analysis was carried out using {\sc XSPEC} version 12.9.0 \citep{xspec}. The energy ranges used for fitting were  0.3--10~keV for XRT and 15--150~keV for BAT. In some cases there were no counts in the XRT above an energy smaller than 10~keV, and in these cases the upper energy range for the fit was adjusted accordingly. 

It should be noted that by using Bayesian blocks to define the time intervals for the spectral analysis, we obtain a significantly higher time resolution than most previous studies. Our approach also differs from previous works in that we do not exclude any time intervals that include flares or other untypical behaviour in the light curve. This is motivated by the fact that the time-resolved analysis ensures that we do not have a problem of photons from different emission phases of the light curve contributing to the same spectrum. In addition, all models considered for the origin of the blackbody component predict a rise in flux followed by a decay, which would be seen as a flare in the light curve if other emission components are not too strong. Previous studies of flares have shown that the spectra are often well described by power laws, but that more complex models, for examle including blackbodies, are required in some cases \citep{Falcone2007,Peng2014}.

Below about 2~keV, Galactic as well as intrinsic H column densities are important. These absorptions are encoded into the \textit{tbabs} and \textit{ztbabs} models in XSPEC \citep{Wilms2000}. When using these models, we set the Solar abundance vector to the values used in \cite{Wilms2000} and the cross section table  to the one used in \cite{Verner1996}. The Galactic column density ($N_{\mathrm{H,Gal}}$) for each burst was calculated using the tool $\rm{N_{H,tot}}$\footnote{\url{http://www.swift.ac.uk/analysis/nhtot/}}, which includes contributions from both atomic and molecular H to the absorption, as described in \cite{Willingale2013}. In bursts where the value of the molecular component differed by more than $\pm 10$ per cent from the $20$ per cent that the \textit{tbabs} model assumes, we used the \textit{tbvarabs} model in order to manually set the value of atomic and molecular H absorption (this was the case for 4 GRBs: GRB~121211A, GRB~130514A, GRB~140114A and GRB~150818A). The Galactic absorption was kept fixed in all fits.

In order to determine the intrinsic column density ($N_{\mathrm{H,intr}}$) we made the assumption of constant column density during the observation (typically less than 1000~s). We assessed different ways of determining $N_{\mathrm{H,intr}}$ by comparing the results for different parts of the light curve, represented by WT and PC mode data. For WT data, $N_{\mathrm{H,intr}}$ was determined by simultaneously fitting all time intervals with $N_{\mathrm{H,intr}}$ tied but all other parameters free to vary. The value of $N_{\mathrm{H,intr}}$ was derived separately depending on the model (power law or power law + blackbody) in order to ensure that the presence of a blackbody component did not result in an inflated value of $N_{\mathrm{H,intr}}$ when fitting with only a power law. In contrast, when deriving $N_{\mathrm{H,intr}}$ from late-time PC data, the time-averaged spectrum was fitted with a power law due to the significantly lower count rates and the fact that less spectral evolution is expected during this phase. 

\begin{figure}
	\includegraphics[width=\columnwidth]{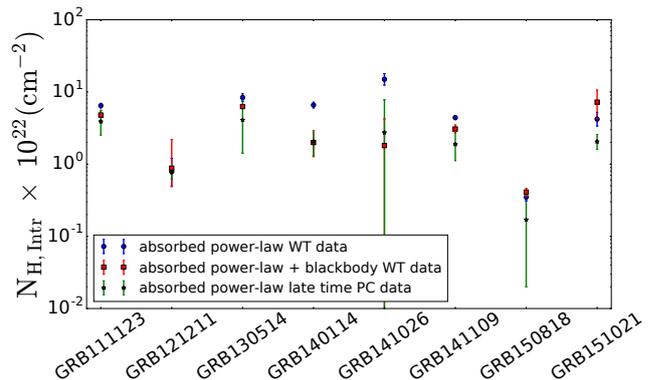}
	\caption{Comparison between intrinsic column densities determined in different ways. Values from time-resolved fits to WT data using an absorbed power-law and an absorbed power-law + blackbody are are shown as blue and red dots, respectively. Green dots show values from fits to late time PC data.}
	\label{intrinsic}
\end{figure}

We tested the different approaches on eight randomly selected bursts and present the resulting values of $N_{\mathrm{H,intr}}$ in Fig. \ref{intrinsic}. From the figure it is clear that $N_{\mathrm{H,intr}}$ derived from the different XRT observing modes are mostly compatible within error-bars, but that the $N_{\mathrm{H,intr}}$ derived from WT data is better constrained due to to the higher count rates in this mode. We also note that there are some small differences in $N_{\mathrm{H,intr}}$ obtained from the two models fitted to WT data. Based on these results, we use the $N_{\mathrm{H,intr}}$ values derived from WT data in the following analysis. The value determined for each model  was kept fixed in the fits. The impact of $N_{\mathrm{H,intr}}$ on the results is discussed further in Section \ref{nh}.  

To search for the presence of a blackbody we fitted all spectra with both an absorbed power law and an absorbed power law together with a blackbody at the redshift of the host galaxy (\textit{pow+zbb}). In XSPEC, the power-law model is defined as $A(E) = K E^{-\Gamma}$, where $\Gamma$ is the photon index and K is the normalization constant. The parameters of the \textit{zbb} model are the redshift, temperature and normalisation. While we base our results on the time-resolved spectra, we also fitted the time-averaged spectra of the entire WT mode for comparison. Differences between time-resolved and time-averaged spectra will be discussed in Section \ref{timav}. 

In cases where BAT data were available simultaneously with XRT (in total six bursts) we fitted the spectra from both detectors together. In these fits we replaced the absorbed power-law model with an absorbed cutoff power-law model in order to account for the possible peak of the prompt emission in the BAT energy range. This was done in fits both with and without an added blackbody. These fits offer a way to investigate possible curvature in the XRT spectrum due to late prompt emission with a low peak energy. We note we were only able to constrain the peak energy of the cutoff power law in the the time-averaged spectra of the overlap region between BAT and XRT. This is because the BAT spectra have very poor statistics when extracted in the Bayesian blocks identified from the XRT data. 

In these fits we found that the peak energy of the cutoff power-law model was always well outside the XRT energy range ($95~\mathrm{keV} \le E_\mathrm{pk} \le 250~\mathrm{keV}$) and hence did not influence the results on the presence of a blackbody component. Therefore, in the rest of the analysis we will discuss only the absorbed power-law vs. the absorbed power-law plus blackbody model, fitted to XRT data. Finally, if the difference in $\chi^2$ between the two models was higher than the difference in the number of degrees of freedom, significance testing was performed as described below.

\subsection{Significance testing}

In order to assess the significance of the blackbody component we performed Monte Carlo simulations (we note that the F-test is not valid for testing the significance of an additional component, see \citealt{protassov}). For each spectrum, we used the XSPEC fakeit command to simulate 10~000 spectra based on the best-fitting parameters of the absorbed power-law model, using the response files of the real data. Background spectra were also simulated based on the original background files. Simulations were carried out for 38 GRBs, i.e. about half of the sample, and in 30 of these, the significance was tested in all time-resolved spectra. 

Just like for the real data, $N_{\mathrm{H,intr}}$ was first determined by simultaneously fitting the time-resolved spectra of each burst with $N_{\mathrm{H,intr}}$ tied, but the other parameters free to vary. For this step we used a subset of the simulated spectra. Specifically, we used 10 randomly selected spectra from 5 time bins, resulting in 50 spectra for each burst.  In the real data the number of time bins per GRB varied between 3 and 32 and $N_{\mathrm{H,intr}}$ had an average one-sigma error of 30~per~cent. The $N_{\mathrm{H,intr}}$ derived from the faked data using 50 spectra had an average uncertainty of 22~per~cent. With 50 simulated spectra we thus determine $N_{\mathrm{H,intr}}$ with at least the same level of confidence as for the real data. We verified that the value of $N_{\mathrm{H,intr}}$ derived in this way was robust by checking that selecting different spectra for the fits gave consistent results. After determining $N_{\mathrm{H,intr}}$, each of the 10~000 simulated spectra was fitted with the two models described above. The significance of the blackbody component was finally determined by comparing the resulting distribution of  $\Delta \chi^2$ with the $\Delta \chi^2$ from the fits to the real data. 

We note that there are some caveats in using $\chi^2$ statistics when analysing XRT data (for details see \citealt{Humphrey2009}). Therefore, we also performed the fitting and significance testing using Cash statistics for a representative sub-sample of bursts. We found that the results did not change when Cash statistics was used.

\section{Results} \label{withthermal}

We have identified in total six GRBs in which a thermal component is significant ( GRB~111123A, GRB~111225A, GRB~121211A, GRB~131030A, GRB~150727A and GRB~151027A). This was based on the criterion that the blackbody component should be significant at $ > 3 \sigma$ in at least three consecutive time bins. We use this rather strict criterion in order to exclude spurious detections and to ensure that we can use the time-evolution of the parameters to draw conclusions about the origin of the emission.  To the best of our knowledge, only GRB 151027A has a previously reported thermal component \citep{Nappo2016}, while the others are presented here for the first time. We note that the detection of the blackbodies rely on the assumption that the underlying spectrum is an absorbed power law. This assumption and the properties of the power law are discussed further in section \ref{discussion}.

\begin{figure*}
    \centering
    \begin{subfigure}[b]{0.49\textwidth}
        \includegraphics[width=\textwidth]{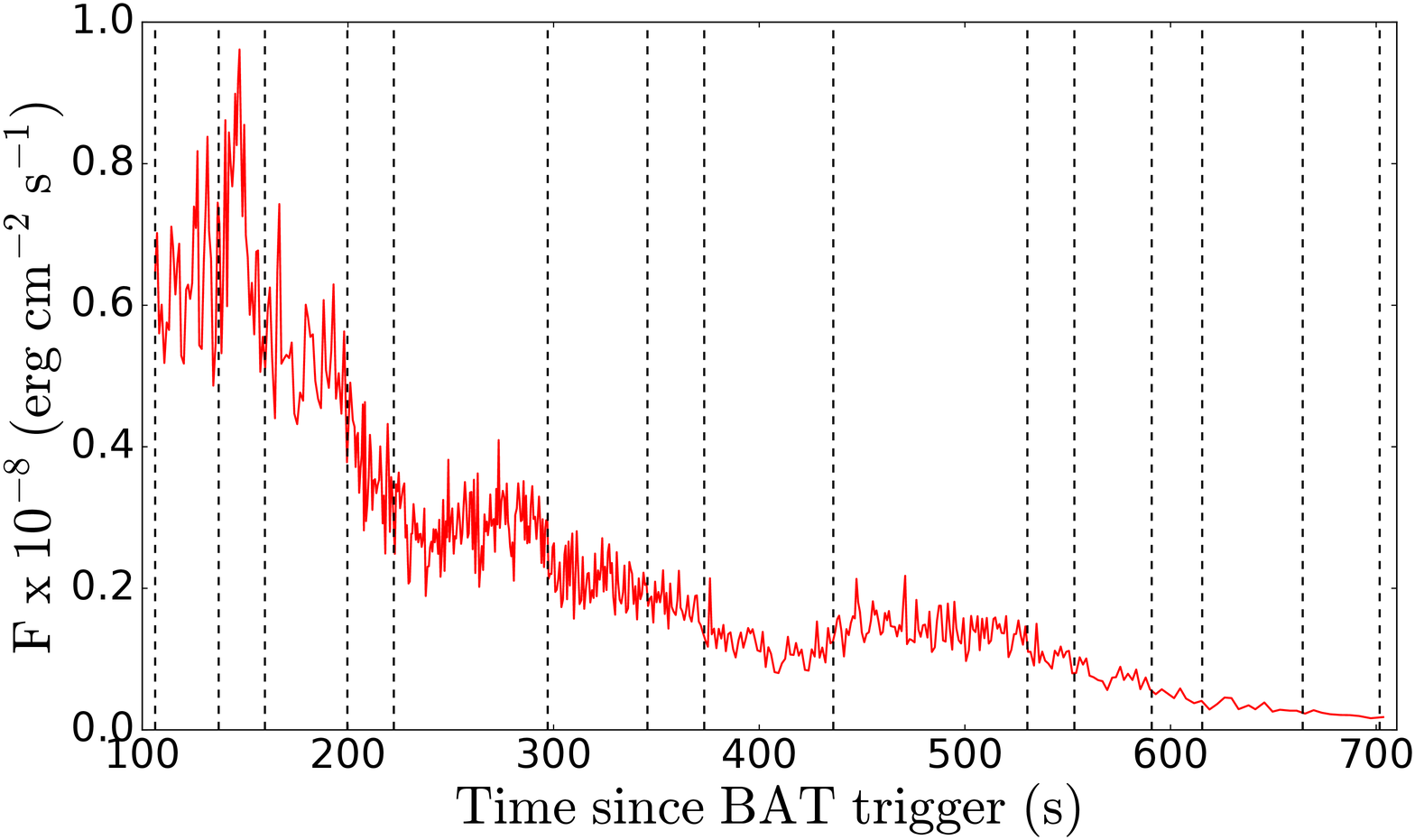}
    \end{subfigure}
    \begin{subfigure}[b]{0.49\textwidth}
        \includegraphics[width=\textwidth]{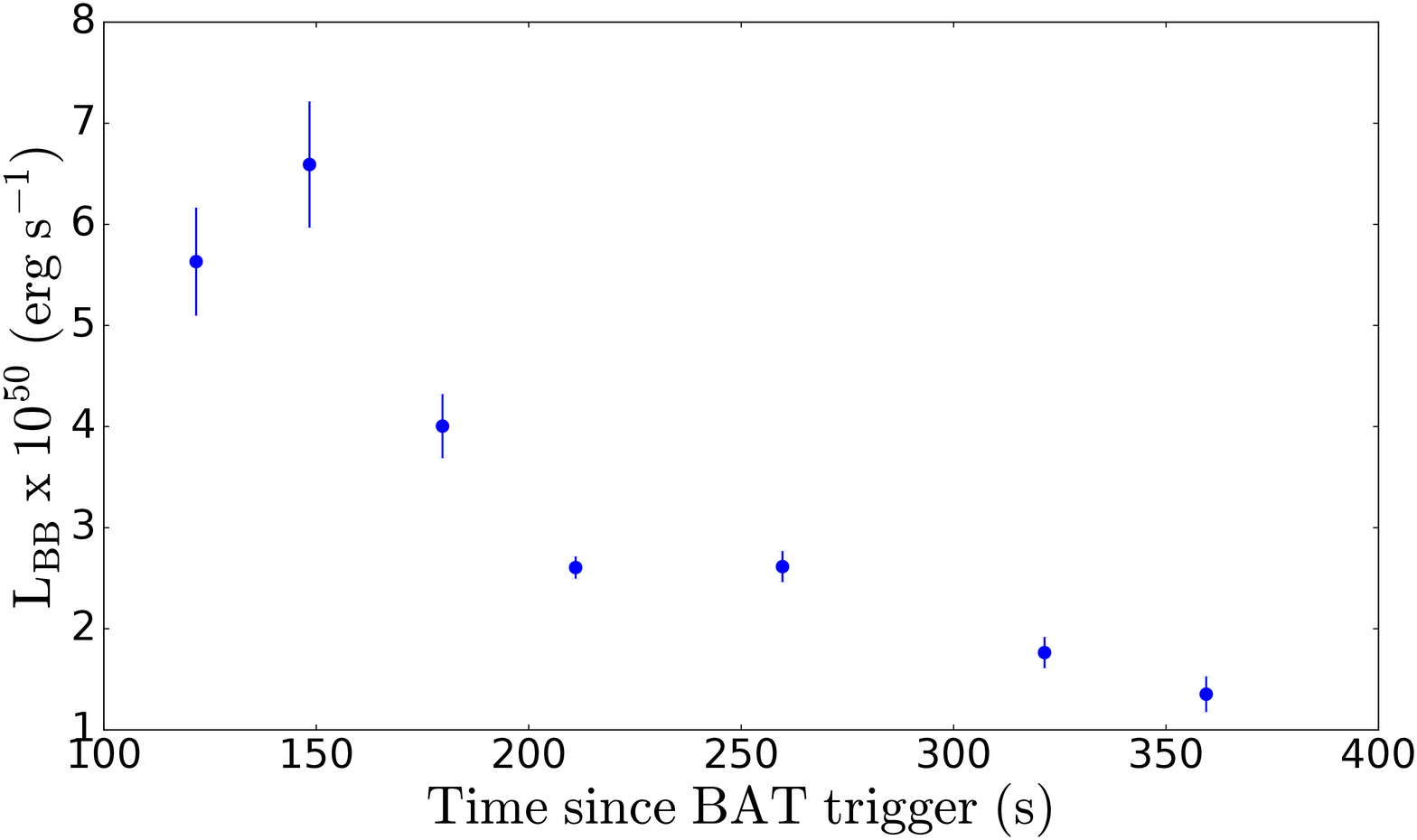}
    \end{subfigure}
        \begin{subfigure}[b]{0.49\textwidth}
        \includegraphics[width=\textwidth]{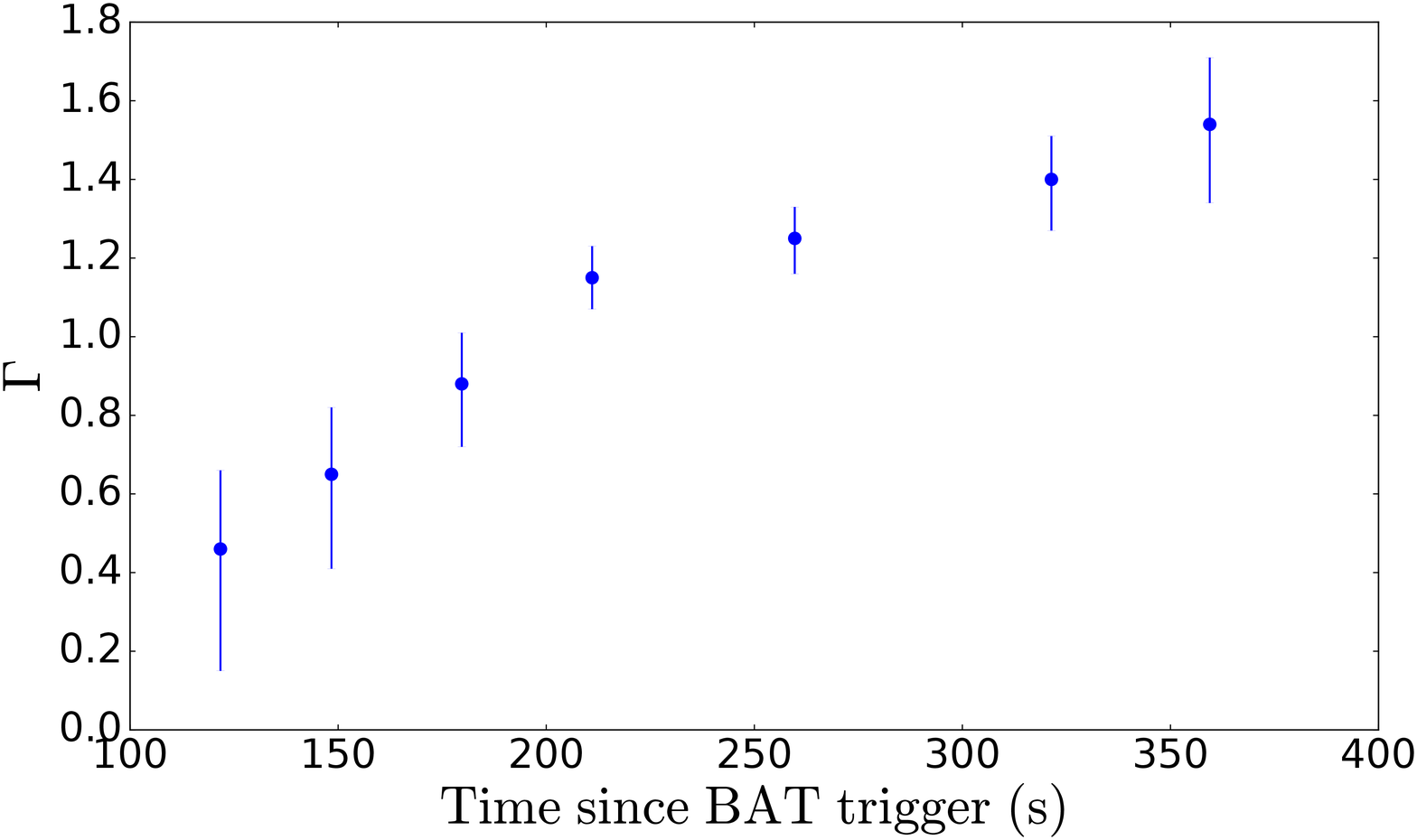}
    \end{subfigure}
        \begin{subfigure}[b]{0.49\textwidth}
        \includegraphics[width=\textwidth]{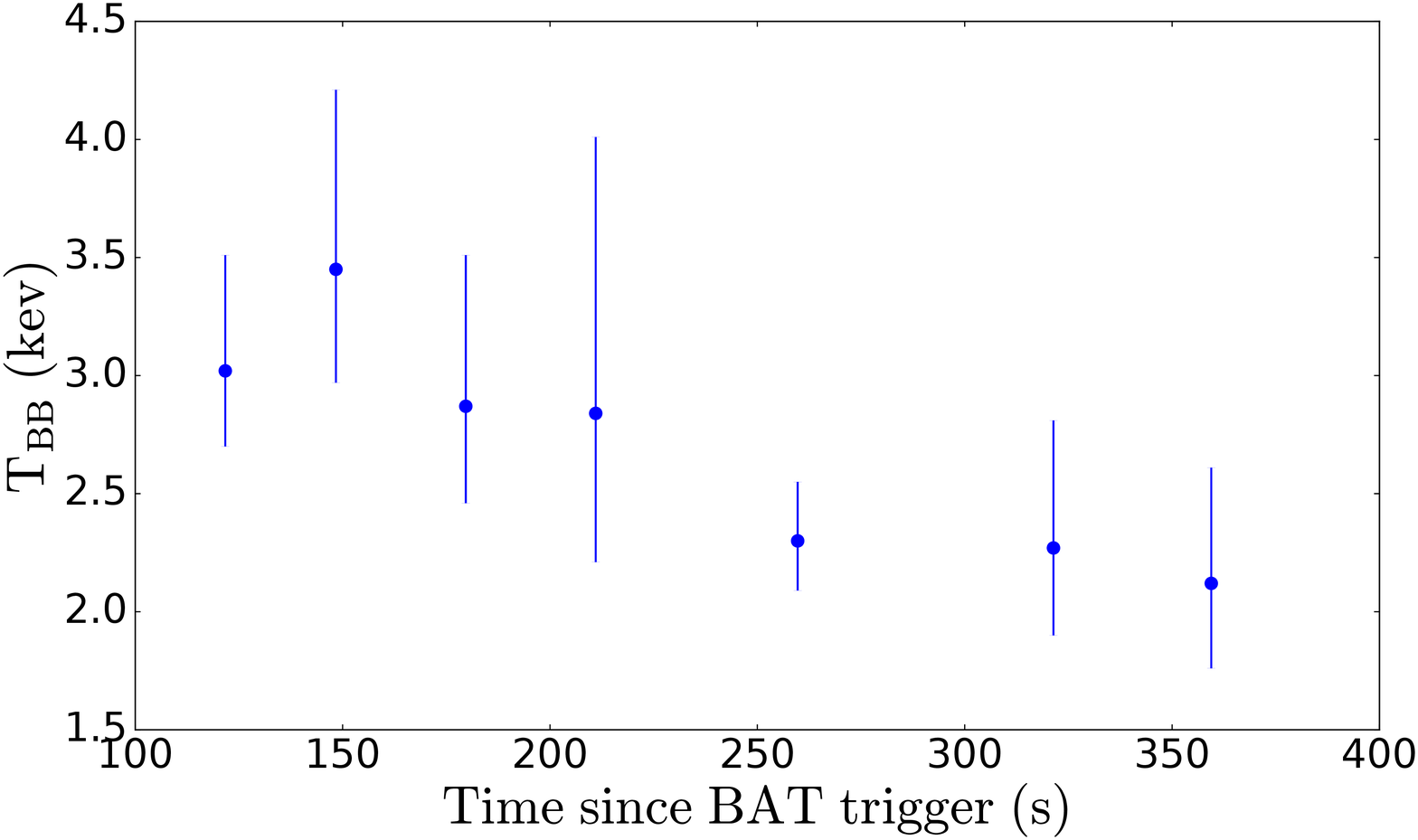}
    \end{subfigure}
    \caption{Light curve and time evolution of the best-fitting parameters for the power law + blackbody model for GRB 111123A. Top left: light curve with Bayesian blocks marked as dashed black lines. Top right: luminosity of the blackbody. Bottom left: photon index. Bottom right: blackbody temperature. }
    \label{111123}
\end{figure*}

\begin{figure*}
    \begin{subfigure}[b]{0.49\textwidth}
    \includegraphics[width=\columnwidth]{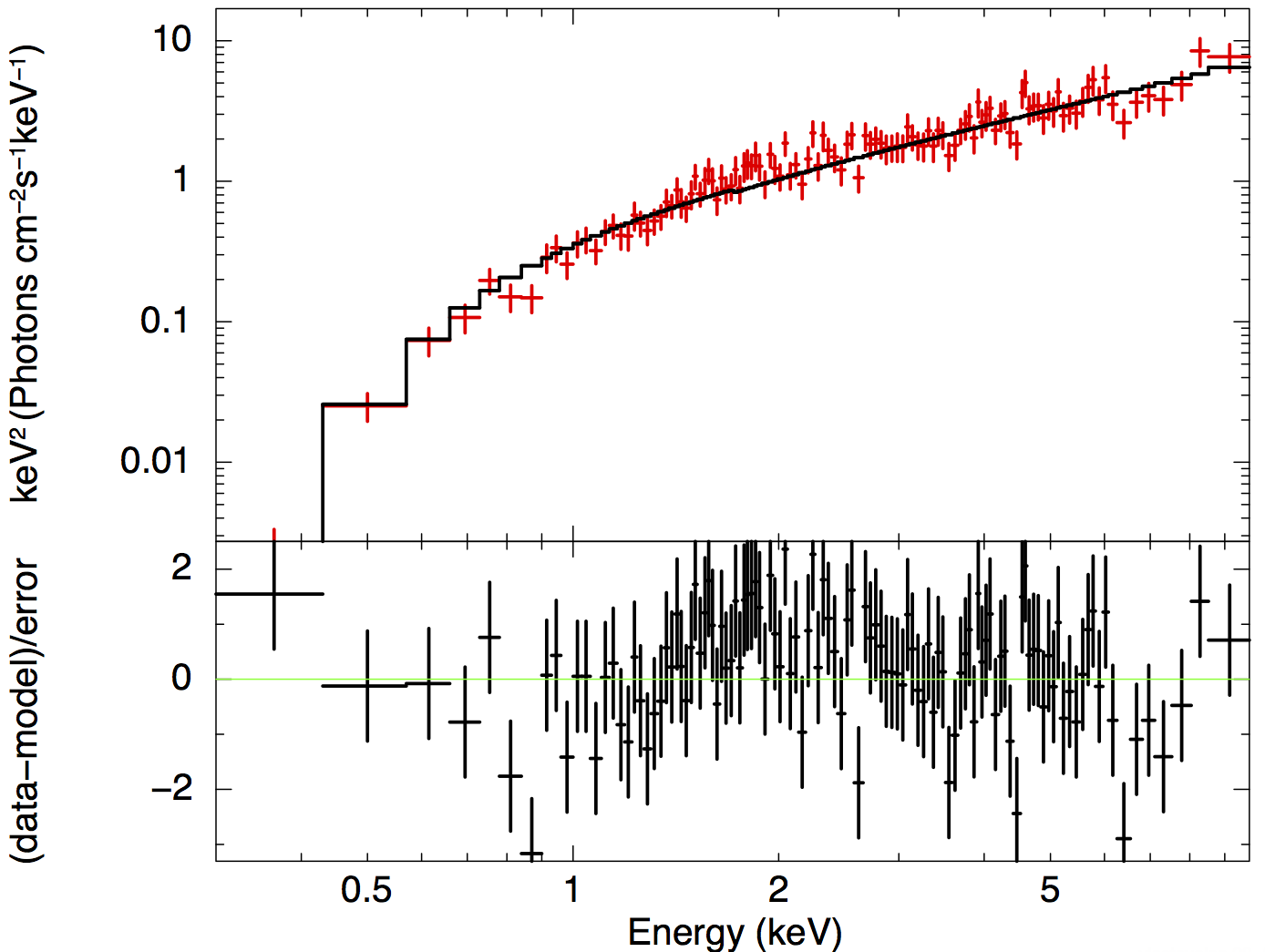}
    \end{subfigure}
    \begin{subfigure}[b]{0.49\textwidth}
     \includegraphics[width=\columnwidth, height = 6.66cm]{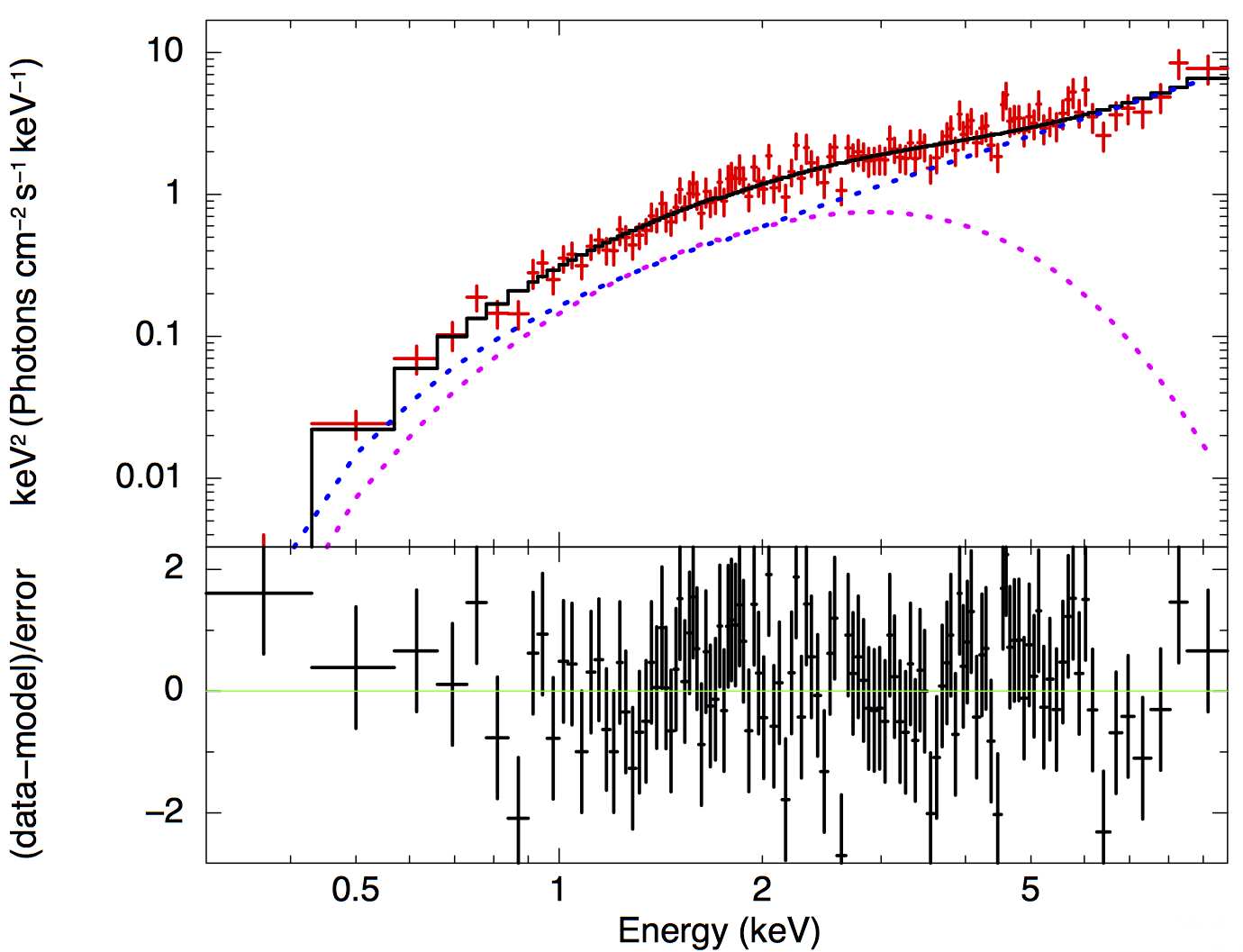}
    \end{subfigure}
    \caption{Fits to the spectrum of GRB 111123A in the time interval 106--137~s. Left panel: Fit to the absorbed power-law model. The lower panel shows the residuals to the fit. Right panel: Fit to the absorbed power law + blackbody model. The individual model components are shown as dashed blue and magenta lines for power law and blackbody, respectively, and the total model is shown in black. The lower panel shows the residuals of the fit.}
	\label{grb111123fit}
\end{figure*}

\begin{figure*}
    \centering
    \begin{subfigure}[b]{0.49\textwidth}
        \includegraphics[width=\textwidth]{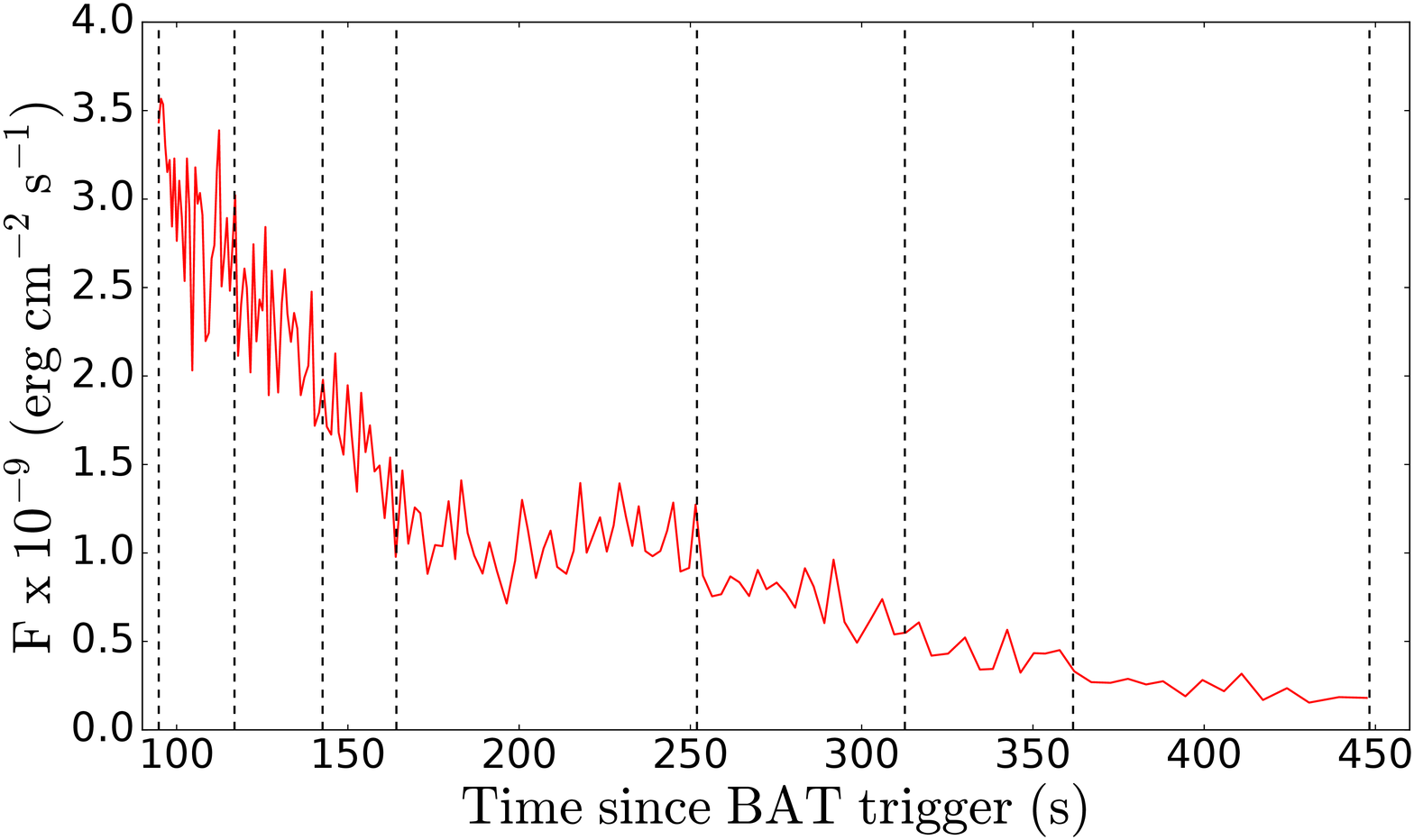}

    \end{subfigure}
    \begin{subfigure}[b]{0.49\textwidth}
        \includegraphics[width=\textwidth]{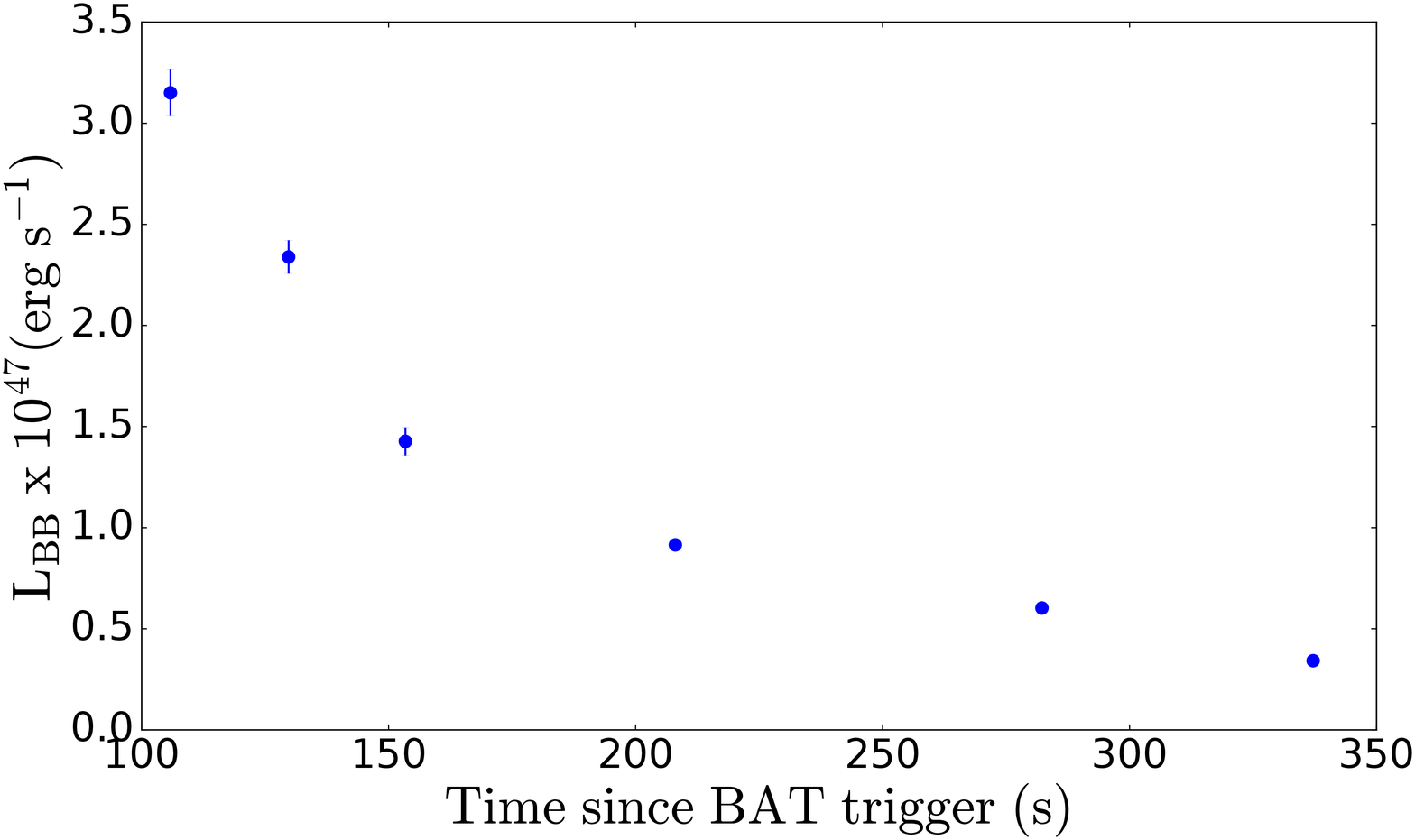}
    \end{subfigure}
        \begin{subfigure}[b]{0.49\textwidth}
        \includegraphics[width=\textwidth]{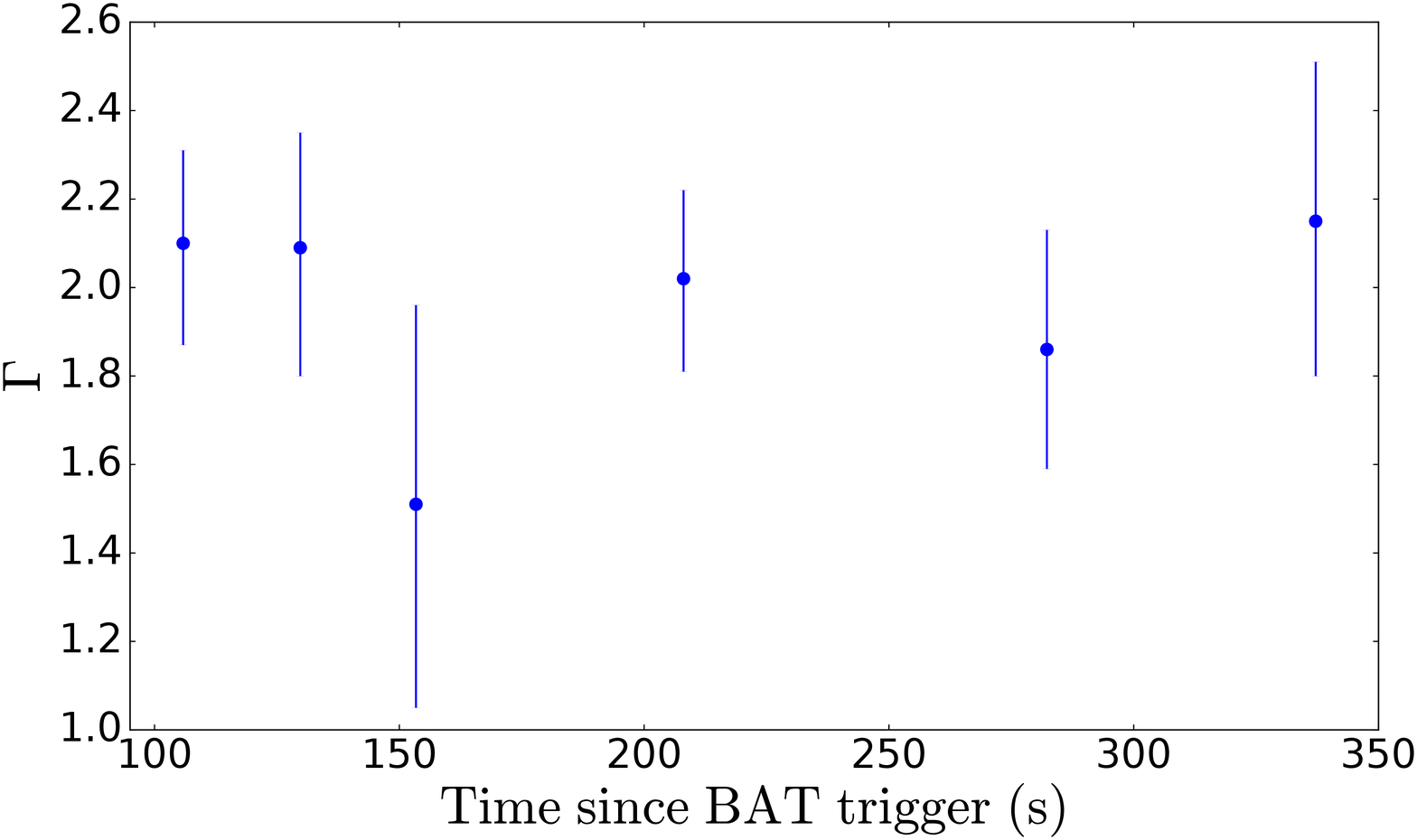}
    \end{subfigure}
        \begin{subfigure}[b]{0.49\textwidth}
        \includegraphics[width=\textwidth]{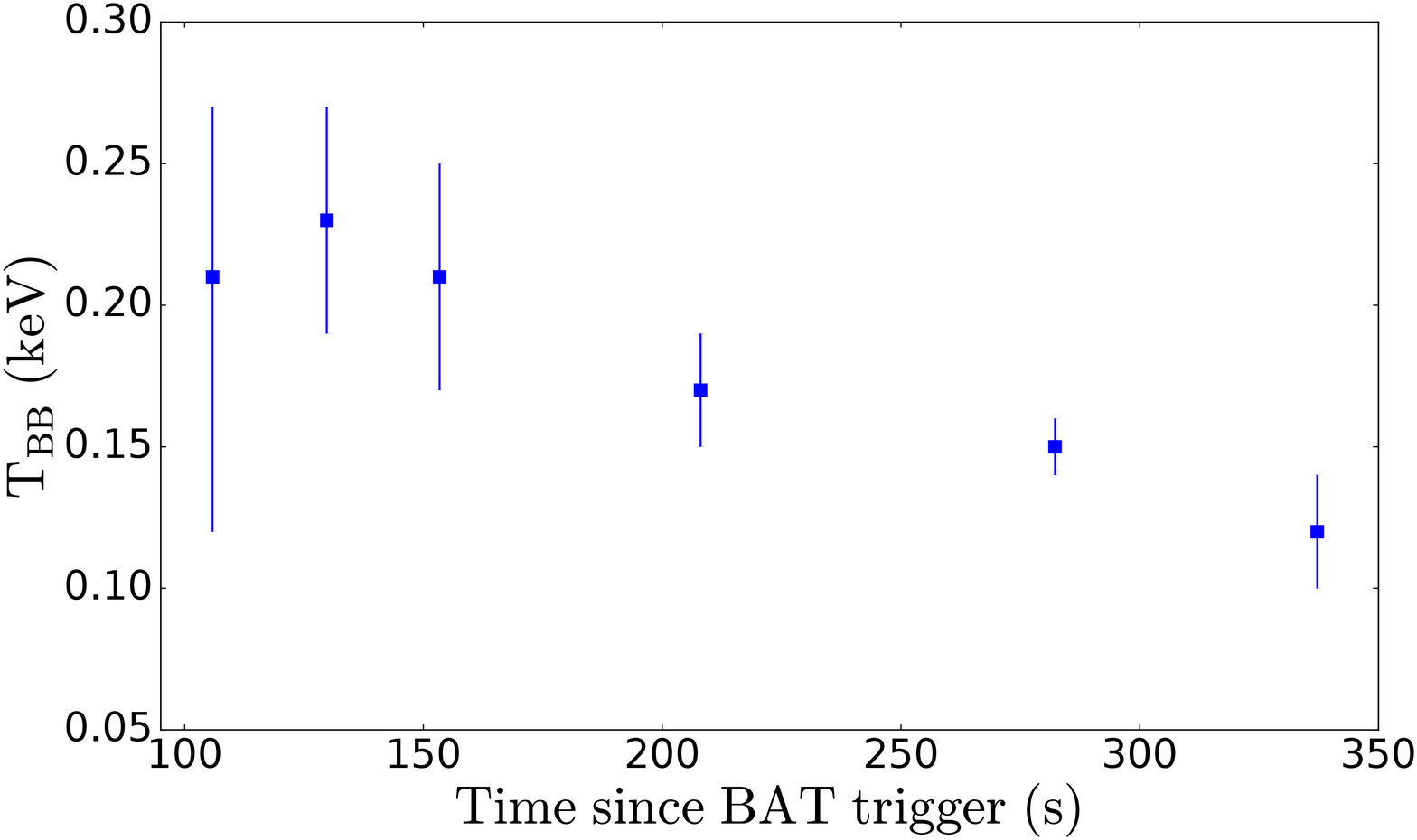}
    \end{subfigure}
    \caption{Light curve and time evolution of the best-fitting parameters for the power law + blackbody model for GRB 111225A. Top left: light curve with Bayesian blocks marked as dashed black lines. Top right: luminosity of the blackbody. Bottom left: photon index. Bottom right: blackbody temperature.}
    \label{111225}
\end{figure*}

\begin{figure*}
    \begin{subfigure}[b]{0.49\textwidth}
    \includegraphics[width=\columnwidth]{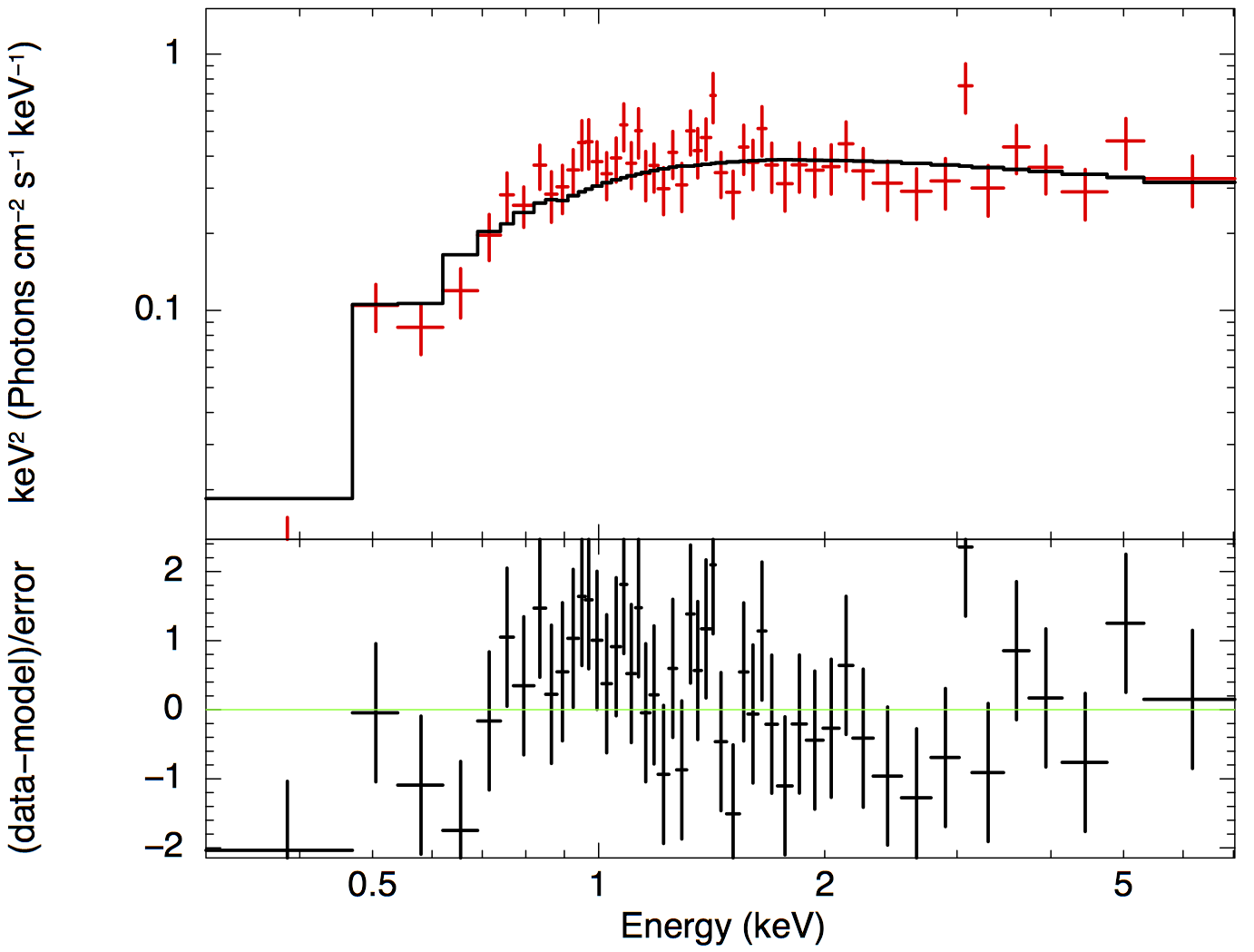}
    \end{subfigure}
    \begin{subfigure}[b]{0.49\textwidth}
     \includegraphics[width=\columnwidth]{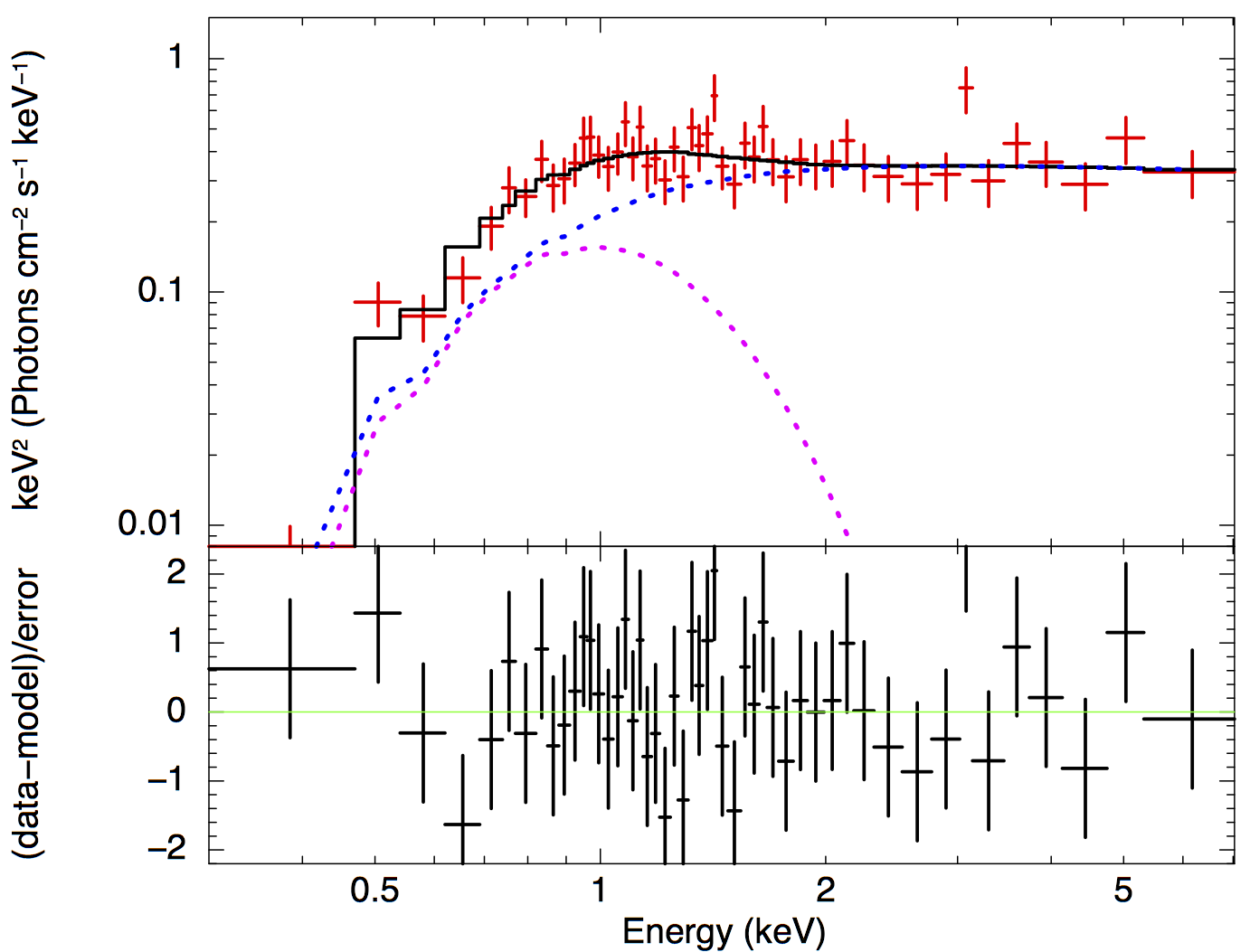}
    \end{subfigure}
    \caption{Fits to the spectrum of GRB 111125A in the time interval 117--142~s. Left panel: Fit to the absorbed power-law model. The lower panel shows the residuals to the fit. Right panel: Fit to the absorbed power law + blackbody model. The individual model components are shown as dashed blue and magenta lines for power law and blackbody, respectively, and the total model is shown in black. The lower panel shows the residuals of the fit.}
	\label{grb111225fit}
\end{figure*}

\begin{figure*}
    \centering
    \begin{subfigure}[b]{0.49\textwidth}
        \includegraphics[width=\textwidth]{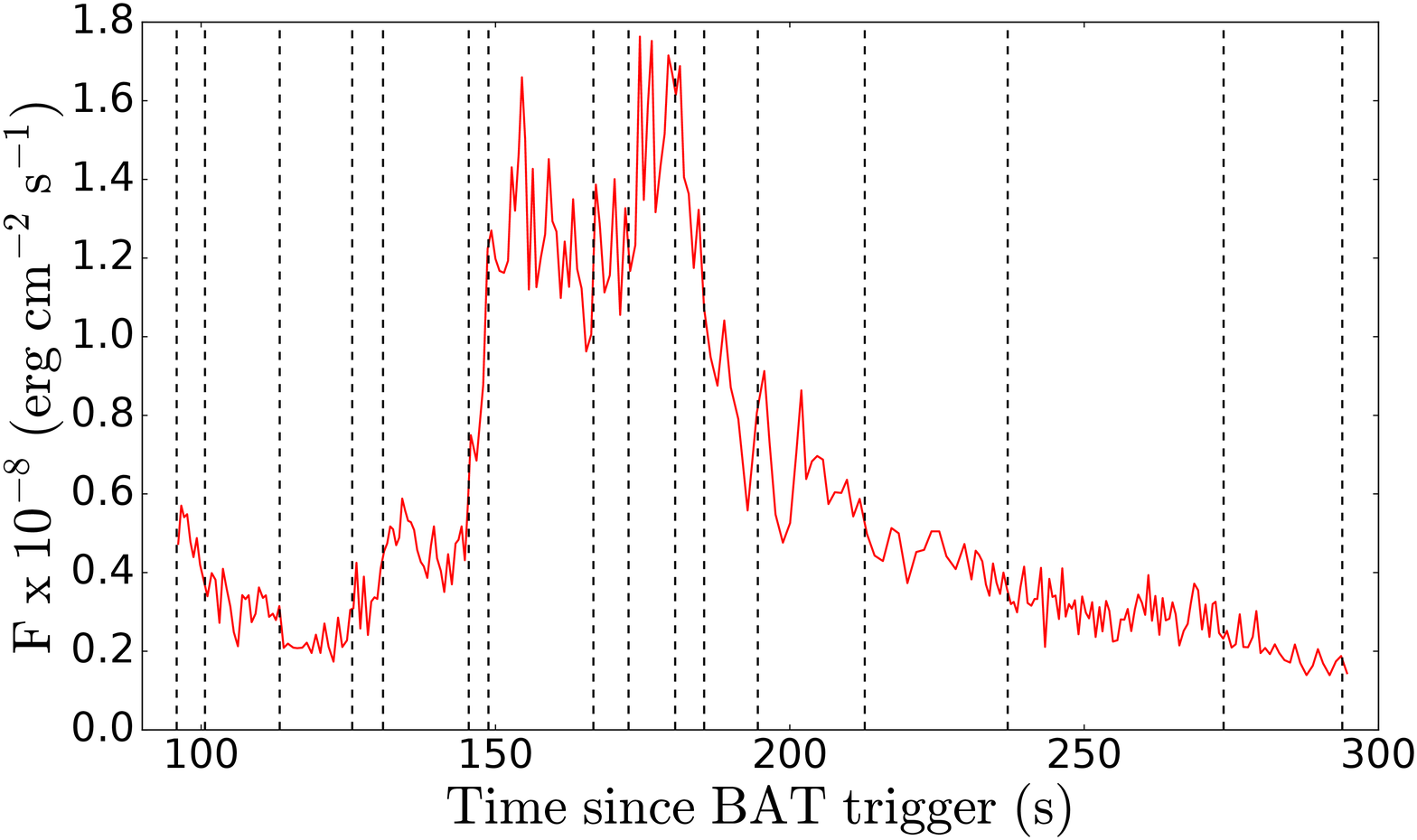}
    \end{subfigure}
   \begin{subfigure}[b]{0.49\textwidth}
        \includegraphics[width=\textwidth]{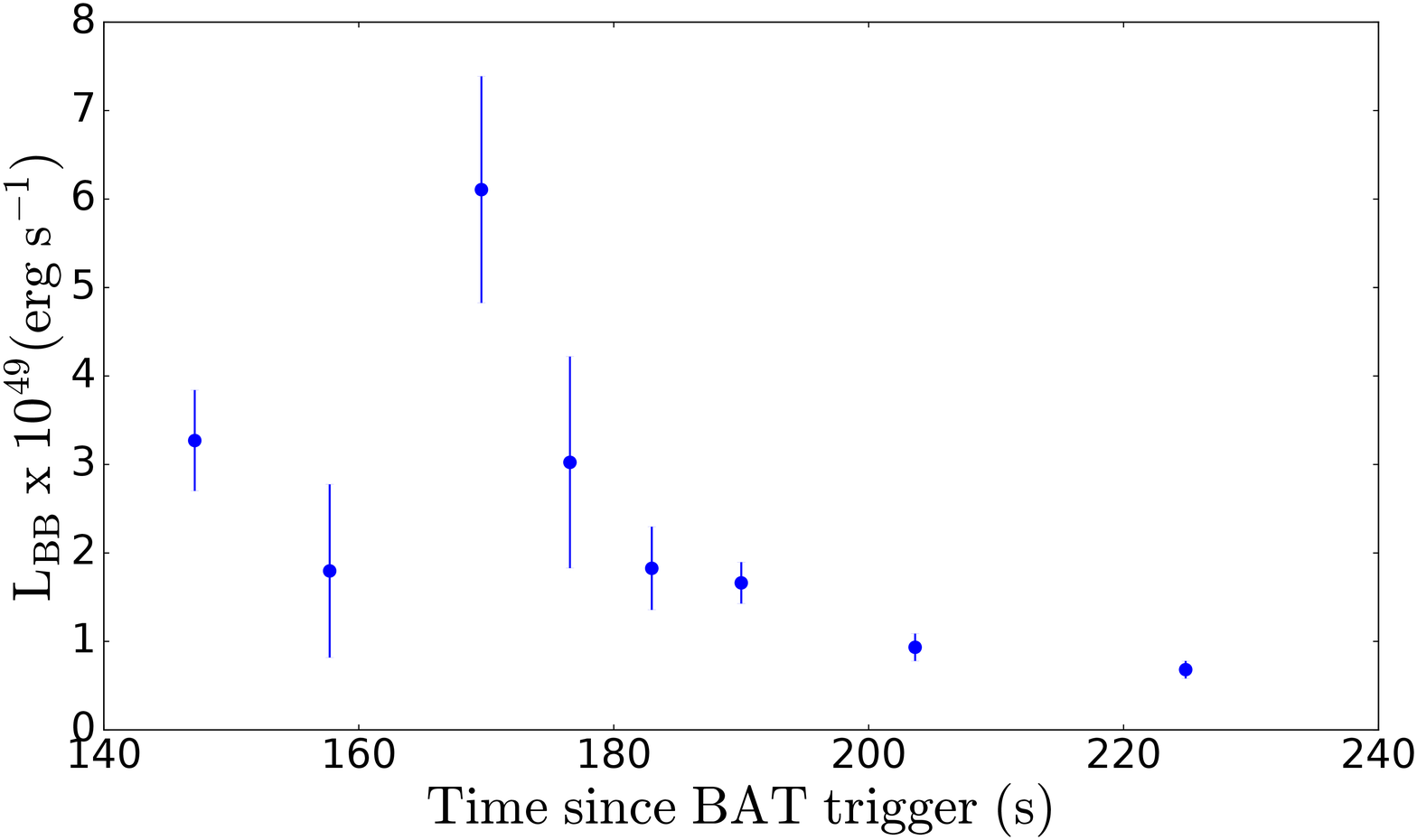}
    \end{subfigure}
        \begin{subfigure}[b]{0.49\textwidth}
        \includegraphics[width=\textwidth]{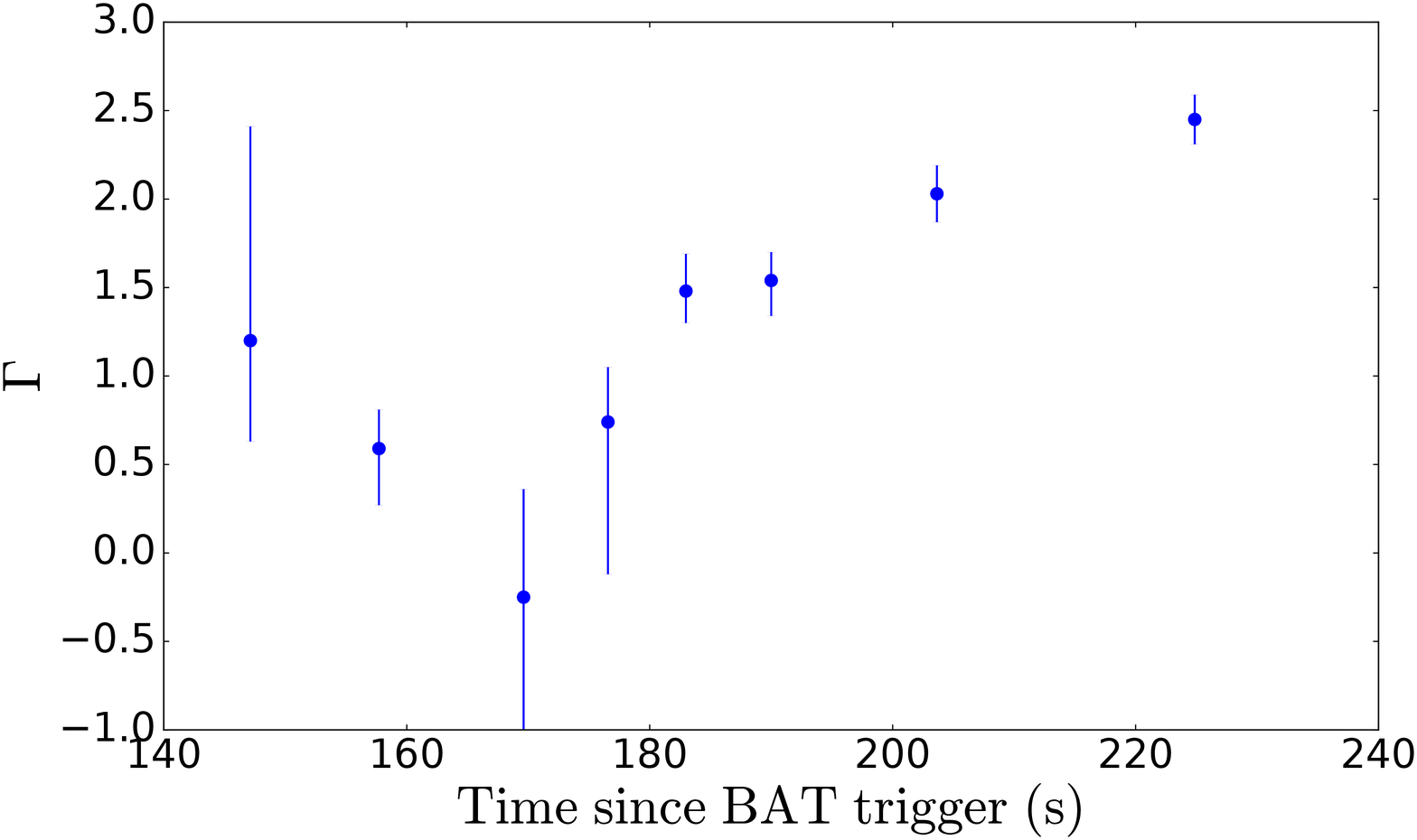}
    \end{subfigure}
        \begin{subfigure}[b]{0.49\textwidth}
        \includegraphics[width=\textwidth]{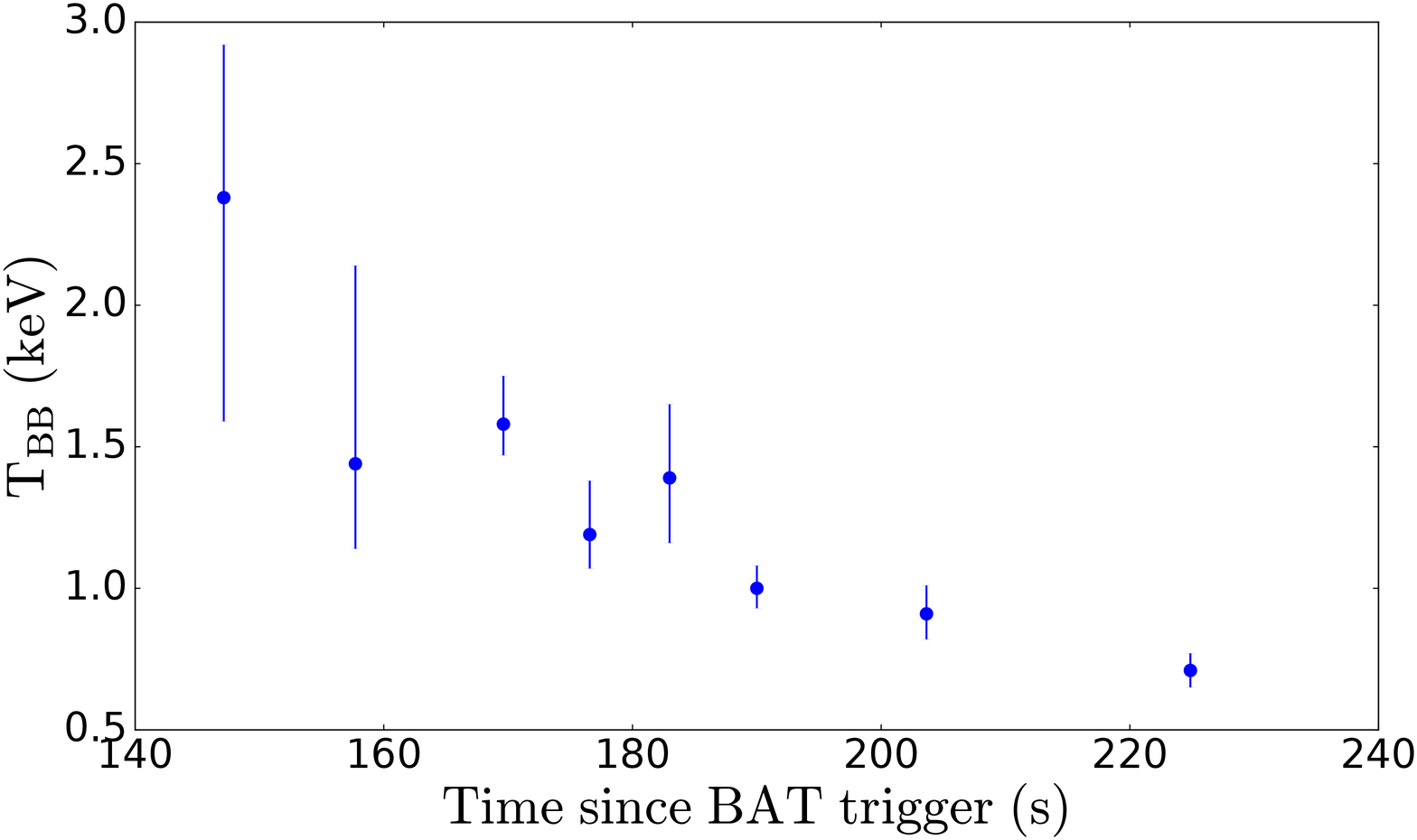}
    \end{subfigure}
    \caption{Light curve and time evolution of the best-fitting parameters for the power law + blackbody model for GRB 121211A. Top left: light curve with Bayesian blocks marked as dashed black lines. Top right: luminosity of the blackbody. Bottom left: photon index. Bottom right: blackbody temperature.}
    \label{121211}
\end{figure*}

\begin{figure*}
    \begin{subfigure}[b]{0.49\textwidth}
    \includegraphics[width=\columnwidth, height = 6cm]{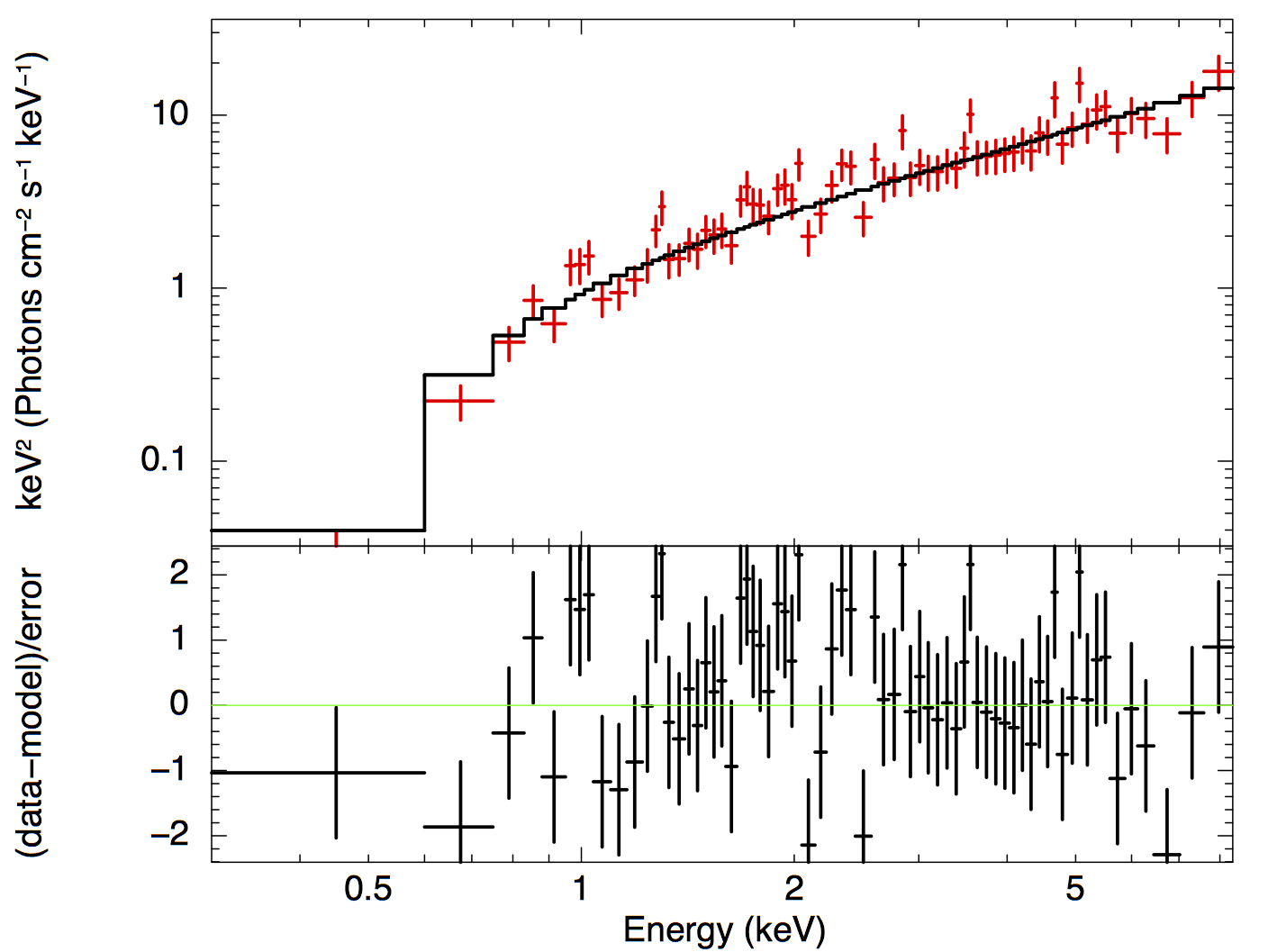}
    \end{subfigure}
    \begin{subfigure}[b]{0.49\textwidth}
     \includegraphics[width=\columnwidth, height = 6cm]{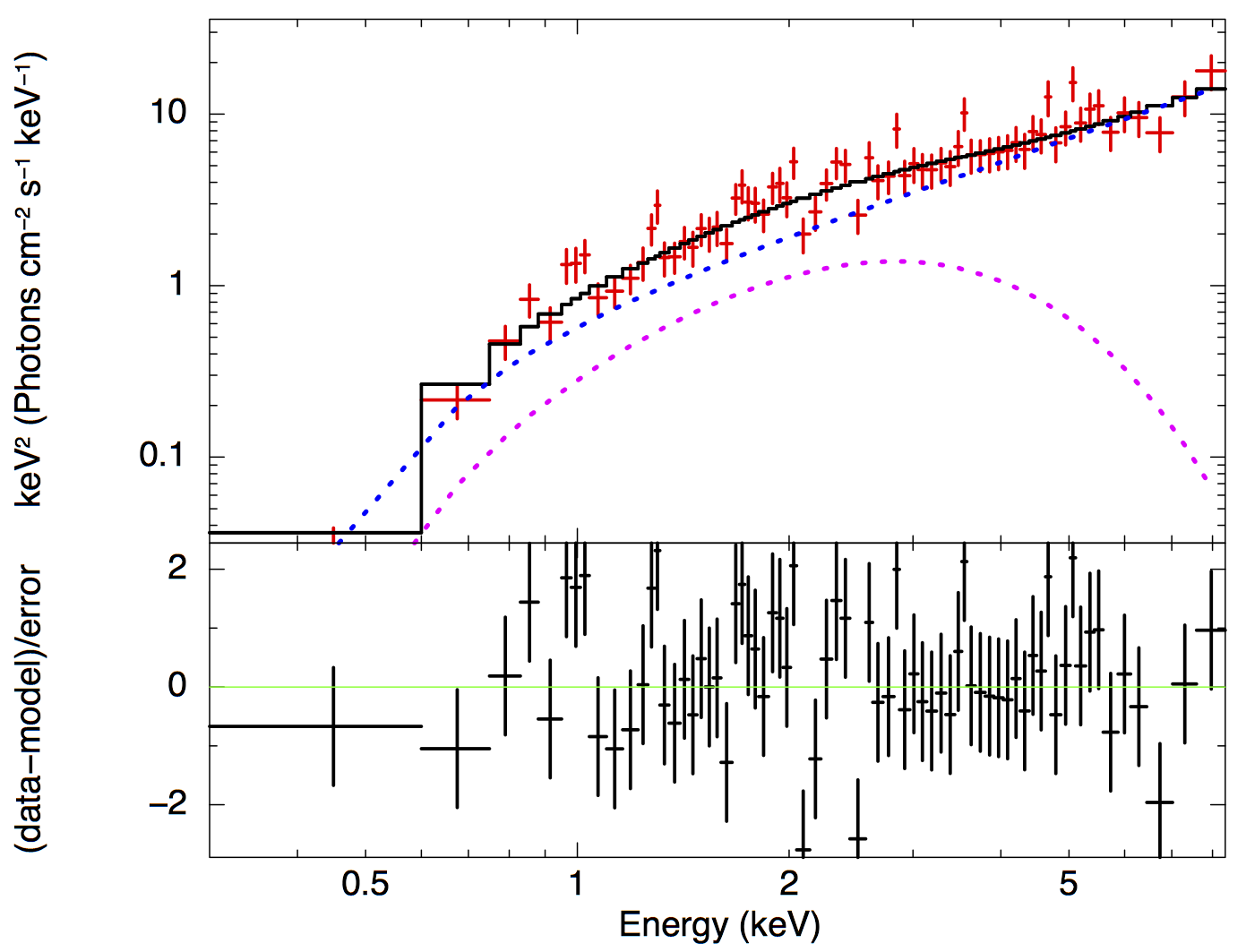}
    \end{subfigure}
    \caption{Fits to the spectrum of GRB 121211A in the time interval 149--166~s. Left panel: Fit to the absorbed power-law model. The lower panel shows the residuals to the fit. Right panel: Fit to the absorbed power law + blackbody model. The individual model components are shown as dashed blue and magenta lines for power law and blackbody, respectively, and the total model is shown in black. The lower panel shows the residuals of the fit.}
	\label{grb121211fit}
\end{figure*}

\begin{figure*}
    \centering
    \begin{subfigure}[b]{0.49\textwidth}
        \includegraphics[width=\textwidth]{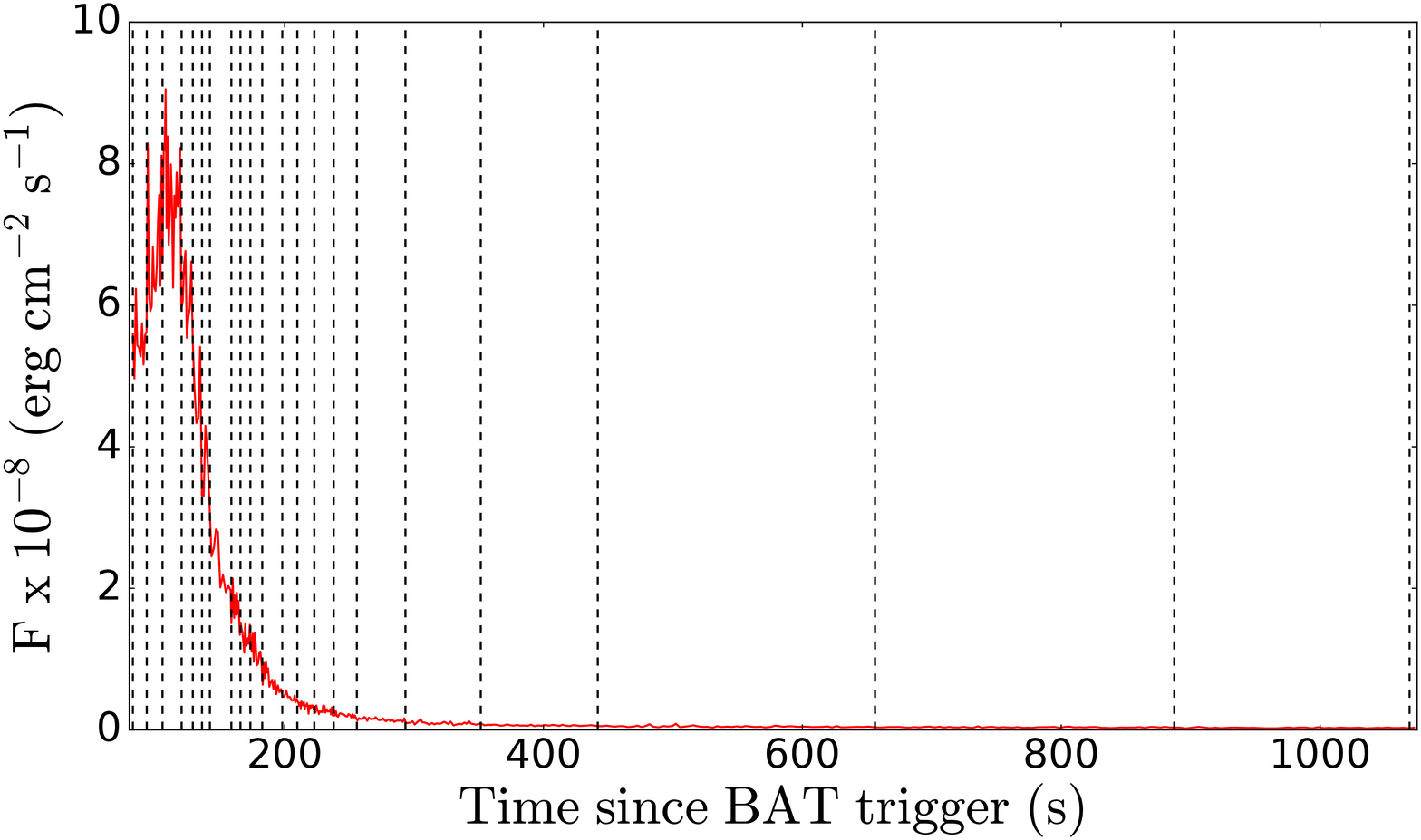}
    \end{subfigure}
    \begin{subfigure}[b]{0.49\textwidth}
        \includegraphics[width=\textwidth]{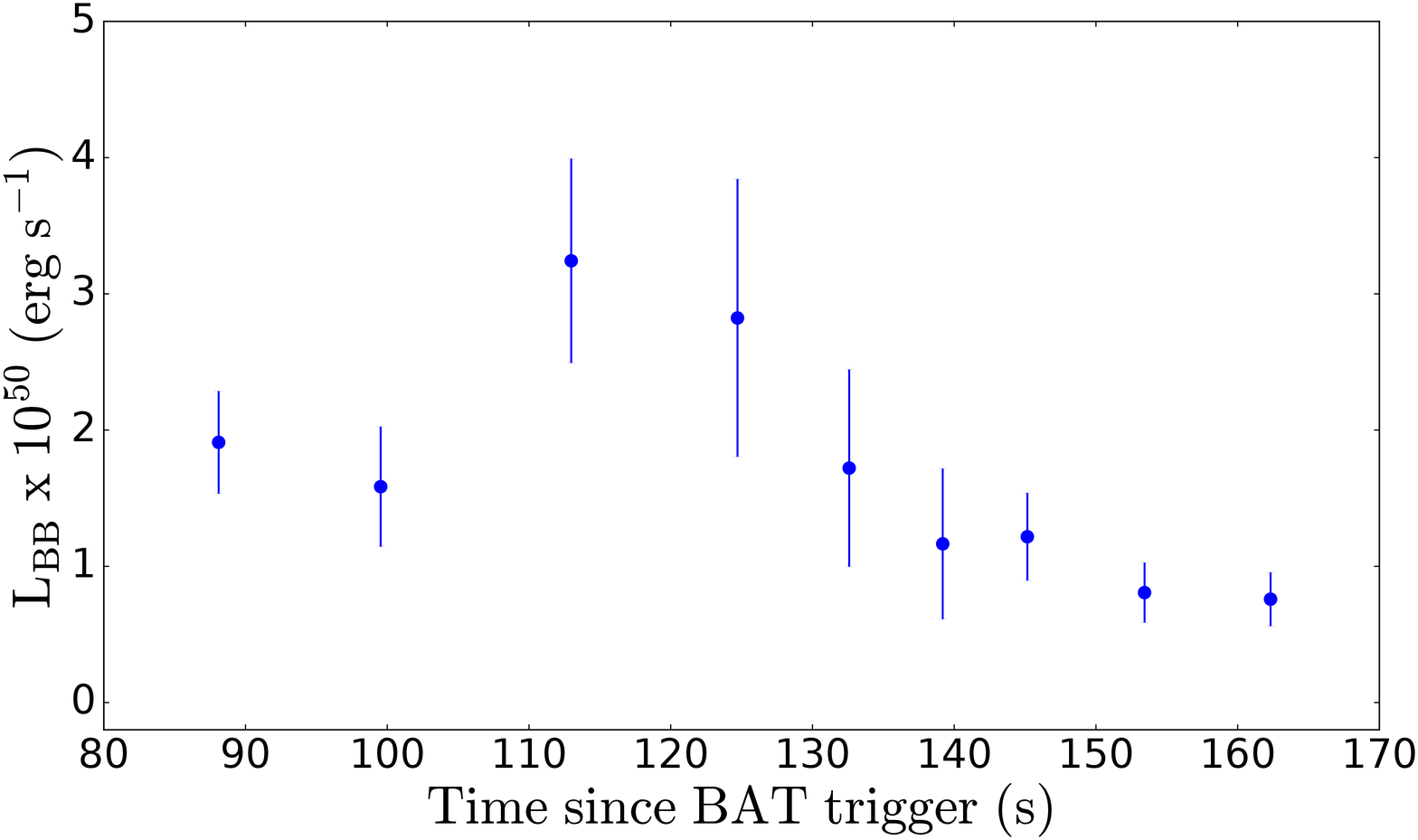}
    \end{subfigure}
        \begin{subfigure}[b]{0.49\textwidth}
        \includegraphics[width=\textwidth]{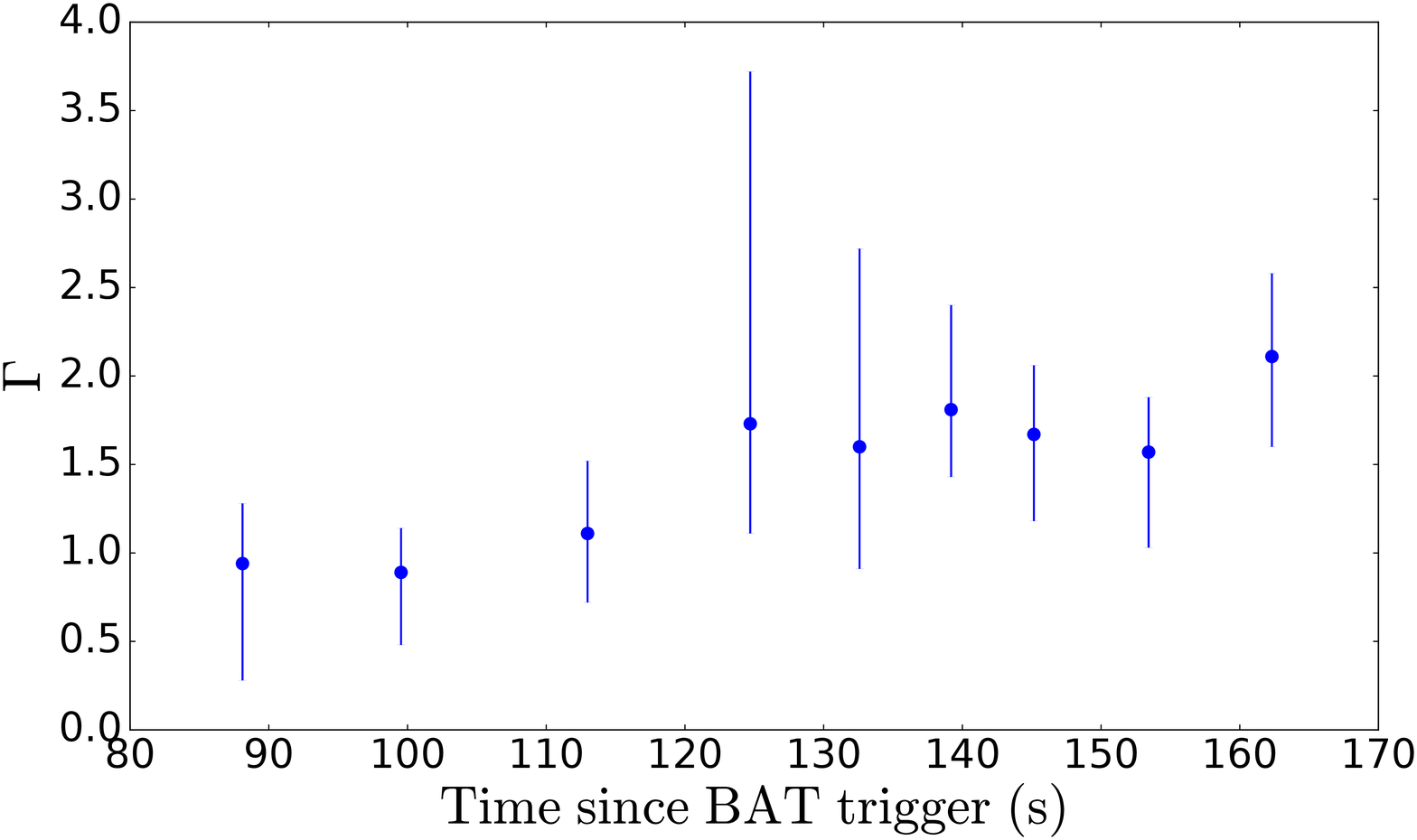}
    \end{subfigure}
        \begin{subfigure}[b]{0.49\textwidth}
        \includegraphics[width=\textwidth]{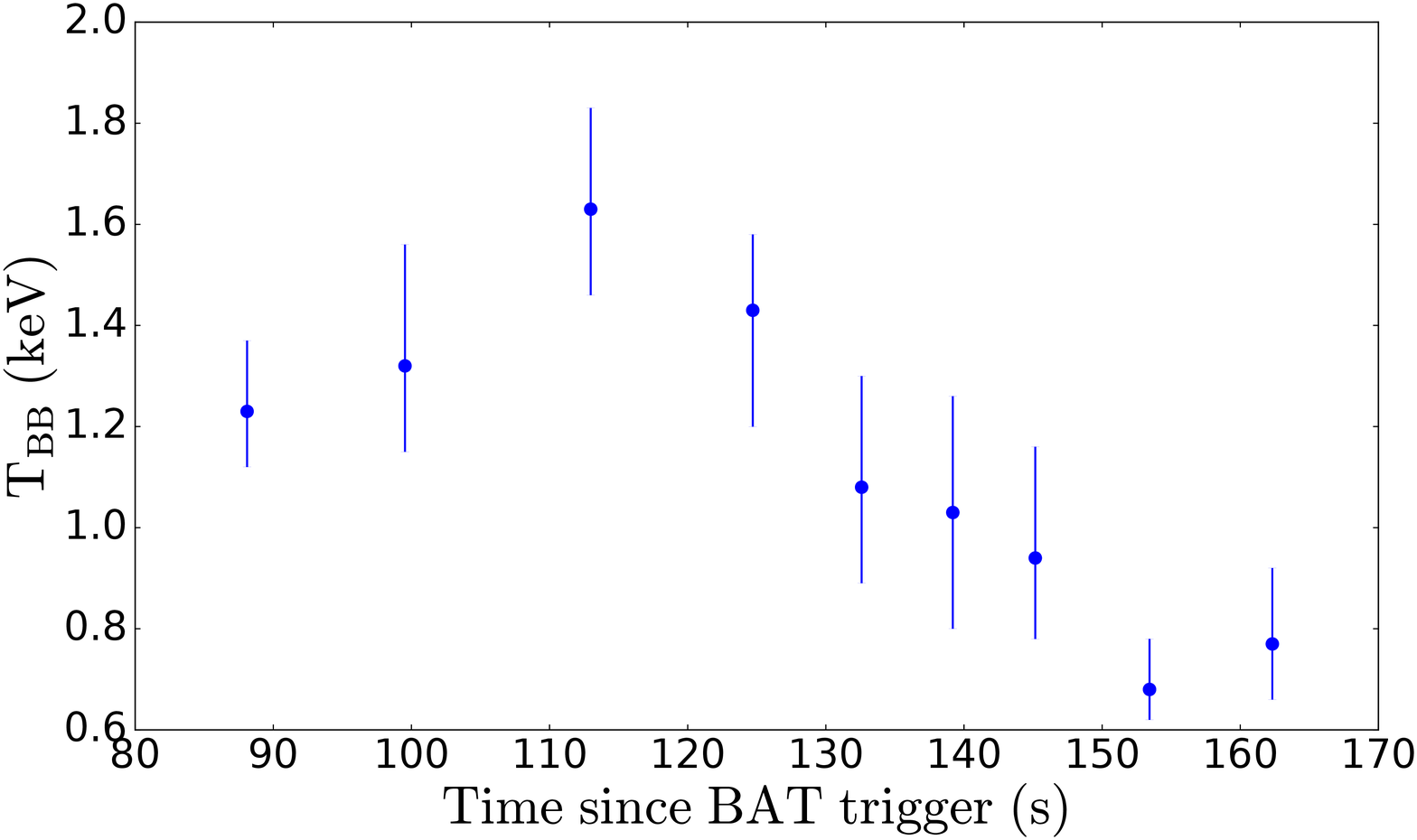}
    \end{subfigure}
    \caption{Light curve and time evolution of the best-fitting parameters for the power law + blackbody model for GRB 131030A. Top left: light curve with Bayesian blocks marked as dashed black lines. Top right: luminosity of the blackbody. Bottom left: photon index. Bottom right: blackbody temperature. }
    \label{131030}
\end{figure*}

\begin{figure*}
    \begin{subfigure}[b]{0.49\textwidth}
    \includegraphics[width=\columnwidth, height = 6cm]{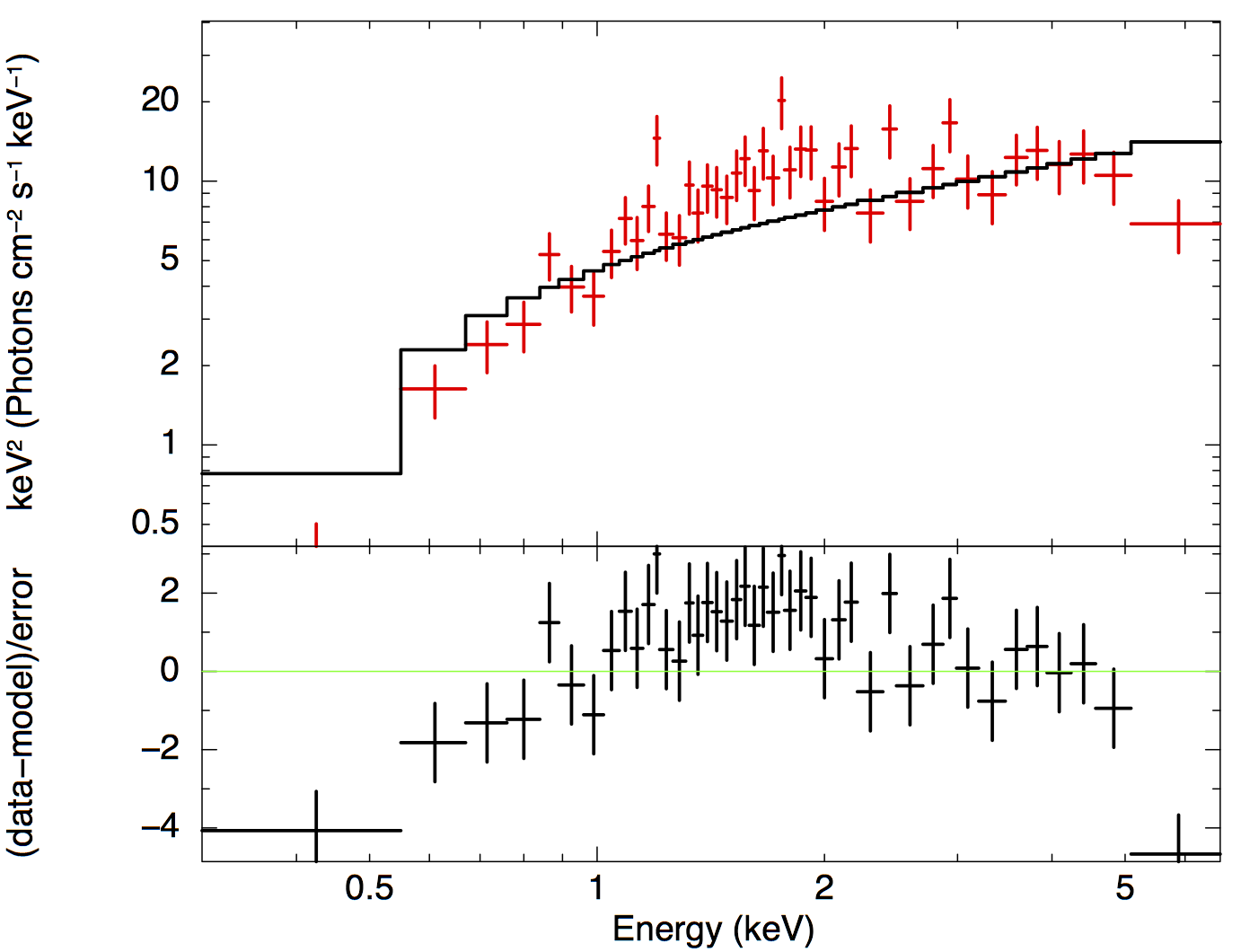}
    \end{subfigure}
    \begin{subfigure}[b]{0.49\textwidth}
     \includegraphics[width=\columnwidth, height = 6cm]{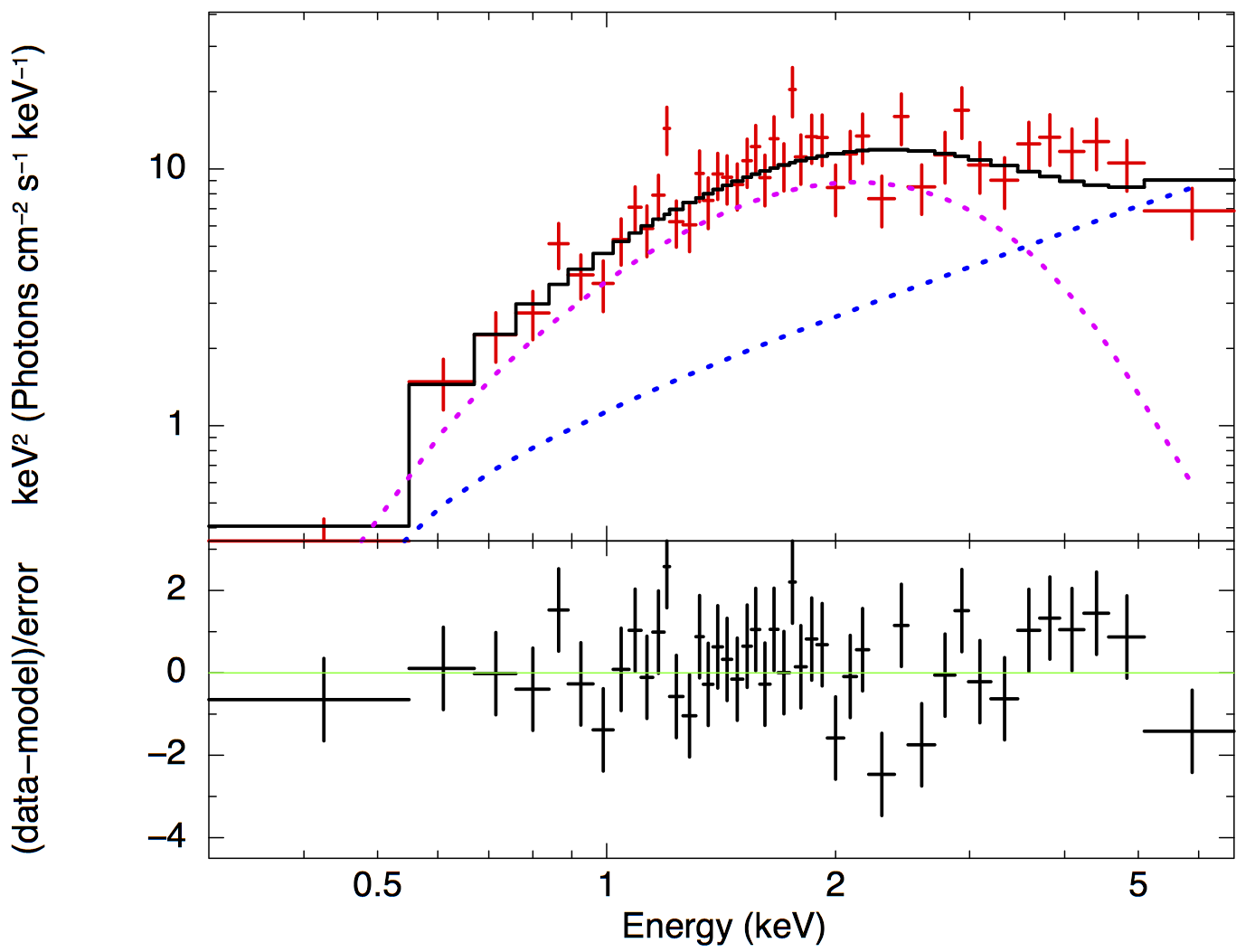}
    \end{subfigure}
    \caption{Fits to the spectrum of GRB 131030A in the time interval 82--93~s. Left panel: Fit to the absorbed power-law model. The lower panel shows the residuals to the fit. Right panel: Fit to the absorbed power law + blackbody model.  The individual model components are shown as dashed blue and magenta lines for power law and blackbody, respectively, and the total model is shown in black. The lower panel shows the residuals of the fit.}
	\label{grb131030fit}
\end{figure*}

\begin{figure*}
    \centering
    \begin{subfigure}[b]{0.49\textwidth}
        \includegraphics[width=\textwidth]{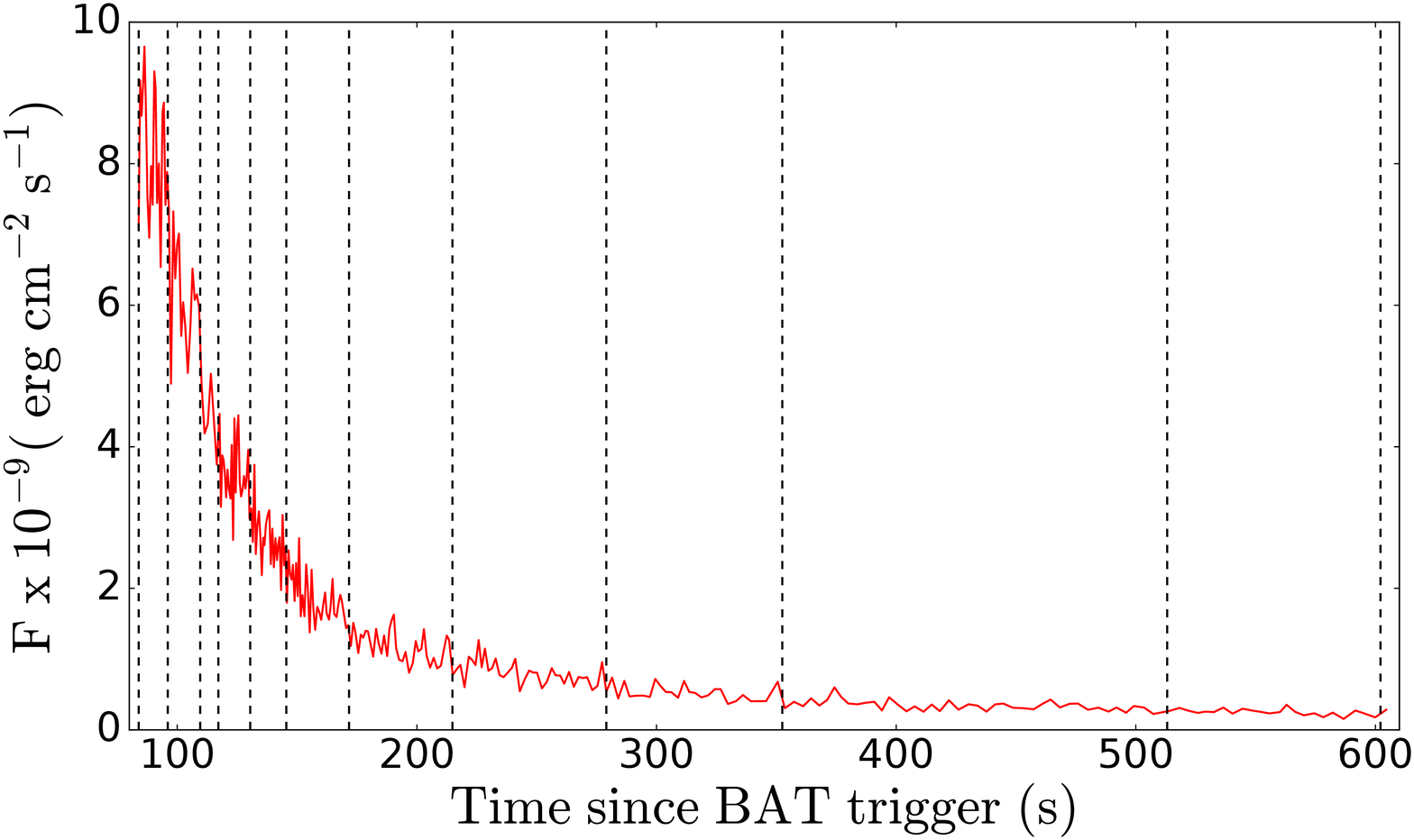}
    \end{subfigure}
    \begin{subfigure}[b]{0.49\textwidth}
        \includegraphics[width=\textwidth]{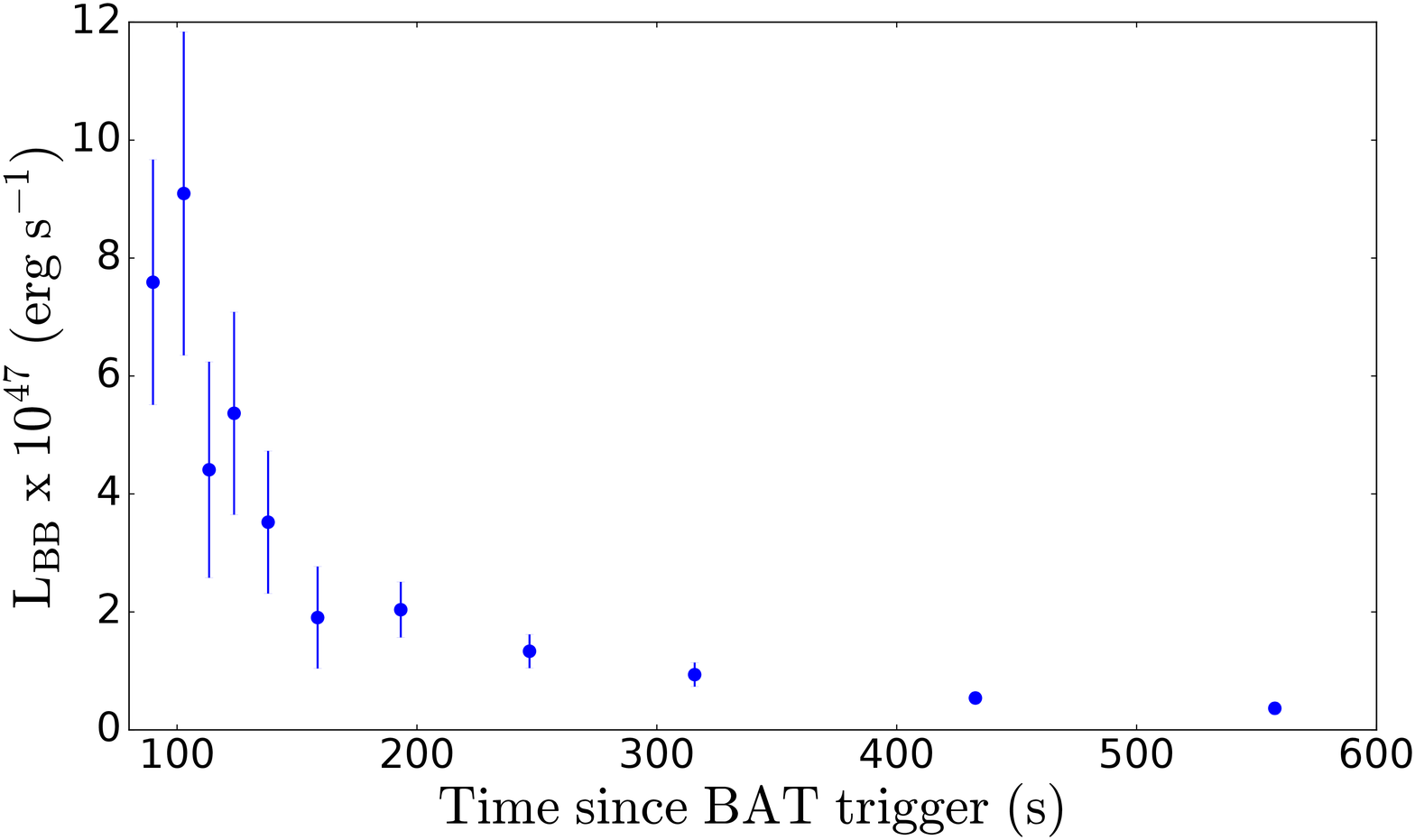}
    \end{subfigure}
        \begin{subfigure}[b]{0.49\textwidth}
        \includegraphics[width=\textwidth]{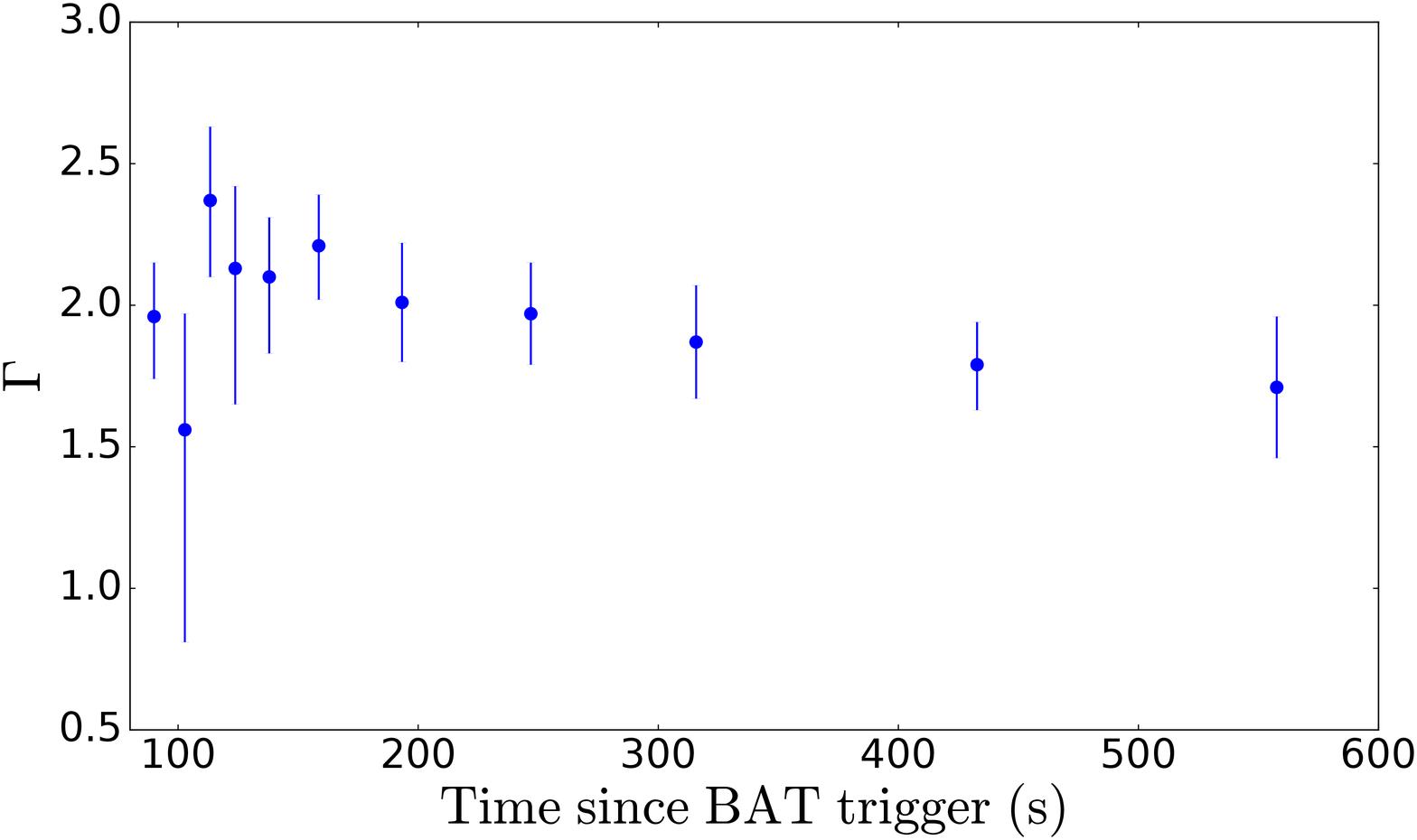}
    \end{subfigure}
        \begin{subfigure}[b]{0.49\textwidth}
        \includegraphics[width=\textwidth]{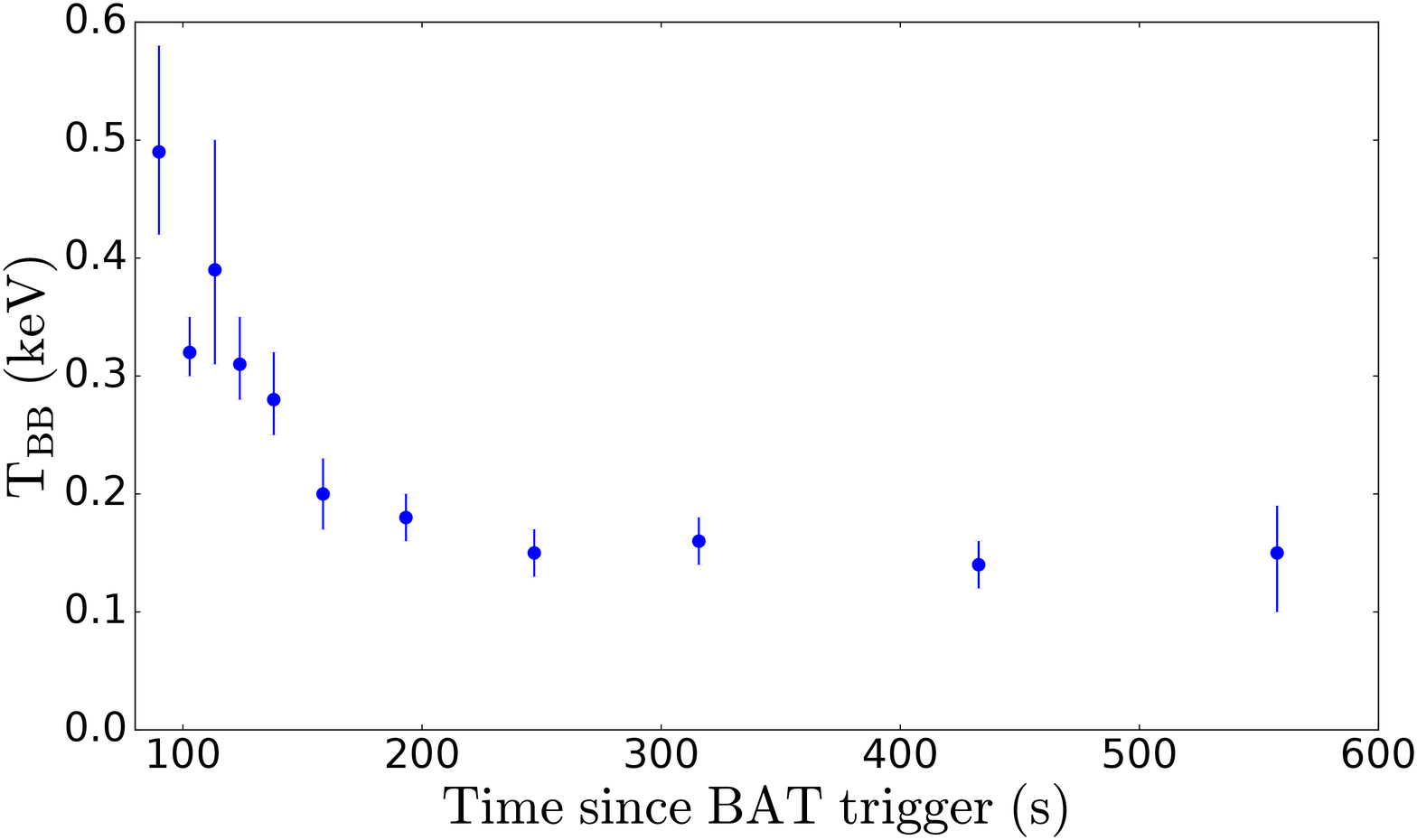}
    \end{subfigure}
    \caption{Light curve and time evolution of the best-fitting parameters for the power law + blackbody model for GRB 150727A. Top left: light curve with Bayesian blocks marked as dashed black lines. Top right: luminosity of the blackbody. Bottom left: photon index. Bottom right: blackbody temperature. }
    \label{150727}
\end{figure*}

\begin{figure*}
    \begin{subfigure}[b]{0.49\textwidth}
    \includegraphics[width=\columnwidth, height = 6cm]{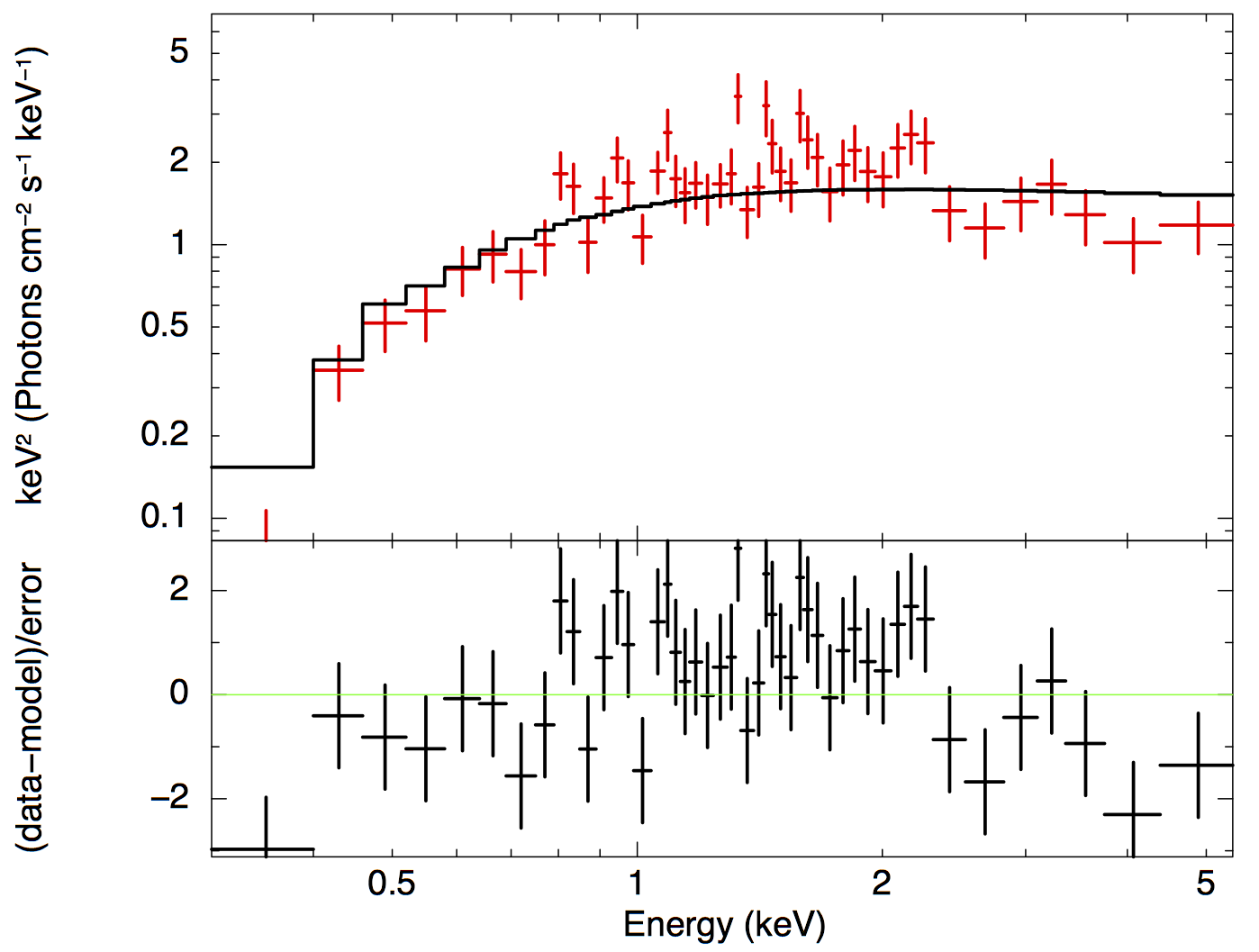}
    \end{subfigure}
    \begin{subfigure}[b]{0.49\textwidth}
     \includegraphics[width=\columnwidth, height = 6cm]{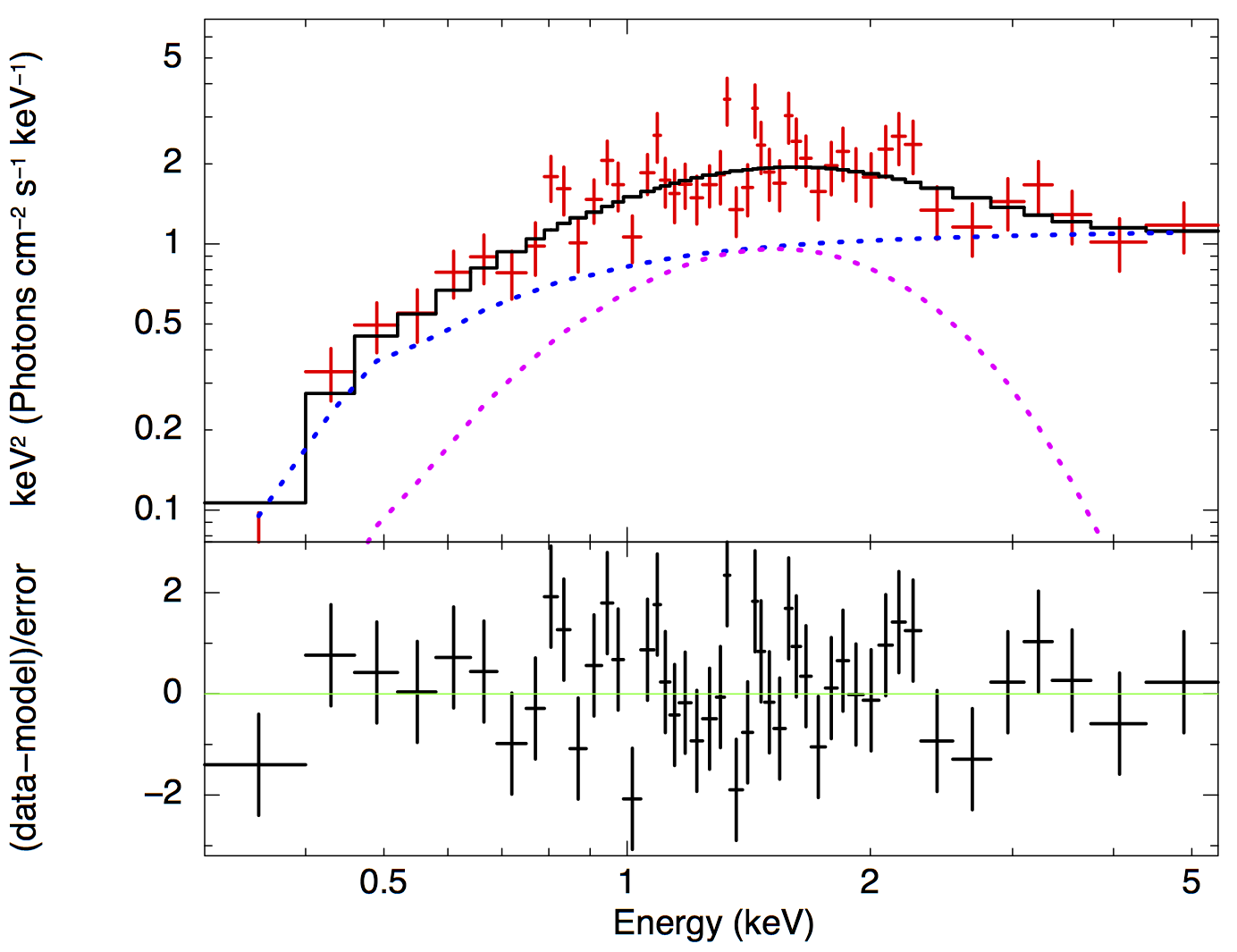}
    \end{subfigure}
    \caption{Fits to the spectrum of GRB 150727A in the time interval 84--96~s. Left panel: Fit to the absorbed power-law model. The lower panel shows the residuals to the fit. Right panel: Fit to the absorbed power law + blackbody model. The individual model components are shown as dashed blue and magenta lines for power law and blackbody, respectively, and the total model is shown in black. The lower panel shows the residuals of the fit.}
	\label{grb150727fit}
\end{figure*}

\begin{figure*}
    \centering
    \begin{subfigure}[b]{0.49\textwidth}
        \includegraphics[width=\textwidth]{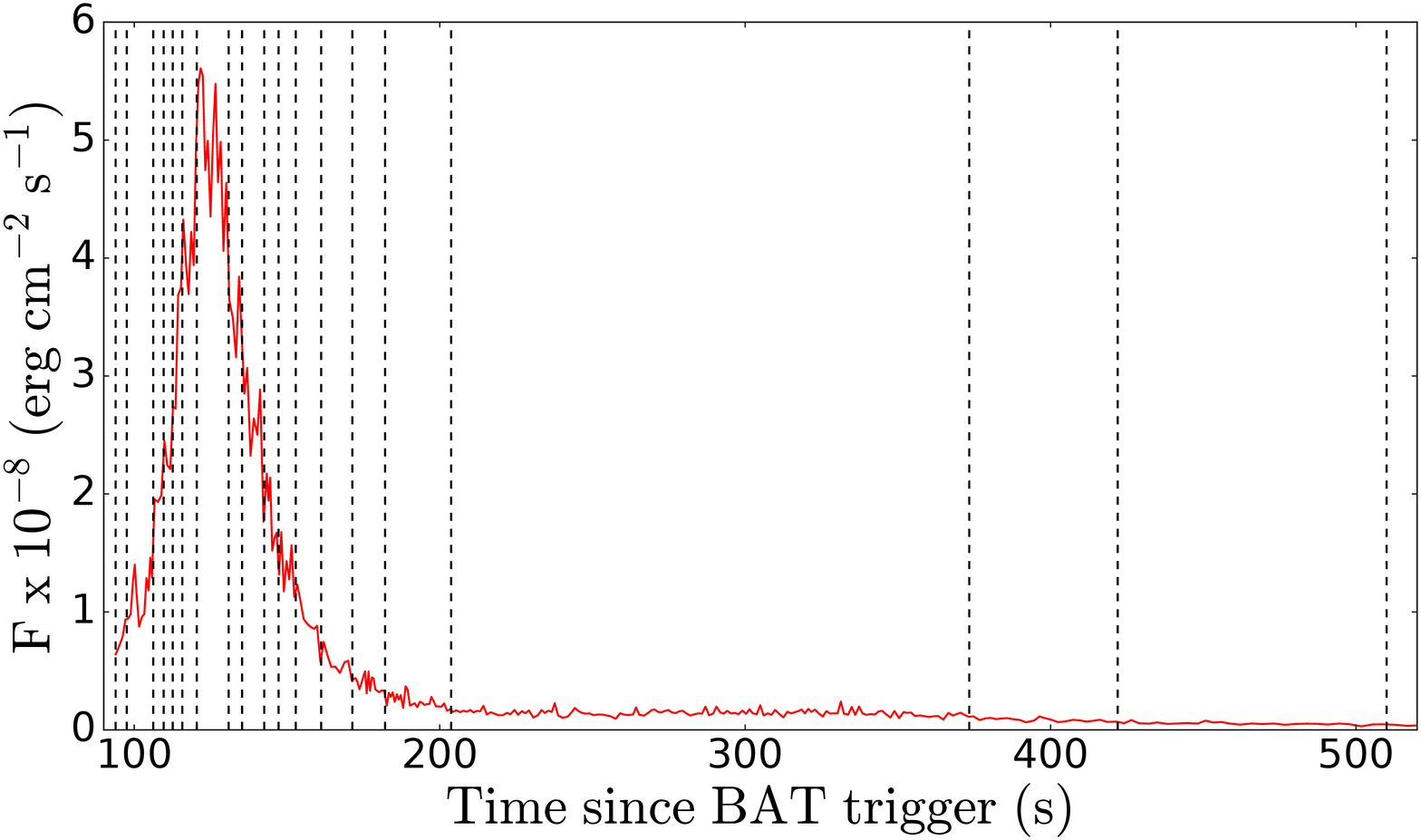}
    \end{subfigure}
    \begin{subfigure}[b]{0.49\textwidth}
        \includegraphics[width=\textwidth]{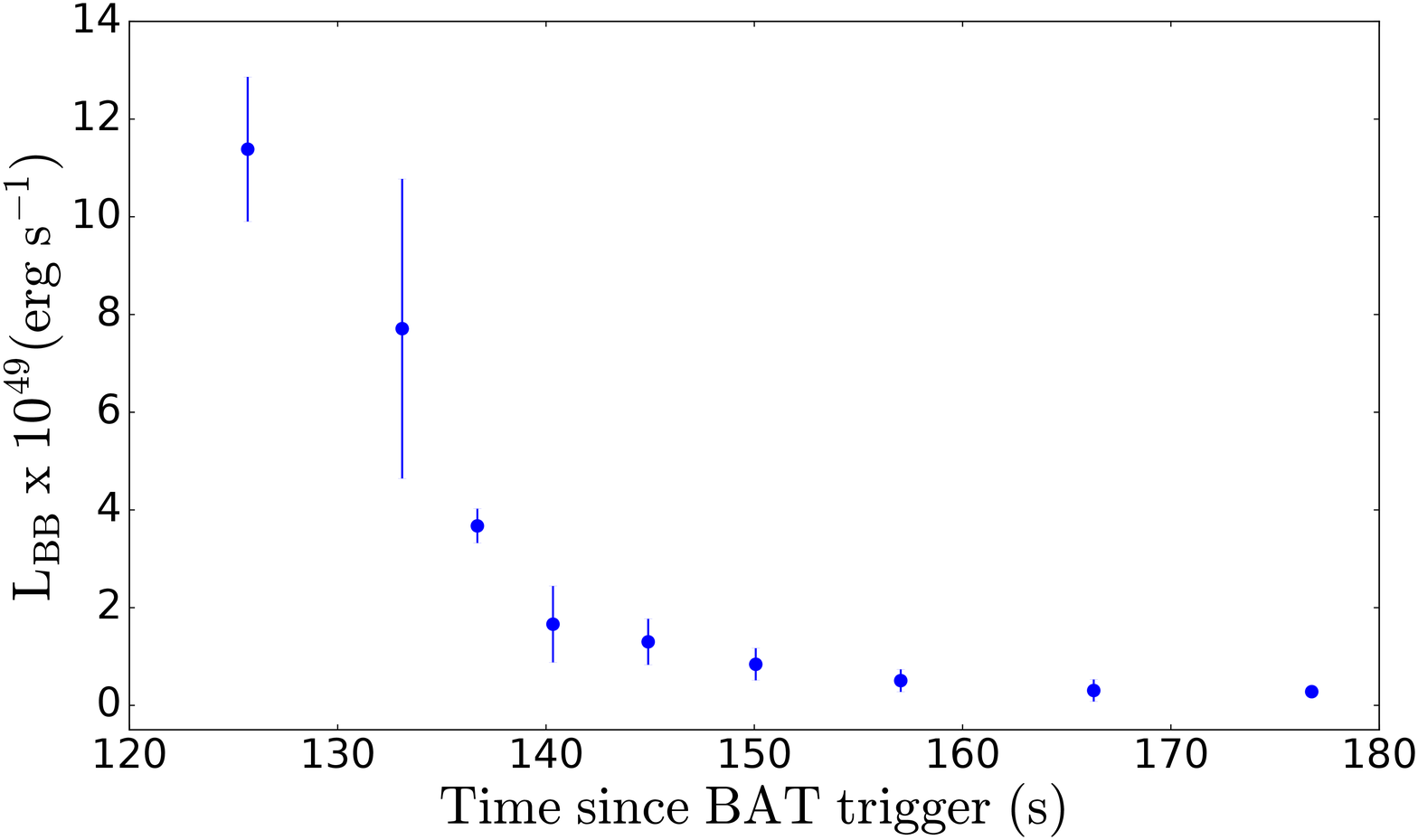}
    \end{subfigure}
        \begin{subfigure}[b]{0.5\textwidth}
        \includegraphics[width=\textwidth]{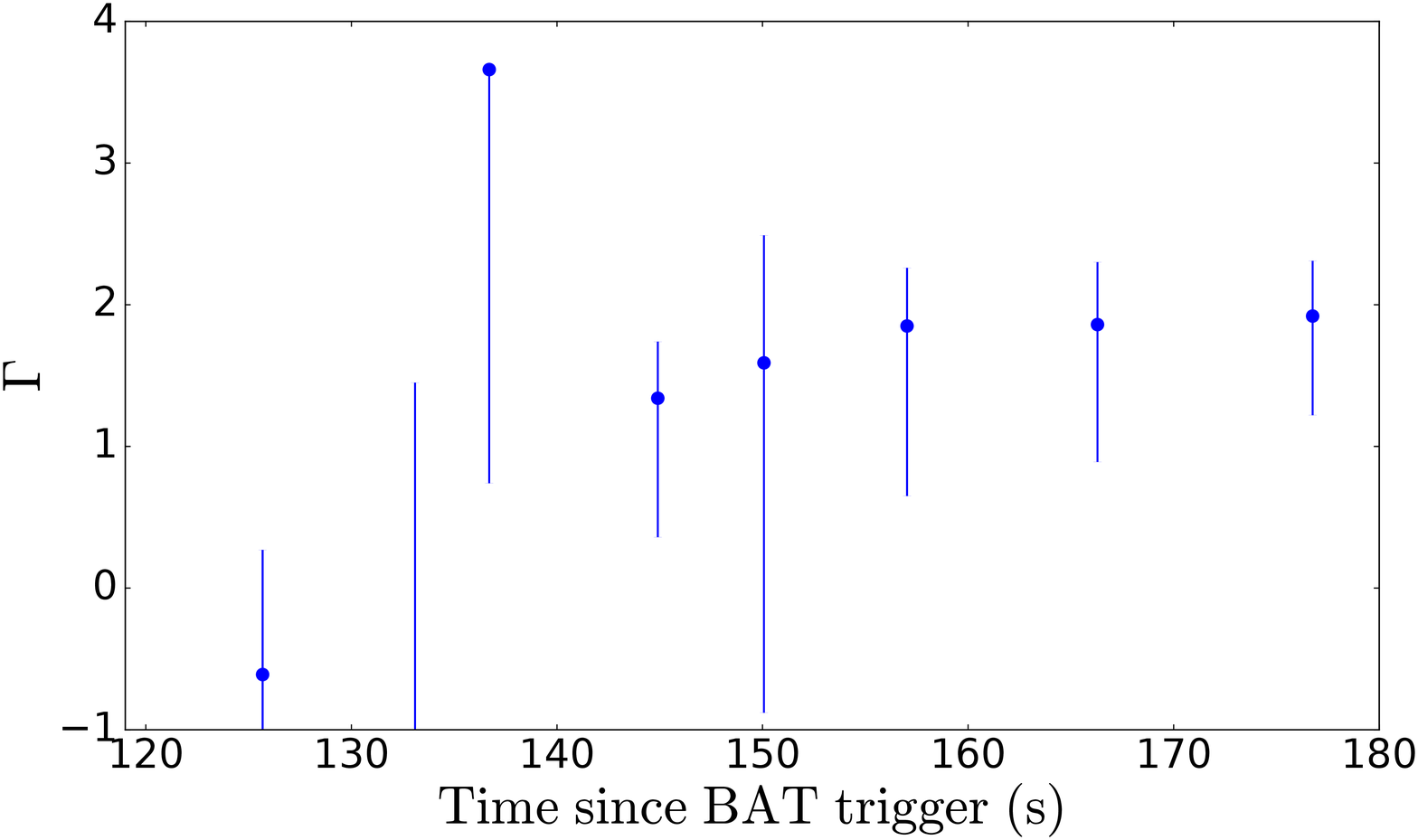}
    \end{subfigure}
        \begin{subfigure}[b]{0.49\textwidth}
        \includegraphics[width=\textwidth]{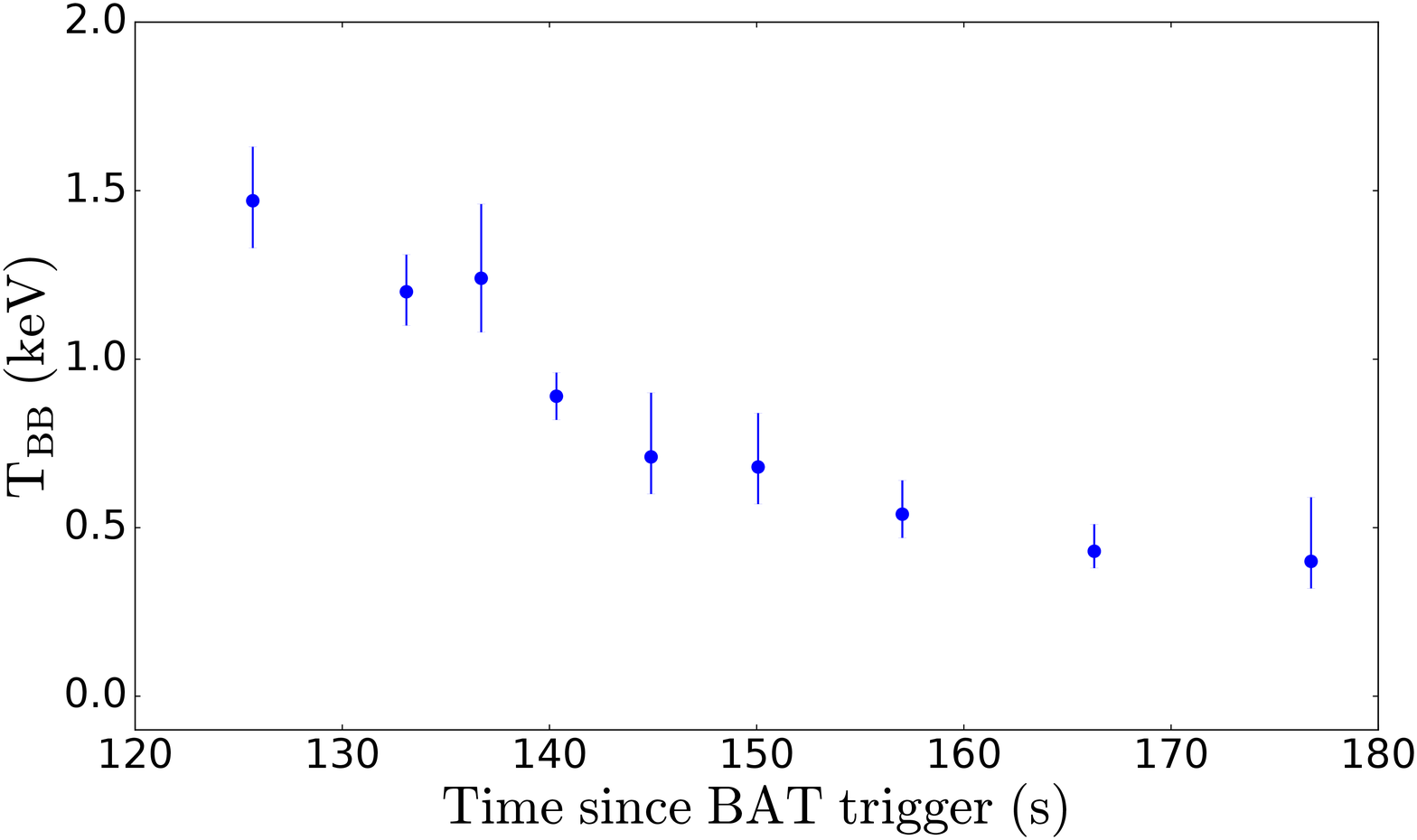}
    \end{subfigure}
    \caption{Light curve and time evolution of the best-fitting parameters for the power law + blackbody model for GRB 151027A. Top left: light curve with Bayesian blocks marked as dashed black lines. Top right: luminosity of the blackbody. Bottom left: photon index. Bottom right: blackbody temperature. }
    \label{151027}
\end{figure*}

\begin{figure*}
    \begin{subfigure}[b]{0.49\textwidth}
    \includegraphics[width=\columnwidth, height = 6cm]{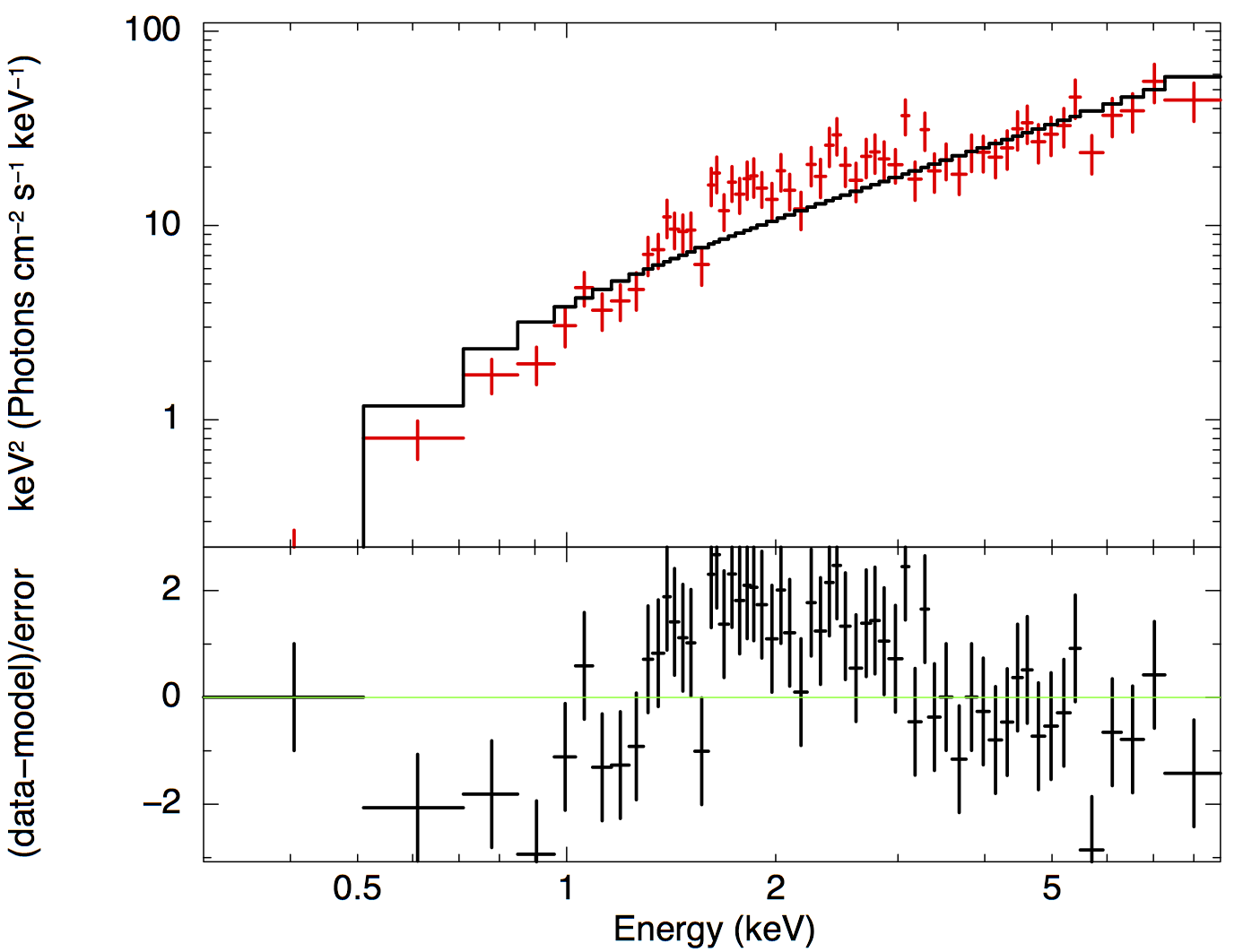}
    \end{subfigure}
    \begin{subfigure}[b]{0.49\textwidth}
     \includegraphics[width=\columnwidth, height = 6cm]{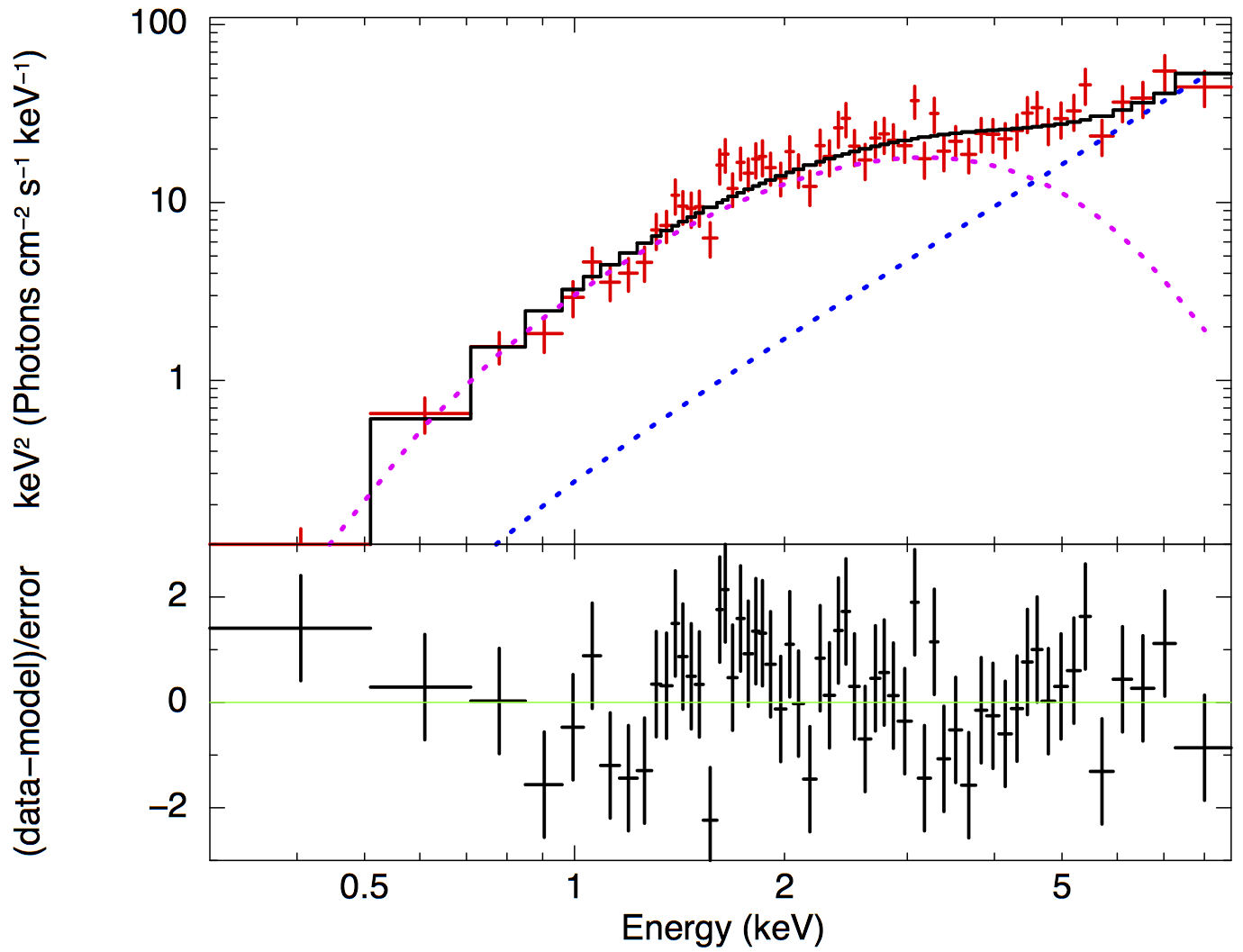}
    \end{subfigure}
    \caption{Fits to the spectrum of GRB 151027A in the time interval 120--131~s. Left panel: Fit to the absorbed power-law model. The lower panel shows the residuals to the fit. Right panel: Fit to the absorbed power law + blackbody model.  The individual model components are shown as dashed blue and magenta lines for power law and blackbody, respectively, and the total model is shown in black. The lower panel shows the residuals of the fit.}
	\label{grb151027fit}
\end{figure*}

Light curves, alongside the time evolution of the inferred blackbody luminosities, temperatures and photon indices are plotted in Figs. \ref{111123}, \ref{111225}, \ref{121211}, \ref{131030}, \ref{150727} and \ref{151027} for each of the bursts. Light curves show the 0.3--10~keV observed flux, while the blackbody luminosities are the full luminosities including parts of the blackbody that may fall outside of the fitted energy range. Light curves are plotted for the entire duration of the XRT WT mode observations (referred to as XRT observations from hereon), while the time evolution of the best-fitting parameters are plotted only in the time interval where the blackbody is significant at $> \ 3 \sigma$. The temperatures of the blackbody are in the rest-frame of the GRBs. Following the procedure described in \cite{Olivares2012} we fit cooling profiles to the temperature using the function $T = T_{\rm{o}} - t^{n}$, where $t$ is the time. Values of the initial temperature ($T_{\rm{o}}$) and decay index ($n$) are reported in the text for each individual burst. Table \ref{fits} contains the best-fitting parameters for all the GRBs (the full table is available online). The fits for the first time-bin where the blackbody is significant at $> \ 3 \sigma$ are presented in Figs. \ref{grb111123fit}, \ref{grb111225fit}, \ref{grb121211fit}, \ref{grb131030fit}, \ref{grb150727fit} and \ref{grb151027fit} for each of the bursts, respectively. The spectra are plotted in the observed energy range. Below we present results and conclusions based only on the time-resolved fits. We discuss the significance of the blackbody in the time-averaged spectra in Section \ref{timav}.

\paragraph*{GRB 111123A (Figs. \ref{111123} and \ref{grb111123fit}):}

The XRT observations for this burst start at 106~s after the BAT trigger and is divided into 14 Bayesian blocks. The light curve is characterized by multiple small flares, with the most prominent one at the beginning of the observation. The blackbody is significant at $> 3 \sigma $ in the first 7 bins during the first flare and until 373~s after the BAT trigger. The temperature cools from $3.45 \pm 0.47$~keV to $2.12 \pm 0.43$~keV in 236~s, with a decay index of $n=-0.52 \pm 0.38$ for an initial temperature of $T_{\rm{o}} = 7.21 \pm 0.15$~keV. We also observe a strong evolution of the photon index, which softens with time from $0.46 \pm 0.27$ to $1.54 \pm 0.19$. 

\paragraph*{GRB 111225A (Figs. \ref{111225} and \ref{grb111225fit}):}
 
The XRT observations for this GRB start at 95 s after the BAT trigger and we have 7 Bayesian blocks in total. The light curve decays smoothly except for a flat part around 200~s. The blackbody becomes significant at $  > 3 \sigma$ at the beginning of the light curve and stays significant until the end of the XRT observation at 361s. The temperature cools from $0.23 \pm 0.07$ ~keV to $0.12 \pm 0.02$ ~keV in 266~s with a decay index of $n=-0.54 \pm 0.39$ for an initial temperature of $T_{\rm{o}} = 0.5 \pm 0.2$ ~keV. The photon index varies between $2.1 \pm 0.22$ to $1.5 \pm 0.45$.

\paragraph*{GRB 121211A (Figs. \ref{121211} and \ref{grb121211fit}):} 
The XRT observations of this burst start at 96~s after the BAT trigger and we have in total 15 Bayesian blocks. The light curve is dominated by a strong flare with a rapid rise around 150~s. The blackbody becomes significant at $> 3 \sigma$  at the beginning of the flare and lasts until 237 s after the BAT trigger. The temperature cools from $2.38 \pm 0.65$ ~keV to $0.91 \pm 0.09$ ~keV in 92~s with a decay index of $n=-2.68 \pm 0.84$ for an initial temperature of $T_{\rm{o}} = 6.15 \pm 1.61$~keV. We also note that the photon index shows a strong evolution: it first hardens to values between $0.6 \pm 0.27 -- 0.7 \pm 0.58$  and then softens to a value around $2 \pm 0.16$.

\paragraph*{GRB 131030A (Fig. \ref{131030} and \ref{grb131030fit}):}

The XRT observations for this burst start at 83~s after the BAT trigger and in total we have 20 Bayesian blocks. The light curve has a single, smooth flare, which rises sharply at the beginning of the observation, reaches a peak at approximately 120~s, and then declines. The blackbody is significant at $> 3 \sigma$ from the beginning of the observation until 175~s after the trigger. The temperature cools from $1.63 \pm 0.18$ ~keV to $0.58 \pm 0.06$~keV in 67~s with a decay index of $n=-2.56 \pm 1.00$ for an initial temperature of $T_{\rm{o}}= 5.5 \pm 2.1$~keV. The photon index softens from $0.94 \pm 0.50$ to $2.11 \pm 0.49$.  

\paragraph*{GRB 150727A (Figs. \ref{150727} and \ref{grb150727fit}):}

The XRT observations start at 83 s after the BAT trigger and continue until 600 s. The light curve is split into 11 Bayesian blocks and is smoothly decaying during the entire observation. The blackbody is significant at $ > 3 \sigma$ throughout the burst. The temperature cools from $0.49 \pm 0.08$ ~keV to $0.15 \pm 0.05$ ~keV in 517~s with a decay index of $n=-0.69 \pm 0.18$ for an initial temperature of $T_{\rm{o}} = 0.91 \pm 0.07$~keV. The photon index remains nearly constant at around $1.97 \pm 0.26$.

\paragraph*{GRB 151027A (Figs. \ref{151027} and \ref{grb151027fit}):}

The XRT observations of this GRB start at 94~s after the BAT trigger and we have in total 19 Bayesian blocks. The light curve has a single, smooth flare, which reaches a maximum at around 150~s. The blackbody becomes significant at $> 3 \sigma$  at 120~s (near the peak of the light curve) and stays significant until 171 s after the BAT trigger. The temperature cools from $1.47 \pm 0.15$ ~keV to $0.43 \pm 0.07$~keV in 51~s with a decay index of $n=-4.35 \pm 1.52$ for an initial temperature of $T_{\rm{o}} = 9.31 \pm 4.91$~keV. The photon index intially shows a strong evolution with no clear pattern, oscillating between -3.0--3.6, before levelling out at a value between $1.34 \pm 0.69$ and $1.86 \pm 0.70$. We point out that the extreme values of the photon index at the beginning are very poorly constrained.

\section{GRBs with previous detections of thermal components}
\label{prevdets}
For completeness we also analysed all GRBs that were previously reported  to have a thermal component during the XRT observations. This gives us a larger sample from which to draw conclusions about possible common properties and the origin of the blackbody emission.  These bursts were analysed using the same methods as above (Section \ref{analysis}). We do not show individual plots of the light curves and parameter evolutions for these bursts since they have been presented in previous publications, but do include the confirmed cases in Fig. \ref{overplot}, where we show the temperatures, luminosities and radii of the blackbodies for all GRBs with thermal components. The fit results are also included in Table \ref{fits}.  We confirm just a small number of the previously reported cases using our methods. As we discuss at the end of this section, this is mainly due to the fact that we employ a finer time resolution and stricter criteria for the significance. 

\subsection{Confirmed highly significant cases}

\paragraph*{GRB 060218}
We confirm the results reported by \cite{campana}. The XRT observations of this GRB start at 150~s after the BAT trigger and continue for 2700~s. The actual duration of the burst is, however, much longer than the WT XRT observation and the light curve continues up to 10~000~s. The light curve has a smooth bump, with a slow rise until approximately 1000~s, followed by a slow decay. The blackbody is significant at $> 3 \sigma $ during the entire duration of the XRT observations and the temperature is nearly constant at around $ 0.16 \pm 0.03$~keV. The photon index is nearly constant around $1.42 \pm 0.06$ during the first 1000~s and then softens from $1.6 \pm 0.02$ to $2.40 \pm 0.09$ in the last 1000~s of the light curve. Due to the temperature being nearly constant it is not possible to fit a cooling profile to the temperature of this GRB. 

\begin{table*}
    \centering
    \caption{Resulting best-fitting parameters for the absorbed power-law+blackbody model. The blackbody luminosity is the full luminosity, including parts of the blackbody that may fall out of the observed energy range. The full table, together with the best-fitting parameters for the absorbed power-law model, is available as online material.}
    \label{fits} 
    {\def\arraystretch{2}\tabcolsep=10pt

\begin{tabular}{c c c c c c c c}
\hline
\multicolumn{1}{p{2cm}}{\centering Time interval $\mathrm{(s)}$} & \multicolumn{1}{p{2cm}}{\centering $kT$  \\ $\mathrm{(keV)}$} & \multicolumn{1}{p{2cm}}{\centering $L_{\mathrm{BB, unabs}}$  \\ $\mathrm{(erg \ s^{-1})}$} & \centering $\Gamma$  &  \multicolumn{1}{p{2cm}}{\centering $F_{\mathrm{obs,0.3-10 keV}}$  $\times 10^{-9}$ \\ $\mathrm{(erg \ cm^{-2} \ s^{-1})}$} & \multicolumn{1}{p{2cm}}{\centering $F_{\mathrm{BB}}/ F_{\mathrm{tot}}$ \\ $(\%)$ } & $\chi^2/ \mathrm{dof}$\\
\hline

GRB 060218 & & & & & & \\
159 - 273 & $0.14_{-0.03}^{+0.03}$ &  $1.81_{-0.5}^{+0.5} \times 10^{45}$ &  $1.48_{-0.07}^{+0.08}$ &  $2.37_{-0.06}^{+0.06}$ & $64 \pm 6$ & 149.98/165 \\
273 - 339 &  $0.15_{-0.03}^{+0.03}$ &  $3.33_{-1.23}^{+1.23}\times 10^{45} $ &  $1.46_{-0.08}^{+0.09}$ &  $3.06_{-0.14}^{+0.13}$ &  $65 \pm 13$ & 127.89/129 \\
339 - 415 & $0.18_{-0.02}^{+0.02}$ &  $2.45_{-0.5}^{+0.5} \times 10^{45}$ &  $1.37_{-0.09}^{+0.08}$ &  $3.66_{-0.10}^{+0.09}$ &  $65 \pm 4$  & 158.28/165 \\
415 - 544 & $0.14_{-0.02}^{+0.02}$ &  $2.54_{-0.5}^{+0.5} \times 10^{45}$ &  $1.39_{-0.05}^{+0.05}$ &  $4.40_{-0.08}^{+0.07}$ &  $66 \pm 3$ & 309.08/299 \\
544 - 632 & $0.16_{-0.02}^{+0.02}$ &  $2.56_{-0.6}^{+0.6} \times 10^{45}$ &  $1.36_{-0.06}^{+0.06}$ &  $5.13_{-0.11}^{+0.09}$ &  $67 \pm 3$ & 220.23/249\\
632 - 820 & $0.16_{-0.01}^{+0.01}$ &  $2.84_{-0.0}^{+0.0} \times 10^{45}$ &  $1.44_{-0.04}^{+0.04}$ &  $5.51_{-0.06}^{+0.08}$ &  $77 \pm 1$ & 392.39/432\\
820 - 1298 & $0.14_{-0.01}^{+0.01}$ &  $3.20_{-0.4}^{+0.4} \times 10^{45}$ &  $1.61_{-0.02}^{+0.02}$ &  $5.50_{-0.04}^{+0.04}$ &  $69 \pm 2$  & 747.05/607\\
 \hline
\end{tabular}
}
\end{table*}

\paragraph*{GRB 090618}

Our results are compatible with the ones presented in \cite{page2011}. The XRT observations start at 125~s after the BAT trigger and last until 2400~s. The light curve shows a general decaying trend with two small flares at 135~s and 150~s. The blackbody is significant from the beginning of the XRT observations until 159~s after the BAT trigger and is significant during the two flares. The temperature cools from $1.16 \pm 0.09$ until $0.74 \pm 0.14$~keV with a decay index of $n=-2.42 \pm 1.38$ for an initial temperature of $T_{\rm{o}} = 5.25 \pm 4.11$~keV. The photon index softens from $0.01 \pm 0.84$ to $1.34 \pm 0.28$.  

\paragraph*{GRB 101219B}
 
We confirm the findings of \cite{starling2011} of the significant thermal component in this GRB. The XRT observations start at 150~s after the BAT trigger and last until 530~s. The light curve is decaying with a bump at around 300~s. The blackbody becomes significant at $> 3 \sigma$ at 225~s and stays significant until the end of the XRT observations. The temperature cools from $ 0.19 \pm 0.02$ to $0.14 \pm 0.03$~keV with a decay index of $n = -0.64 \pm 0.90$ for an initial temperature of $T_{\rm{o}}= 0.86 \pm 0.75$. The photon index stays constant around $1.48 \pm 0.23$. 

\subsection{Marginal cases and non-detections}

\paragraph*{GRB 100316D}
This GRB was reported by \cite{starling2011} to have a thermal component present in its afterglow, but a later study by \cite{margutti2013} with newly released calibration files showed that the statistical significance of the thermal component was substantially lower than previously reported. Here we also find that the blackbody is not statistically significant and therefore exclude this burst from further analysis.    

\paragraph*{GRB101225A}

The famous "Christmas burst" is an ultralong GRB. \cite{Thone2011} reported finding a blackbody component present in the XRT data. According to our criteria, this GRB is not classified as having a significant thermal component since the significance is not at the $> 3 \sigma$ level in 3 consecutive time-bins. However, we note that there are time-bins where it is significant at $> 3 \sigma$. 

\paragraph*{GRB 120422A}
This GRB was previously reported as a possible detection of a thermal component by \cite{starling2012}, who found a significance level at $> \ 4\sigma$. Using our method we did not find the blackbody to be significant through Monte Carlo simulations.

Apart from the bursts presented above, there are eleven additional GRBs for which  significant thermal components have been reported \citep{sparre,mette}. In our analysis of these bursts, we found that only three (GRB 060418, GRB 061110A, GRB 100621A) show some evidence for a blackbody. These bursts have a blackbody component that is significant at $> \ 3 \sigma$ in at least 1 time-bin,  but not in 3 consecutive ones as we require.

\subsection{Possible reasons for discrepancies with previous works}  
In those cases where we confirm the presence of a blackbody component, our best-fitting parameters are compatible with previous works. However, we do not confirm the majority of the previously reported cases using our approach. The main reasons for this are discussed below: 

\begin{itemize}

\item the time intervals used for the spectral analysis: while we are not excluding any time intervals, \cite{sparre} excluded time intervals with flares and assessed the presence of the blackbody towards the end of the WT observations. On the other hand, \cite{mette} selected time intervals with 10 000 counts in the WT observations. Depending on the data, those intervals can be very long and contain significant spectral evolution. 

\item the criteria for a detection: Our requirement that the blackbody should be significant at $> 3 \sigma$ in 3 consecutive time-bins is stricter than in previous studies.

\item the assessment of significance: while we used Monte Carlo simulation to assess the significance of the blackbody component, \cite{sparre} reported detections based on the F-test. Although the authors perfomed simulations based on GRB~101219B to motivate the use of the F-test, it remains as a possible reason for discrepancies for the other bursts.

\item the models used for the comparison: while in this work we are comparing an absorbed power-law model with an absorbed power-law plus a blackbody model, \cite{mette} compared a Band model with Band plus a blackbody model. 

\end{itemize}

We have excluded differences in $N_{\mathrm{H,intr}}$ as an explanation for not confirming cases reported in \cite{sparre}, since the values we obtained and the values reported in that paper are consistent within error bars. For the bursts studied in \cite{mette}, we cannot make such a comparison since $N_{\mathrm{H,intr}}$ is not reported.

\section{Discussion}
\label{discussion}
\subsection{Common properties of GRBs with thermal components} \label{common}

\begin{figure*}
    \centering
    \begin{subfigure}[b]{0.49\textwidth}
        \includegraphics[width=\textwidth]{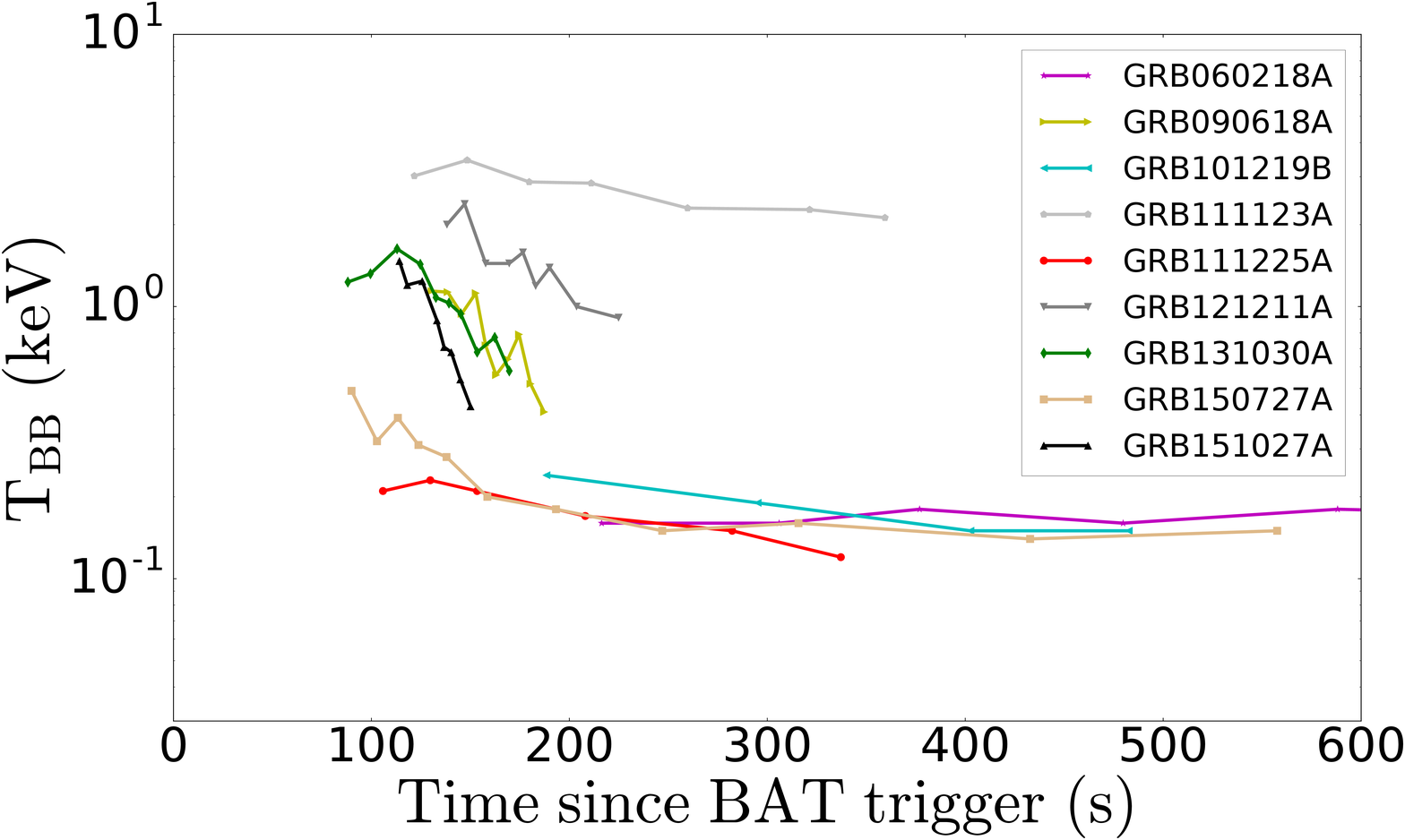}
    \end{subfigure}
        \begin{subfigure}[b]{0.49\textwidth}
        \includegraphics[width=\textwidth]{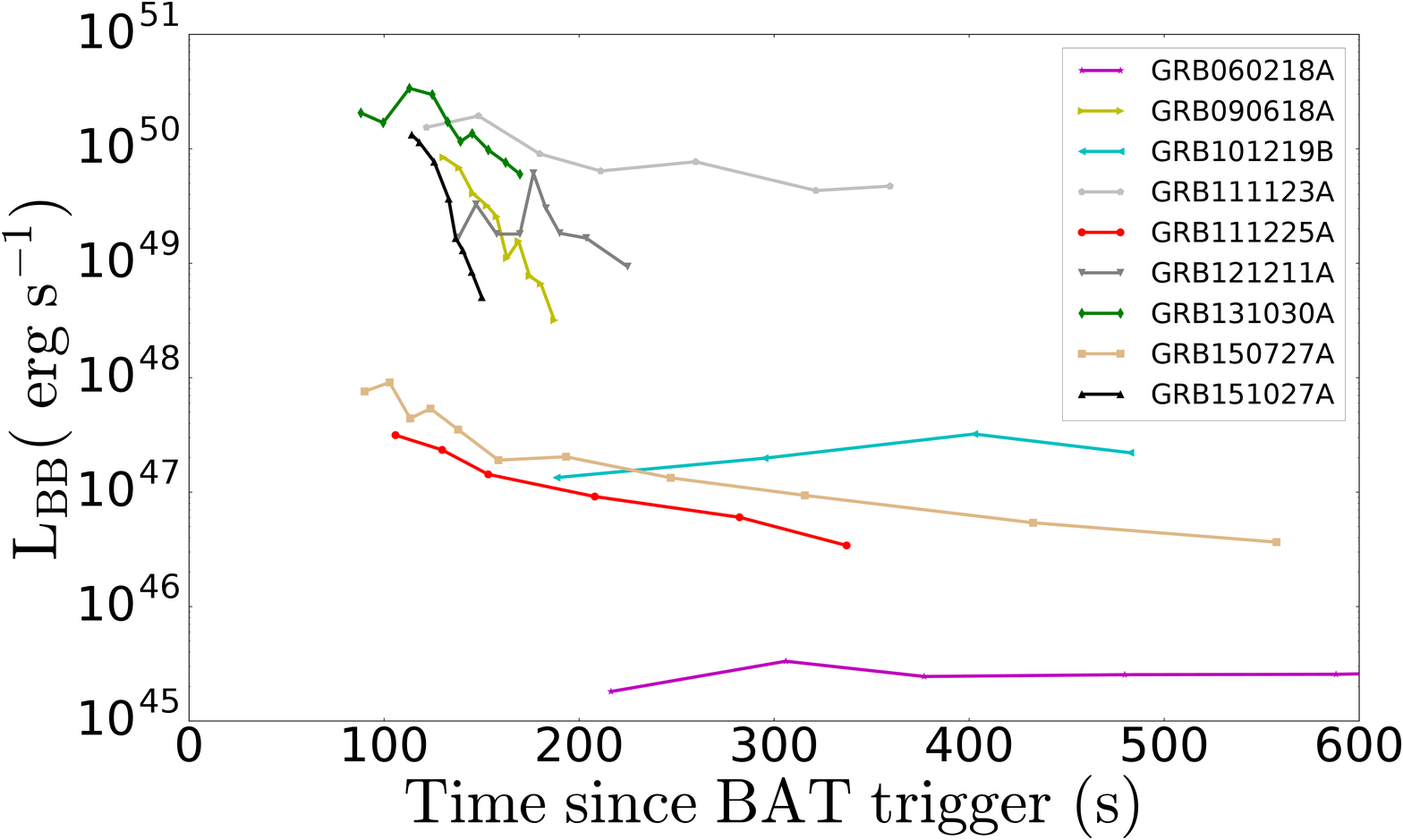}
    \end{subfigure}
     \begin{subfigure}[b]{0.49\textwidth}
        \includegraphics[width=\textwidth]{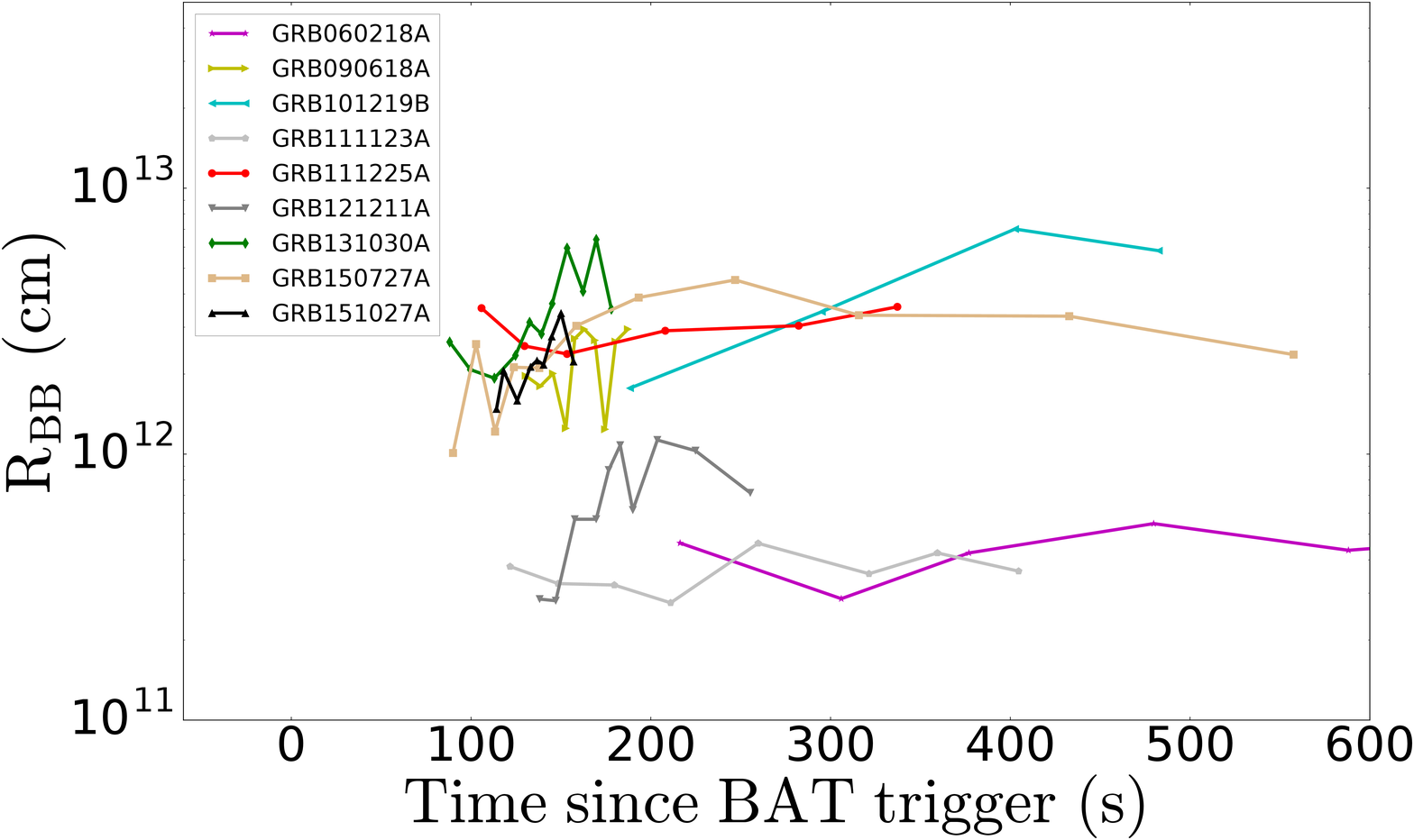}
    \end{subfigure}
        \begin{subfigure}[b]{0.49\textwidth}
        \includegraphics[width=\textwidth]{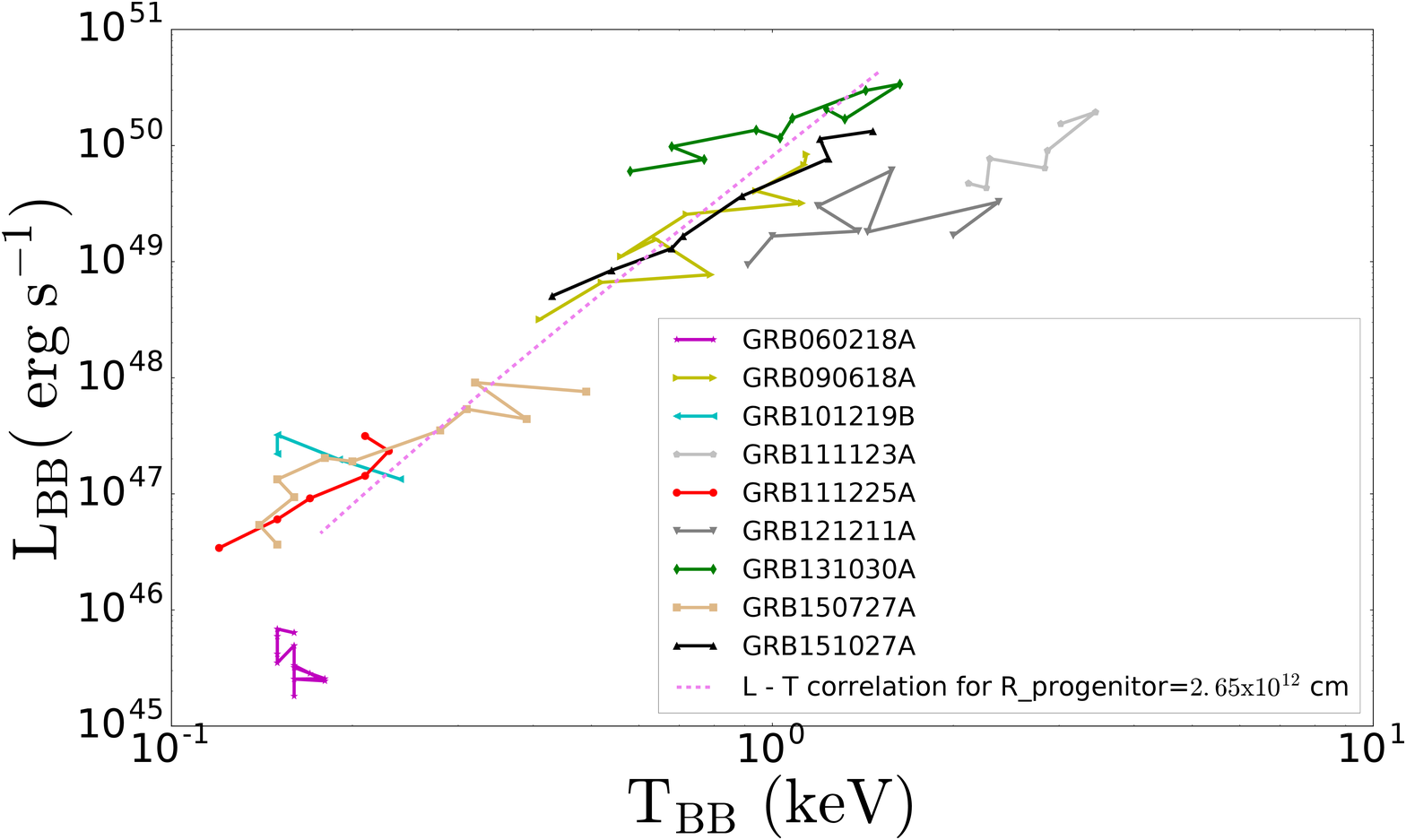}
    \end{subfigure}
    \caption{Properties of the blackbody parameters in GRBs with significant detections. Top left: time evolution of the temperature. Top right: time evolution of the luminosity. Bottom left: time evolution of the radius. Bottom right: the correlation between temperature and luminosity. Error bars are omitted from this figure for clarity. They are shown in Figs.~\ref{111123} - \ref{151027}.}
    \label{overplot}
\end{figure*}

To compare the properties of all the detected thermal components we plot the blackbody parameters together in Fig.~\ref{overplot}. The top panels show the time-evolution of the temperatures and luminosities. GRB~101219B, GRB~111125A and GRB~150727A all have temperatures that decay in the range $\sim 0.4-0.1$~keV, as well as slowly decaying or constant luminosities around $10^{47}\ \mathrm{erg\ s^{-1}}$. GRB~060218 has a similar temperature,  but a significantly lower luminosity of the order $10^{45}\ \mathrm{erg\ s^{-1}}$. On the other hand, GRB~090618, GRB~131030A and GRB~151027A all have higher temperatures that decay faster ($\sim 1.5-0.5$~keV, decaying as $t^{-n}$, with n~$\sim 2-4$). The peak luminosities of these bursts are in the range  $8 \times 10^{49}- 5 \times 10^{50}\mathrm{erg\ s^{-1}}$  and they decay in a similar manner ($t^{-n}$, n~$\sim 4-9$). GRB~111123A clearly stands out from the other bursts by having a high and slowly evolving temperature and luminosity ($T\sim 3$~keV and $L \sim 10^{50}\ \mathrm{erg\ s^{-1}}$). GRB~121211A also stands out in terms of a higher temperature at late times (1.7--0.7~keV), but its luminosity is not among the highest ($L_{\mathrm{peak}} \sim 6 \times 10^{49} \mathrm{erg\ s^{-1}})$. 

The bottom panels of Fig. \ref{overplot} show the relation between the temperatures and luminosities as well as the blackbody radii. The lower right panel shows that the majority of the bursts roughly fall along a $L\propto T^{4}$ correlation. For comparison, the relation $L=\sigma 4\pi R^2 T^4$, with $R=2.65 \times 10^{12}\ $cm is overplotted on the data (note that it is not a fit). This specific value of radius is the median value of the radius calculated not taking into accounts the outliers (GRB~060218A, GRB~111123A and GRB~121211A). The individual bursts show a somewhat flatter evolution than $L\propto T^{4}$, as expected if the radii are expanding. GRB~060218 and GRB~111123A, GRB~121211A are clear outliers to this relation at the low and high luminosity end, respectively. As expected from these results, the majority of GRBs occupy a fairly narrow range of blackbody radii $2 \times 10^{12} \mathrm{cm} \le R_{\mathrm{BB,av}} \le 5 \times 10^{12} \mathrm{cm}$ (lower left panel of Fig. \ref{overplot}). GRB~060218 and GRB~111123A have smaller radii around a few times $10^{11}$~cm, while GRB~121211A has an average radius of $7\times 10^{11}$~cm. In all GRBs, the radii are either staying approximately constant or increasing with time.

In Table \ref{thermalwithz} we present redshifts, $N_{\mathrm{H,Gal}}$, $N_{\mathrm{H,intr}}$ and the total energy of the blackbody for all the GRBs. The latter was calculated as the energy released by the blackbody during the time when it is significant  $> 3 \sigma$, and is thus a lower limit. The emitted energies are in the range $E_{\mathrm{BB}} \sim 10^{49}  - 10^{52}$~erg. We also calculated the light travel time from the average radius of each burst and compared it to the time for the flux of the blackbody to half (as a proxy for the time-scale of the light curve). The latter time scale could not be defined in GRB~060218, since we observe just a small part of the entire duration of the burst, GRB~111225A due to large error-bars and GRB~101219B since we do not observe a decay of the flux. We found that the two time-scales are mostly compatible. Light travel times are in general of the order $~100$ s ($61\  \mathrm{s} \le t_{\mathrm{light}} \le 150\  \mathrm{s}$) except for GRB~060218 and GRB~111123A where it is $t_{\mathrm{light}} < 20$ s. The time for flux to half is, for cases where this time-scale can be defined, also of the order $~100$ s ($107\  \mathrm{s} \le t_{\mathrm{f/2}} \le 200\  \mathrm{s}$), except for GRB~090618 where $t_{\mathrm{f/2}} \approx 24 \mathrm{s}$. Finally, we note that GRB~111123A and GRB~121211A show more erratic light curves, with lower amplitude variability on a shorter time-scale that is not captured by these half times.

\begin{table*}
    \centering
    \caption{Redshifts, galactic and intrinsic H column densities (derived by fitting an absorbed power-law + blackbody) for GRBs with thermal components. \\
    * We note that the energy of the blackbody for GRB~060218 is a lower limit from the XRT WT mode data only.}
    \label{thermalwithz}
    {\def\arraystretch{2}\tabcolsep=10pt
    \begin{tabular}{c c c c c} % four columns, alignment for each
        \hline
        GRB  & z  & \multicolumn{1}{p{2cm}}{\centering $N_{\mathrm{H,Gal}}$ \\ $\times \ 10^{22}$ \\ $\mathrm{(cm^{-2})}$} & \multicolumn{1}{p{2cm}}{\centering $N_{\mathrm{H,intr}}$ \\ $\times \ 10^{22}$ \\ $\mathrm{(cm^{-2})}$} & \multicolumn{1}{p{2cm}}{\centering $E_{\mathrm{BB,tot}}$ \\ $\mathrm{(erg)}$ }\\ 
        \hline
   
		060218A & 0.0331  & 0.142 & $0.60_{-0.02}^{+0.02}$ & $ 1.1 \times 10^{49 *} $ \\
        090618A & 0.54 & 0.076 & $0.28_{-0.03}^{+0.04}$ & $ 1.41 \times 10^{51}$ \\
        101219B & 0.5519 & 0.033 & $0.16_{-0.09}^{+0.11}$ & $ 4.56 \times 10^{49}$ \\ 
        111123A & 3.1516 & 0.069 &  $4.79_{-0.35}^{+0.73}$ & $ 6.68 \times 10^{51}$ \\
        111225A	& 0.297  & 0.275 & $ 0.17_{-0.07}^{+0.08}$ & $2.3 \times 10^{49}$ \\
        121211A	& 1.023  & 0.00949 & $ 0.58_{-0.17}^{+0.18} $ & $1.04 \times 10^{51}$ \\
        131030A & 1.295  & 0.0562  &  $0.42_{-0.10}^{+0.13}$ & $7.24 \times 10^{51}$ \\   
        150727A	& 0.313 & 0.0981 & $ 0.08 \pm 0.05$ & $ 5.94 \times 10^{49}$ \\
	    151027A	& 0.810 & 0.0375 & $0.33_{-0.05}^{+0.07}$ & $ 1.40 \times 10^{51}$ \\
        \hline 
    \end{tabular}
}
\end{table*}

We also investigated the power-law properties for the GRBs with thermal components. The average photon indices in these bursts are in the range $ 1.2 < \Gamma < 2.2$, which is fully consistent with the GRBs without thermal components. In some time bins, $\Gamma$ deviates from these values, and both very hard and soft values are seen. In most of these cases, the error bars on $\Gamma$ are very large. This is expected in time bins with limited statistics and a very large contribution from the blackbody. An example of this is seen in the first time bins of GRB~151027A  (see bottom left panel on Fig. \ref{151027}). In most bursts, the time-evolution of $\Gamma$ is either consistent with staying constant or showing some moderate softening with time. The strongest time-evolution  is seen in GRB~090618, GRB~11123A and GRB~121211A. In these bursts $\Gamma$ is very hard at the beginning ($\lesssim 0.5$) and then softens with time, though we note that the hard values at the beginning of GRB~090618 are poorly constrained.

Three GRBs out of the nine GRBs with significant thermal components also have an accompanying SN. These are GRB 060218 \citep{campana}, GRB 090618A \citep{cano2011} and GRB 101219B \citep{sparre2011}. GRB 151027A was reported to have a bump in the optical light curve, but it was not firmly concluded that this is due to a SN \citep{Nappo2016}.

\subsection{Systematic effects and uncertainties in the analysis} \label{nh}

When deriving $N_{\mathrm{H,intr}}$ we assumed that it will stay constant during the WT observation time. However, we allowed it to be different for the two models that we fit to  data in order to avoid biasing the detection of a blackbody (see Section \ref{analysis}). For each GRB we therefore have two different values of $N_{\mathrm{H,intr}}$, depending on the model. These values are compatible within error bars in the majority of cases. However, in twelve bursts the values differ significantly and in all these cases the values are high ($N_{\mathrm{H,intr}}> 1.5 \ \times \ 10^{22} \ \mathrm{cm^{-2}}$). Out of these twelve bursts only one had a significant detection of a blackbody  (GRB 111123A). \cite{sparre} showed that it is unlikely to detect a thermal component if $N_{\mathrm{H,intr}}$ is  of the order $0.4 \times 10^{22} \ \rm{cm}^{-2}$ or higher. While half of the GRBs with detections have $N_{\mathrm{H,intr}}$ close to or above this value (see Table \ref{thermalwithz}), we find that only one case, GRB~111123A, has $N_{\mathrm{H,intr}} > 1 \times 10^{22} \ \mathrm{cm^{-2}}$.

We have also tested for degeneracies between pairs of fit parameters using the "steppar" command in XSPEC. While  $kT$--$N_{\mathrm{H,intr}}$ do not show any systematic trends, we do find a general positive degeneracy between $N_{\mathrm{H,intr}}$--$\Gamma$. By determining $N_{\mathrm{H,intr}}$ from the joint fits to all the time intervals in a given burst and then keeping it fixed, we reduce the impact of this degeneracy. We also find some degeneracies between $\Gamma$-$kT$, but they are not consistent from time-bin to time-bin. When they are present they always show a positive correlation, which is contrary to the negative one observed for these parameters in the case of GRB~111123A (Fig. \ref{111123}) and GRB~121211A (Fig. \ref{121211}). Therefore, the anti-correlation seen between $\Gamma$-$kT$ for these two GRBs is not caused by degeneracies. 

In our fits we have kept the redshifts of the GRBs as fixed parameters. Two GRBs had errors reported on their redshifts: GRB~110422A at $z=1.77 \pm 0.001$ \citep{Ugarte2011} and GRB~130427B at $z=2.78 \pm 0.02$ \citep{Flores2013}. These error-bars are  small and do not significantly affect the results.

We also investigated if any of the bursts with detected blackbodies are affected by the redistribution uncertainties mentioned in Section \ref{datareduction}. We found that the 'turn up' at low energies is present in three GRBs: GRB~121211A, GRB~131030A and GRB151027A (see Table \ref{RMFs}). For these cases we performed the fits with the best position-dependent RMF as described in Section \ref{datareduction}. We note that this did not significantly alter the results. As a precaution, we also performed the fits ignoring the data below 0.5~keV. This changed the total observed fluxes of the models by a few per cent, but not the values of the photon indices or blackbody temperatures.

In our analysis we have not investigated if the spectra can be fitted by other (combinations of) models than a power law + blackbody.  As a result we cannot exclude other possible explanations for the spectra. Indeed, cases with atypical behaviour of the photon index could indicate that the power law + blackbody may be mimicking another spectral shape. However, the fact that the majority of bursts with significant detections fall close to an $L \propto T^4$ relation for a single radius  (Fig. \ref{overplot}) is a strong argument in favour of a blackbody component. We also stress that the our main conclusions regarding the significance and properties of the blackbodies do not change if we exclude time intervals where $\Gamma$ is unusually hard or soft.

A related point is that most theoretical models for the origin of this component predict a multicolour blackbody rather than a pure blackbody (Section \ref{origin}). We have not performed such fits, but we note that the residuals of the power law + blackbody fits do not indicate that there is any additional spectral curvature not accounted for. In addition, spectra with very good statistics would be needed in order to discriminate between a pure blackbody and a slightly broadened blackbody.  

Some of the previous systematic searches for thermal components in the early X-ray afterglows excluded flaring periods from the analysis \citep{sparre,starling2012}. We did not make any such cuts (see Section \ref{analysis}) and found that four of the GRBs with blackbody components have light curves that exhibit flares (GRB~111123A, GRB~121211A, GRB~131030A and GRB~151027A). The former two have erratic light curves with multiple flares, while the latter two are dominated by single, smooth pulses. GRB~111123A, GRB~131030A and GRB~151027A are consistent with the blackbody dominating the luminosity of the flares, although with large uncertaintes for GRB~151027A, which has a poorly constrained photon index in many bins (see Table \ref{fits}). On the other hand, in GRB~121211A the power-law comonent dominates in most time bins. The fact that a significant blackbody component is seen during a flare that is dominated by a power-law indicates that both components originate from the same region (likely the jet, see Section \ref{thermal_prompt}). Alternativley, as already discssed above, the power-law+blackbody model may be mimicking another spectral shape.

\subsection{Properties of the prompt emission}
\label{prompt}
Five GRBs out of the nine with significant thermal components also have {\it Fermi} GBM data during the prompt phase (GRB~090618, GRB~101219B, GRB~121211A, GRB~150727A and GRB~151027A). With its wide energy range (8~keV -- 40~MeV), {\it Fermi} GBM gives a much better view of the prompt emission than {\it Swift} BAT (which covers 15--150~keV). We analysed the GBM data to address the question of whether the bursts with detected thermal components also share some common properties during the prompt phase. Bayesian blocks were used to define the bins for our time-resolved analysis and all spectra were fitted with a Band function (a smoothly broken power law with power-law indices $\alpha$ and $\beta$ below and above the peak energy, respectively\footnote{Note that the definition of these indices differ by a minus-sign compared to the photon index ($\Gamma$) of the power law fits to the XRT data.}). We find no systematic behaviour of the best-fitting parameters among these bursts. GRB~101219B exhibits hard values of $\alpha$,  with $ \alpha \sim 0.3$ and $\alpha \sim 0.8$, whereas the others are found to have more typical values as compared to the full population GRBs detected by GBM \citep{Yu:2016bn}. The bursts are also not exceptional in duration or total energy, with $5 < t_{90} < 124$ s, and isotropic equivalent energies $3\cdot 10^{50}  \le E_{\mathrm{iso}} \le 8\cdot 10^{52}$ erg s$^{-1}$. Finally, there is no particular similarity or systematic behaviour in their light curves. Hence, apart from GRB~101219B, which is known to have a hard spectrum that is close to a blackbody \citep{josefin}, the bursts appear to be typical in terms of their prompt emission.

Additionally, we checked for correlations in the overlap region between the prompt and afterglow phases. Only GRB~090618 and GRB~151027A have such overlapping regions.
In these regions we used the time bins defined by the XRT analysis and compared the values of $\alpha$ from the GBM fits to those of $\Gamma$ from the XRT fits. These values should be consistent if the power law at soft X-rays is a simple extrapolation of the prompt gamma-ray spectrum. However, we find that we cannot draw any conclusions regarding this due to the limited statistics in many of the time bins. Finally, we compared the total energy of the prompt emission, $E_{\mathrm{iso}}$, to that of the blackbody component of the afterglow, $E_{\mathrm{BB}}$, and found no correlation. However, if GRB~121211A is excluded due to likely being of a different origin than the others (see discussion in section \ref{origin}) there is a positive trend between $E_{\mathrm{iso}}$ - $E_{\mathrm{BB}}$.  A significantly larger sample would be needed to draw any firm conclusions about correlations between the energetics in the two phases. 

\subsection{Origin of the thermal components}
\label{origin}
Three possible origins have been proposed for the blackbody components:  late-time prompt emission from the jet, SN shock breakout, and emission from the cocoon surrounding the jet. Below we discuss how the predictions for each of these scenarios compare with our observations and also consider the resulting implications for the progenitor systems. In all scenarios the underlying power-law component is assumed to originate from some combination of late prompt emission and the onset of the afterglow from the interaction with the CSM (\citealt{Obrien2006}) (however see caveats in Section \ref{nh}).

Our observations show that six of the nine bursts in the sample have a small range of radii ($2 \times 10^{12}\  \mathrm{cm} \le R_{\mathrm{BB,av}} \le 5 \times 10^{12}\  \mathrm{cm}$) as well as smooth light curves characterized by either a simple decay or a single pulse. The small range of radii is especially noteworthy given that the bursts span a large range of peak luminosities ($\sim 10^{47} - 10^{51}\ \mathrm{erg\ s^{-1}}$). The bursts at the low-luminosity end of the distribution decay more slowly than the others, and we assume that their light curves peaked before the start of the observations. Out of the three GRBs that do not fit entirely into the above picture, GRB 060218 is a well-known low-luminosity GRB, while  GRB~111123A and GRB~121211A show more erratic light curves, as well as best-fitting parameters that differ somewhat from the other bursts, with higher temperatures at late times and smaller radii.

\subsubsection{Prompt emission}
\label{thermal_prompt}

The early X-ray light curves of GRBs commonly display flares \citep{Chincarini2007,Chincarini2010}. The spectra of the flares are often well fitted by simple power laws, but a subset require more complex models, such as a Band function or a power-law + blackbody \citep{Falcone2007,Peng2014}. Studies of the spectral evolution during flares reveal a hard-to-soft evolution, similar to the prompt emission phase (e.g., \citealt{Margutti2010}). Indeed, the properties of the flares point to them being due to late, weak central engine activity and hence connected to the prompt emission.

Two of the bursts in our sample stand out as likely being due to flaring prompt emission: GRB~111123A and GRB~121211A. Compared to the other bursts, they exhibit more irregular light curves with rapid variability (see Figs. \ref{111123} and \ref{121211}). They also deviate in terms of the evolution of $L_{\mathrm{BB}}$ and $T$, which have higher values at late times, and by having smaller radii (see Fig. \ref{overplot}). The photon indices also show a particularly strong evolution in these two bursts. Within the interpretation of prompt emission, both the blackbody and the power law likely originate from the jet. This is analogous to gamma-ray observations of the prompt phase where blackbody components have been observed together with non-thermal components (e.g., \citealt{Guiriec2011, Axelsson2012}). GRB~111123A and GRB~121211A were also included in the study of bright X-ray flares by \cite{Peng2014}, who found them to be well fitted by a Band model. Their results are not directly comparable to ours since they are based on time-averaged fits to the BAT+XRT spectra during the entire flare.

An origin from the jet is possible also in the case of smooth light curves, since a small fraction of GRBs exhibit single-peaked light curves with blackbody spectra at gamma-ray energies during the prompt phase \citep{Ryde2004}. A jet origin for the blackbody emission has also been suggested for GRB~060218 \citep{Irwin2016} and GRB~151027A \citep{Nappo2016}, both of which are in our sample. In the case of GRB~151027A, \cite{Nappo2016} suggest a re-born fireball scenario, whereby  dissipation of energy that occurs when the ejecta interacts with slower material causes a re-acceleration of the fireball. Such a scenario is in principle a possibility for the rest of the bursts in our sample. However, the fact that GRB~151027A together with GRBs 090618, 101219B, 111125A, 131030A and 150727A, have a narrow range of radii, point to an origin that is connected to a characteristic radius of the progenitor, such as SN shock breakout or cocoon emission.

\subsubsection{Shock breakout}

Shock-breakout in a supernova occurs when photons from the expanding shock wave start escaping from the star (see \cite{Waxman2016} for a recent review). This produces a flash of UV/X-ray radiation with a blackbody spectrum, lasting seconds to tens of minutes depending on the progenitor. The cooling envelope subsequently produces optical emission on a longer time-scale. The initial breakout phase is hard to observe and only a few cases have been reported (\citealt{Soderberg2008,Schawinski2008,Garnavich2016}). The thermal emission in GRB 060218 has also been attributed to shock-breakout (\citealt{campana}, although the large released energy and the long time scale present problems for this interpretation \citep{ghisellini2007b,ghisellini2007}.  

A typical WR star, which is assumed to be the progenitor of long GRBs, has a radius of about $10^{11}$~cm (e.g., \citealt{Sander2012}). The shock-breakout from such a star is expected to occur on a time-scale corresponding to $R/c$, which is of the order of a few seconds. The expected energy is $E_{\mathrm{bo}} \sim 10^{44}$~erg and the spectrum is expected to peak at keV energies (\citealt{Waxman2016}). The blackbody components found in our sample do peak in the X-rays, but clearly have significantly longer durations, larger radii and higher luminosities than expected for shock-breakout. The observed radii and durations may still be compatible with shock-breakout through winds. Indeed, there is growing evidence that stars undergo strong mass-loss before exploding as supernovae (e.g., \citealt{GalYam2014,Ofek2014}). In addition, asymmetries in the explosions can affect the time-scales and luminosities. However, the observed energies in the blackbody components are many orders of magnitude too high ($\sim 10^{49} - 10^{52}$~erg, see Table \ref{thermalwithz}), which rules out this scenario as a possible explanation.

\subsubsection{Cocoon emission} \label{cocoon}

The interaction with stellar material when a jet propagates through a star results in a hot cocoon surrounding the jet (\citealt{Ramirez2002}). This has been investigated through hydrodynamical simulations by several authors (e.g., \citealt{Zhang2003, Morsony2007, Bromberg2016}). The cocoon is mildly relativistic and starts expanding spherically when it emerges from the star together with the jet. When it breaks out it produces a signal that is similar to SN shock breakout, but more energetic. 

\cite{Suzuki2013} and \cite{DeColle2017} have calculated  light curves and spectra expected from the cocoon by post-processing hydrodynamical simulations. While \cite{Suzuki2013} focus on varying the properties of the wind surrounding the progenitor, \cite{DeColle2017} consider two different progenitors. The details of the predictions vary with the progenitor and CSM properties, but the basic features of the light curves and spectra are similar. The light curves have a rapid rise, reaching a peak luminosity of $\sim 10^{47}\ \mathrm{erg\ s^{-1}}$, followed by a slower decay on a time-scale of $\sim 100\ $s of seconds. The spectra are quasi-thermal, peaking in the soft X-rays. This is broadly consistent with the observed signal for GRB~101219B, GRB~111125A and GRB~150727A, although we note that we only catch these bursts during the decaying phase. \cite{starling2012} also showed that the general properties of the light curve and spectra of GRB 101219B can be produced in the cocoon model by \cite{Peer2006}, where an analytical treatment of the cocoon dynamics was combined with numerical modelling of the radiation.

Compared to these bursts, GRB~090618, GRB~131030A and GRB~151027A are significantly brighter, with peak luminosities in the range $10^{50} - 10^{51}\ $erg. Given that all six bursts share some common properties, we speculate that the latter three may still be due to cocoon emission, but in a scenario with a more energetic jet. The fact that the more energetic blackbodies have higher energies also during the prompt phase (Section \ref{prompt}) supports this picture. We also note that \cite{cocoon} predict a high-luminosity signal ($L_{\mathrm{peak}} \sim 10^{51}\ \mathrm{erg\ s^{-1}}$) from part of the cocoon in their analytical calculations, but in their model the spectrum peaks at gamma-rays and the signal is very short. No emission similar to our observations is predicted. However, these results are strongly dependent on the amount of mixing between the different parts of the cocoon. We note that all energies discussed above are isotropic equivalents. If (part of) the cocoon is beamed and collimated near the jet axis, the actual energy emitted by the cocoon is reduced (see further \citealt{cocoon}).

If the scenario of the cocoon emission is correct, it has some important implications for our understanding of GRB progenitors. In particular, the large radii imply that the cocoon breaks out from a thick wind rather than the surface of the star, while the narrow range of radii point to similar progenitors for all GRBs.

\subsection{Rates and detectability} \label{timav}
\begin{figure}
	\includegraphics[width=\columnwidth]{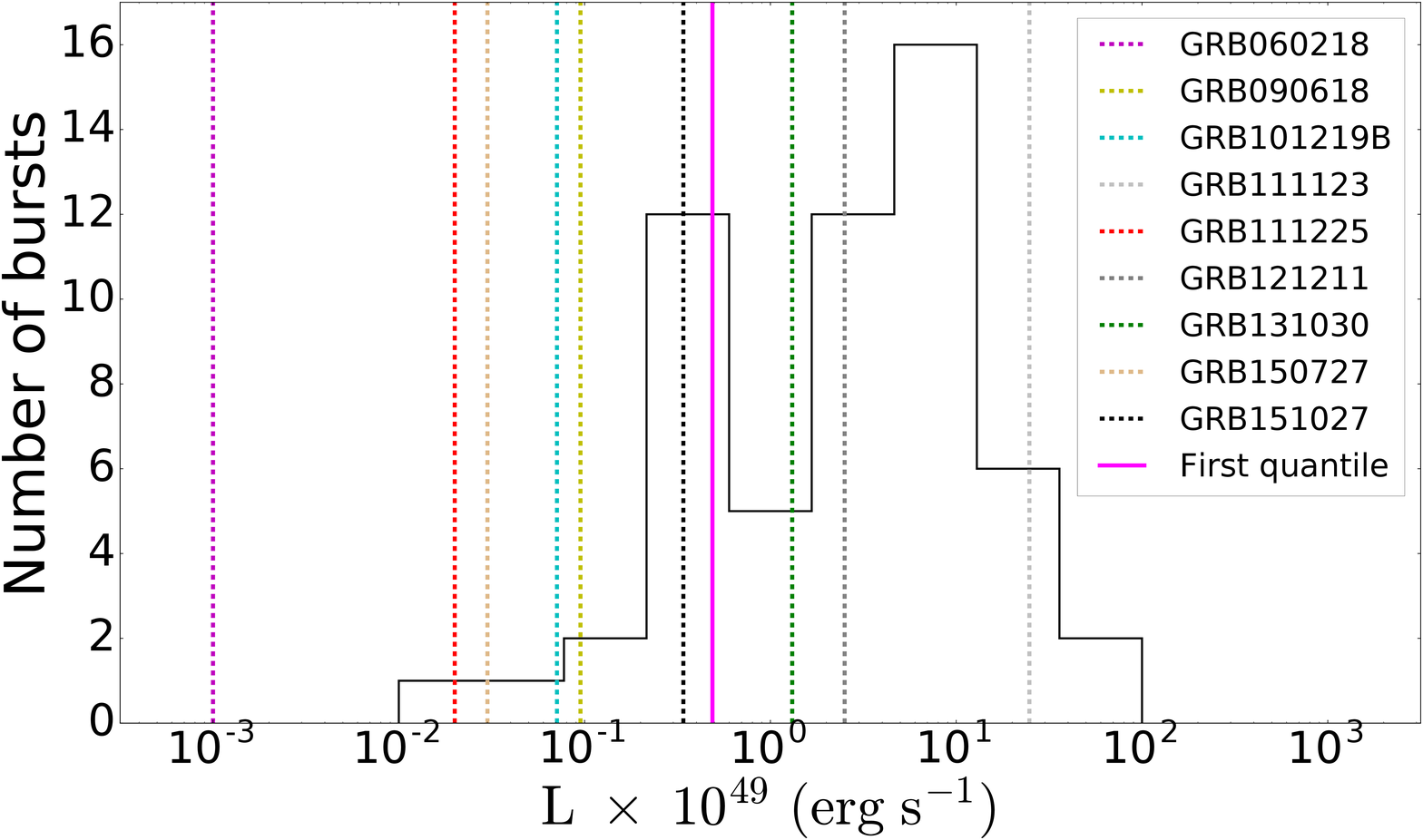}
	\caption{Histogram of average X-ray luminosities for the sample of GRBs. The luminosities of GRBs with detected thermal components are indicated by dashed lines. The solid fuchsia line represents the first quantile.}
	\label{lumin}
\end{figure}

The detection of six GRBs with thermal components in our sample of five years of data correspond to a detection rate of 8~per~cent. If we exclude  GRB~111123A and GRB~121211A on the basis that they are likely due to flaring prompt emission and only consider the four GRBs that we interpret as emission from a cocoon (see Section \ref{cocoon}), the detection rate is instead  5~per~cent. In order to understand the conditions that lead to the detection of a blackbody, it is instructive to look at the values of $F_{\mathrm{BB,\ 0.3-10\ keV}}/F_{\mathrm{tot,\ 0.3-10\ keV}}$. We find that the minimum value of this ratio when the blackbody is significant is $\sim$ 13~per~cent, and that it can reach as high as $\sim$ 98~per~cent. In  line with this, the majority of the GRBs for which we detect a blackbody are in the lowest 25~per~cent of the luminosity distribution of all the GRBs in our sample (see Fig. \ref{lumin}). This result implies that the afterglow component is intrinsically faint in those GRBs where a blackbody can be detected. In addition, all our detections (with the exception of GRB~111123A) have low redshifts ($z \lesssim 1$) and intrinsic column densities $N_{\mathrm{H,intr}} <  1 \ \times \ 10^{22} \ \mathrm{cm^{-2}}$, as presented in Table \ref{thermalwithz}. 

It is interesting to explore differences between the significance of the blackbody in the time-averaged and time-resolved spectra. In the case of GRB prompt emission it was shown by \cite{Burgess2015} that spectral evolution can introduce an artificial blackbody in the time-averaged spectra. While we are considering a different energy range and light curve phase, a similar situation may occur here.

In our sample there are 11 GRBs in which the thermal component is significant in the time-averaged spectrum, but not in the time-resolved analysis (e.g. GRB 110808A, GRB 130514A and GRB 151021A). However, we also have three cases where the opposite is true: GRB 121211A, GRB 111123A and GRB 151027A. This shows that spectral evolution may also remove the signature of a blackbody in the time-averaged spectrum. This work adds to extensive literature that shows the importance of time-resolved spectral analysis for GRBs.

\subsection{Ultra-long GRBS} \label{pop}
There are four ultra-long GRBs included in our sample: GRB 111209A, GRB 121027A, GRB 130925A and GRB 141121A. GRB 111209A is connected to a SN, SN 2011kl, which is an unusual superluminous SN \citep{greiner2015}. GRB 121027A has a somewhat peculiar afterglow with a bump in the  X-ray light curve that lasts more than $10^4\  \mathrm{s}$ \citep{wu2013}. In the case of GRB 130925A, there are disputes about whether a thermal component is present in the afterglow (\citealt{piro,evans2014,basak2015}). GRB 141121A was extensively studied by \cite{cucchiara2015}, where they proposed that a double jet is needed in order to explain the observed light curve. In our study of previously reported detections (see Section \ref{prevdets}), we also analysed the  ultra-long "Christmas burst" GRB 101225A  \citep{Thone2011}. The observational properties of this GRB (thermal component, long duration and lack of a standard afterglow) have been interpreted in terms of a merger of a neutron star and the He core of a massive star \citep{Thone2011, Cuesta2015a,Cuesta2015b}. In this scenario the thermal emission arises as a result of the jet interacting with the ejected envelope of the massive star.

In the majority of these GRBs, the spectra show clear deviations from a pure power law. The addition of a blackbody component generally improves the fit, but the significances do not reach our conditions for a detection (see Section \ref{withthermal}). However, GRB 101225A, GRB 111209A and GRB 130925A come close, as they have a significance level  $>3 \sigma$ in two consecutive time-bins. We also note that in our finely time-resolved analysis, these bursts either have poorly constrained best-fitting parameters with irregular time-evolution, or, as in the case of GRB~130925A, an unusual behaviour where the blackbody flux follows the two peaks in the light curve. For these reasons, the only firm conclusion that we can draw regarding these bursts is that their spectra are usually more complex than single power laws.
In general, this population of GRBs show a diverse range of properties and their origin is not well understood.

\section{Conclusions}

We have presented an analysis of \textit{Swift} XRT observations of 74 long GRBs observed between the beginning of 2011 and the end of 2015. These observations probe the soft X-ray (0.3--10~keV) emission starting approximately 100 seconds after the triggers. With the aim of identifying thermal components in this emission we have performed a time-resolved spectral analysis, addressing the question of whether an additional blackbody component is significant compared to a simple absorbed power law. We detect six GRBs with significant thermal components: GRB~111123A, GRB~111225A, GRB~121211A, GRB~131030A, GRB~150727A and GRB~151027A, where the latter one has previously been reported by \cite{Nappo2016}. We also investigated all cases that had previously been reported in the literature and find that three are classified as significant in our time-resolved analysis: GRB~060218, GRB~090618 and GRB~101219B. 

 In the total sample of nine bursts, we find some clear common behaviour. In particular, six of the bursts have a small range of blackbody radii ($2 \times 10^{12}\ \mathrm{cm} \le R_{\mathrm{BB,av}} \le 5 \times 10^{12}\ \mathrm{cm}$) as well as smooth light curves characterized by either a simple decay or a single pulse. The narrow range of radii is especially noteworthy given that the bursts span a large range of peak luminosities ($\sim 10^{47} - 10^{51}\ \mathrm{erg\ s^{-1}}$), and points to an origin connected to a characteristic radius of the progenitors. We suggest that the observations may be explained by a jet cocoon breaking out from a thick wind that surrounds the progenitor. An explanation in terms of SN shock breakout is ruled out by the high observed luminosities. The three bursts that deviate from this picture are GRB~060218, GRB~111123A and GRB~121211A. The first is a well-known low-luminosity GRB with unusual properties, while the latter two have more irregular light curves with flares, higher temperatures at late times as well as smaller inferred radii. The emission from these two bursts is most likely due to late prompt emission from the jet itself. 
 
The number of thermal components identified in our sample of 4 years of data correspond to a detection rate of  8~per~cent (or 5~per~cent if the cases that are likely due to late prompt flares are excluded). As previously noted by \cite{sparre}, we find that thermal components are preferentially detected at relatively low redshift (z $\lesssim 1$) and low absorption ($\le 1 \ \times \ 10^{22} \ \mathrm{cm^{-2}}$). We also find that the GRBs with detections have low X-ray luminosities compared to the sample as a whole. This suggests that thermal components may in fact be very common, but are often hidden by bright afterglows. Although our results point to an intriguing connection with the progenitors, it should be emphasized that the total sample of nine bursts is small. It will be critical to examine if the different "populations" and narrow range of blackbody radii that we observe persist when more data is added. On the theoretical side, it remains to be seen if the highest luminosities that we observe can be explained by emission from a cocoon.

\section*{Acknowledgements}

We thank M. Friis for helpful discussions. This work was supported by the Knut \& Alice Wallenberg Foundation and the G\"{o}ran Gustfasson Foundation. This work made use of data supplied by the UK Swift Science Data Centre at the University of Leicester.

%%%%%%%%%%%%%%%%%%%% REFERENCES %%%%%%%%%%%%%%%%%	%
\bibliographystyle{mnras}
\bibliography{thermal_component.bib} 

%%%%%%%%%%%%%%%%%%%%%%%%%%%%%%%%%%%%%%%%%%%%%%%%%%
\appendix

\section{Pile-up correction}
\label{pileupapp}
 Our analysis is primarily time-resolved and therefore all time-resolved spectra were individually tested for pile-up. In Table~\ref{pile-up} we provide the complete list of GRBs and the time intervals in which they were piled-up, together with the number of pixels that were excluded from the source to correct for the pile-up. 

\begin{table*}
    \centering
    \caption{List of GRBs with time intervals in which they exhibited pile-up, together with the number of pixels excluded from the centre of the PSF in order to correct for the pile-up effect. The full table is available as online material.}
    \label{pile-up} 
    {\def\arraystretch{1}\tabcolsep=8pt

\begin{tabular}{cc|cc|cc}
\hline
\multicolumn{1}{p{2cm}}{\centering Time interval $\mathrm{(s)}$}  & \multicolumn{1}{p{2cm}}{\centering Number of \\ excluded \\ pixels} & \multicolumn{1}{p{2cm}}{\centering Time interval $\mathrm{(s)}$}  & \multicolumn{1}{p{2cm}}{\centering Number of \\ excluded \\ pixels} & \multicolumn{1}{p{2cm}}{\centering Time interval $\mathrm{(s)}$}  & \multicolumn{1}{p{2cm}}{\centering Number of \\ excluded \\ pixels}\\
\hline
\\
{\bf GRB~110709B} & & {\bf GRB~120922A} & & {\bf GRB~130907A} & \\
592 - 622 & 1  & 121 - 124 & 7 & 476 - 511 & 6 \\
622 - 637 & 1  & 124 - 135 & 8 & 511 - 535 & 5 \\
637 - 647 & 2  & 135 - 150 & 7 & 535 - 577 & 5 \\
647 - 655 & 5  & 150 - 161 & 6 & 577 - 614 & 5 \\
655 - 660 & 4  & 161 - 167 & 5 & 614 - 738 & 5 \\
660 - 678 & 3  & 167 - 179 & 4 & 738 - 765 & 4 \\
678 - 703 & 2  & 179 - 206 & 3 & 765 - 896 & 4 \\
703 - 712 & 2  & 206 - 221 & 1 & 896 - 927 & 3 \\

\hline
\end{tabular}
}
\end{table*}

\section{RMF issues at low energies}
\label{rmfissues}
\begin{figure}
	\includegraphics[width=\columnwidth]{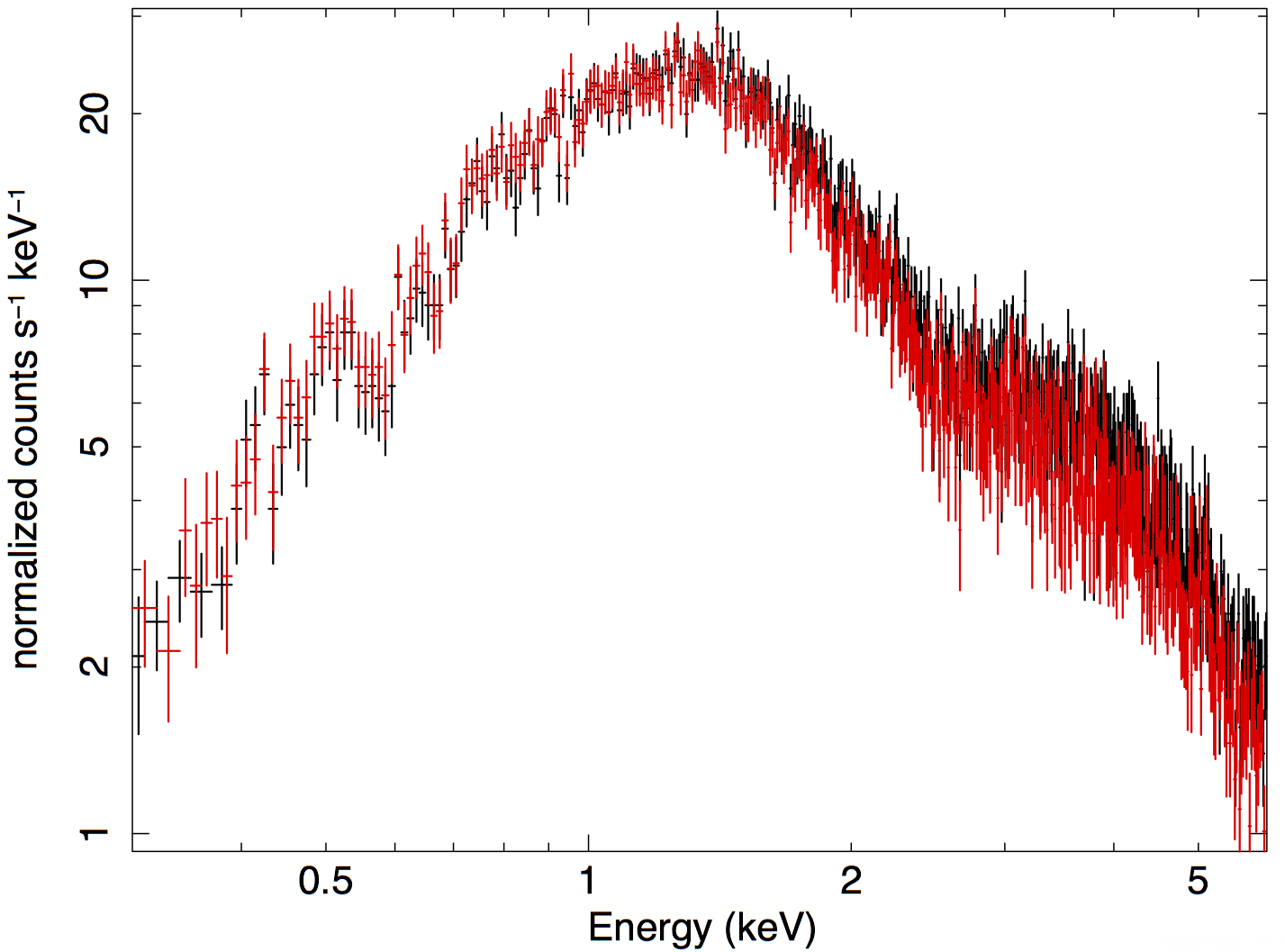}
	\caption{Time-averaged spectra of GRB~111123A produced using Grade 0 only (red) and Grade 0-2 events (black). The data at low energies do not differ significantly for the two spectra.}
	\label{nobump}
\end{figure}

As described in Section \ref{datareduction}, XRT sources that are moderately to heavily absorbed are prone to certain redistribution issues (see "XRT calibration digest"\footnote{\url{http://www.swift.ac.uk/analysis/xrt/digest_cal.php}}). One of the possible problems is the appearance of a bump at low energies and it is advised to always check if the spectra extracted using only Grade 0 and Grade 0-2 differ. If the spectra extracted with Grade 0-2 show an excess at low energies then Grade 0 spectra must be used. In Fig.~\ref{nobump} we show the time-averaged spectra extracted using Grade 0 and Grade 0-2 for GRB~111123A, which is one of the most heavily absorbed sources in our sample. In the entire sample presented in Section~\ref{sample} we did not have any sources that showed the presence of a bump at low energies. This was also confirmed by fitting the spectra. The best-fitting parameters obtained using Grade 0-2 and Grade 0 only were consistent within the $1 \sigma$ error bars in all cases.

On the other hand, 12 GRBs showed a turn-up below $\sim 0.6$~keV in their spectra. These bursts were fitted with position-dependent RMFs as described in Section \ref{datareduction}. Table \ref{RMFs} presents these 12 GRBs as well as the $\chi^2$ value for each of the RMFs tested. The test was done using the time-averaged spectra and the absorbed power-law model. In all further analysis, the RMF that gave the smallest value of $\chi^2$ was used. This value is shown in bold in the table. It is also worth noting that some of the fits have very high $\chi^2$ per degree of freedom which could imply that the power-law model is not sufficient in fitting these time-averaged spectra due to spectral evolution which is occurring.

\begin{table*}
    \centering
    \caption{$\chi^2$ values for the absorbed power-law model fits to the spectra of the 12 GRBs using different RMFs. $\chi^2_{\rm{ps1}}$ stands for the value of $\chi^2$ for the RMF with psf1, $\chi^2_{\rm{ps2}}$ for the RMF with psf2, etc., while $\chi^2_{\rm{std}}$ stands for the $\chi^2$ obtained using the standard RMF, while dof represents the number of degrees of freedom. Values in bold indicate the RMFs that were used in the analysis.}
    \label{RMFs} 
    {\def\arraystretch{2}\tabcolsep=10pt

\begin{tabular}{c c c c c c}
\hline
GRB & $\chi^2_{\rm{psf1}}$ & $\chi^2_{\rm{psf2}}$ & $\chi^2_{\rm{psf3}}$ & $\chi^2_{\rm{std}}$ & dof\\
\hline

111123A & 645.51 & 644.68 & 659.49 & {\bf 638.30} & 544 \\
121128A & {\bf 69.28} & 69.73 & 69.86 & 69.51 & 71\\
121211A & 504.95 & 527.53 & {\bf 465.74} & 498.00 & 373 \\
130514A & {\bf 929.84} & 944.10 & 952.91 & 937.36 & 454\\
130907A & {\bf 3785.15} & 3901.28 & 4007.38 & 3786.16 & 872 \\
130925A & {\bf 1491.14} & 1826.35 & 12132.11 & 1716.79 & 532 \\
131030A & 1574.84 & 1570.85 & {\bf 1569.46} & 1570.19 & 682 \\
140114A & {\bf 994.80} & 1016.15 & 1054.78 & 996.49 & 532 \\
140430A & {\bf 525.65} & 567.44 & 587.77 & 556.74 & 348\\
141026A & {\bf 319.86} & 328.54 & 333.08 & 324.88 & 185\\
151021A & 791.38 & 796.58 & {\bf 763.86} & 769.16 & 412\\
151027A & {\bf 1066.26} & 1136.07 & 1172.26 & 1169.49 & 592 \\

\hline
\end{tabular}
}
\end{table*}

% Don't change these lines
\bsp	% typesetting comment
\label{lastpage}
\end{document}